\documentclass[12pt,a4paper]{article}
\usepackage[utf8]{inputenc}
\usepackage[T2A]{fontenc}
\usepackage[hyperfootnotes=false]{hyperref}
\usepackage{xcolor}
\usepackage{babel}
\usepackage{amsmath}
\usepackage{amsfonts}
\usepackage{amssymb}
\usepackage{graphicx}
\usepackage{changes}
\usepackage{comment}
\usepackage{verbatim}
\usepackage{lineno}
\usepackage{authblk}
\usepackage{orcidlink}
\usepackage{enumitem}
\usepackage[small,sc,center]{titlesec}
\usepackage{indentfirst}
\usepackage[labelsep=period,labelfont=bf,small]{caption}
\usepackage{setspace}
\def\actaa{Acta. Astron.}
\def\aj{AJ}
\def\apj{Astrophys. J.}

\def\apjs{Astrophys. J. Suppl. Ser.}
\def\apss{Astrophys. Space Sci.}
\def\aap{Astron. Astrophys.}

\def\mnras{MNRAS}

\def\pasj{Publ. Astron. Soc. Pacific}
\def\pasp{Publ. Astron. Soc. Pacific}

\def\zap{Zs. f. Astrophys.}
\def\apgt{\ {\raise-.5ex\hbox{$\buildrel>\over\sim$}}\ }
\def\aplt{\ {\raise-.5ex\hbox{$\buildrel<\over\sim$}}\ }
\def\m{^m\kern-7pt .\kern+3.5pt}
\def\d{^d\kern-7pt .\kern+3.5pt}
\newcommand{\ms}{\mbox {$\rm {M_\odot}$}}
\newcommand{\msun}{\mbox {$\rm {M_\odot}$}}
\newcommand{\ls}{\mbox {$L_{\odot}$}}
\newcommand{\rs}{\mbox {$R_{\odot}$}}

\newcommand{\myr}{\mbox {~${\rm M_{\odot}~yr^{-1}}$}}

\newcommand{\mdot}{\mbox {$\dot{M}$}}
\newcommand{\teff}{\mbox {$T_{\mathrm eff}$}}

\setlist{nosep}
\author[1]{L.R.~Yungelson\orcidlink{0000-0003-2252-430X}\thanks{lev.yungelson@gmail.com}}
\author[1,2]{A.G.~Kuranov}
\author[1,3]{A.V.~Mishakina}
\affil[1]{Institute of Astronomy, Russian Academy of Sciences}
\affil[2]{Sternberg Astronomical Institute\\ M.V.~Lomonosov Moscow State University}
\affil[3]{Moscow Institute of Physics and Technology}

\title{\bf Elusive helium stars in the gap between subdwarfs and Wolf-Rayet stars~III.\\ Formation of the Galactic population of stripped helium stars }
\date{\small \today}
\begin{document}
    \maketitle
\textbf{Abstract} --- 
Population synthesis based on the grid of precomputed evolutionary sequencies for close binaries is applied
to the problem of formation of the Galactic population of stripped helium stars, the remnants
of $5 \leq M_{1,0}/\ms \leq 24$ primary components of close binaries after mass loss due to  Roche lobe overflow. Stellar rotation is taken into account. 
For the current star formation rate in the Galaxy $\mathrm {2\,M_\odot yr^{-1}}$ the estimate of the number of 
1 to 7\,\ms\ objects with \teff$ \apgt $25000\,K ranges from 14700 to 28500, 
while for the number of 
objects with masses from 2\,\ms\ to 7\,\ms\ it is 3200 to 5500, depending on the assumptions 
about formation of common envelopes and their outcomes. 
Other factors influencing the estimate are adopted star formation rate,
distributions of close binaries over initial masses of primary components, mass ratios of
components, orbital periods.
Observational selection may  severely limit the number of detectable objects. 
Angular velocities of rotation of companions to stripped helium stars are 60 to 80 per cent of the critical one. 
Using the models of initially homogeneous helium stars as approximations for the  
remnants after mass loss results in the underestimate  of presupernovae masses.     


\newpage
\section{introduction}
\label{s:intro}

The paper continues the study of close binary systems (CBS) with masses of 
primary components  $M_{\rm 1.0} \approx (4 - 24)$\,\ms,
that aims at investigation of their characteristics after mass exchange and
computation of their evolution from the zero-age main-sequence up to formation
of bound or single compact objects.
The first parts of the study were published by Yungelson et al. (2024, Paper I) 
and Fadeyev et al. (2025).
Significance of close binary stars in this mass range stems from their 
potential to produce binary white dwarfs (possible precursors of Type Ia supernovae),
hydrogen-deficient supernovae, Be- and Wolf-Rayet stars, Be/X and 
$\gamma$\,Cas type X-ray sources, binary and single neutron stars, sources of 
ionizing radiation, and still hypothetical Thorne-{\.Z}ytkow objects.
These CBS produce gravitational wave sources potentially detectable by LISA. 
The remnants of their primaries which filled Roche lobes
on the main-sequence or during hydrogen-shell burning stage, dubbed ``stripped helium 
stars'' (henceforth, HeS), are helium-burning cores of about (0.5 - 7)\,\ms,
surrounded by compact ($\Delta M \ll 1$\,\ms) hydrogen-helium envelopes (Kippenhahn and Weigert, 1967).

The remnants of the donors with $M \apgt $7\,\ms\ are identified
with Wolf-Rayet stars (WR, Paczy{\'n}ski, 1967).
Estimates of the WR population in the Galaxy range from $\simeq 6000$ 
(Shara et al., 2009) to $1200 \pm 200$ (Rate and Crowther, 2020).
About 700 objects have been identified 
(\verb https://pacrowther.staff.shef.ac.uk/WRcat/ ).
Binarity rate of WR stars is estimated as $\sim 50\%$ (D'Silva et al., 2024).

The remnants with $M \lesssim 2$\,\ms\ are identified with helium subdwarfs
(Mengel et al., 1976; Iben and Tutukov, 1987; Tutukov and Iungelson, 1987).
Subdwarfs can be either single stars or components of binaries.
Vast majority of known subdwarfs in close binaries have masses $\aplt$(0.6 -- 0.8)\,\ms\
(Schafenroth et al., 2022; Lei et al., 2023).
However, only about 20 massive subdwarfs ($\approx (1-2)$\,\ms) are known in the Galaxy in binary systems.
There are several stars that are thought to transform into subdwarfs after completion of  mass exchange
(Shenar et al., 2020a; Lennon et al., 2021; El-Badri et al., 2022; Wang et al., 2023; Klement et 
al., 2024; Gabitova et al., 2025).
Their companions are typically Be stars.

In the Hertzsprung-Russell diagram (HRD) the models of donor remnants with 
${\rm Z=Z_{\odot}}$ and 
masses (2 -- 7)\,\ms\ occupy the region $\log (T_\mathrm{eff})\approx (4.5 - 5.0)$,
$\log (L/L_\odot)\approx (2.5 - 5.0)$, between subdwarfs and Wolf-Rayet stars.
In this region of the HRD the lifetime of HeS stars of ${\rm Z=Z_{\odot}}$
with $M \gtrapprox 1$\,\ms\ 
is $\simeq 10$\% of the lifetime of their main-sequence progenitors.
Applying population synthesis based on 
detailed calculations of the evolution of close binaries Yungelson et al. (2024) and 
Hovis-Afflerbach et al. (2025) estimated the number of  
2\,\ms\ to 7\,\ms\ HeS stars in the Galaxy as $\simeq 3000$ and $\simeq 7000$, respectively 
(for star formation rate 
SFR=2\,\ms\,$\mathrm {yr^{-1}}$).

However, at the time of writing only one star has been identified as a Galactic HeS: 
WR2-1, with M=(3.2 -- 5.8)\,\ms\ (Müller-Horn et al., 2026a).
Within the measurements uncertainty, the hot component
of the MWC~656 system is closest to the 2\,\ms\ threshold:
$M=1.48^{+0.55}_{-0.46}$\,\ms\ (Müller-Horn et al., 2026b).
Given  anomalously high luminosity of the star for its mass
($\log(L/\ls)=4.0\pm 0.2$),
one can assume that it is
burning He in the shell (see Fig.~\ref{f:structure} below).
As well, another star, $\gamma$ Col (see Fig.~\ref{f:hestobserved} below),
is known, but its interpretation is controversial.
Considering its position in the Hertzsprung-Russell diagram and the chemical composition
of its surface layers, Irrgang et al. (2022) suggested  that it is a remnant of a star with initial  mass of 
several \ms, transforming into HeS.
However, based on the same chemical composition, Jin and Langer (2026) consider it
to be a former accretor in a close binary. Several stars observed in the young open cluster NGC 663 with $M  \lesssim 2.8\ms$ with an UV excess 
may belong to  HeS (Nedhath et al., 2025).

Given the paucity of massive subdwarfs, we will classify {\it all}
donor remnants with masses (1 - 7)\,\ms\ as helium stars.
For SFR=2\,\ms\,$\mathrm {yr^{-1}}$ the estimated number of the systems with remnant masses (1 - 7)\,\ms\ is $\simeq$30000 (Yungelson et 
al., 2024) and $\simeq$60000 (Hovis-Afflerbach et al., 2025).

Donor remnants in the HeS mass range in the relatively short stage of contraction after detachment
from the Roche lobe, may represent a new class of variable stars with nonlinear
radial pulsations with periods from 0.17 to 3.9 day and amplitudes up to
$\Delta M_{\rm bol} = 0.8^m$ (Fadeyev et al., 2025).

Stripped helium stars have been found in the Magellanic Clouds
(Villase{\~n}or et al., 2023; G{\"o}tberg et al., 2023; Drout et al., 2023; Ramachandran et al., 2023, 2024; Ludwig et al., 2026), where binarity rate of massive stars
does not differ fundamentally from that of the Galactic stars
(Sana et al., 2025; Villase{\~n}or et al., 2025).
Effective temperatures and luminosities of the remnants
of the donors in close binaries of the Milky Way and Magellanic Clouds during core He-burning stage 
also do not differ significantly (Dutta and Klencki, 2024).
However, conditions for detection of HeS in the Magellanic Clouds are more favorable than in the Galaxy.
The mass estimates of the stars found in MC within errors range are (1 - 7)\ms.

Discovery of helium stars confirms the soundness of the modern theory of stellar evolution and
prompts a thorough investigation of the possible reasons for
the paucity of helium stars in the Milky Way, which may be due  both to the evolutionary processes
and observational selection effects.
The estimates of the number of HeS stars may be influenced by assumptions about formation of common envelopes and evolution of binary components in them, which may result 
in a significant decrease in the distance between components and merger (Paczy{\'n}ski, 1976).
Observational selection is due primarily to the significant difference of 
\teff\ between helium stars and their companions (G{\"o}tberg et al., 2018, 2023; Blomberg et al., 2026).
In the optical range, the companion may ``outshine'' HeS star.
Unlike Wolf-Rayet stars, HeS are expected to have optically thin stellar winds.
This may also hamper their detection, as emission lines should be observed  in the spectra of the 
most massive objects only (G{\"o}tberg et al., 2018).
Blomberg et al. (2026), who computed  a mock catalog of the HeS population in the Magellanic Clouds, 
estimated that, for instance, using the data obtained with the Swift-UVOT telescope
(Roming et al., 2005), it would be possible to detect 11\% and 7\% of the HeS in the SMC and LMC, respectively.

In this paper, we calculate a grid of evolutionary tracks for
close binaries with  masses of primaries $\rm {M_{1.0}}$=(4 -- 24)\,\ms.
The set of their  remnants contains the most massive subdwarfs, HeS, and the least massive WR stars.
Particular attention is paid to HeS.
Assumptions used in the calculations are described in detail, as results of them
are expected to be used to study further evolution of close binary systems, up to the formation
of bound or single compact objects.
At difference to Paper I, rotation of both components is taken into account, different
parameters of the  mixing and stellar wind are adopted, another IMF for the primaries and initial
distribution of binaries over  
orbital periods is  assumed.
In most cases, the calculations are extended to the exhaustion of He in the cores of the stars. 
In Section \ref{s:ce} stability of mass exchange in close binaries and conditions for formation of common 
envelopes are discussed.
In Section \ref{sec:evol_comp} we present the main assumptions adopted in
the calculations and population synthesis.
Results of the calculations and population synthesis are presented in Section \ref{s:results}.
In Section \ref{s:disc} we compare results of the study with those of other authors, discuss the 
factors influencing the estimates, and briefly examine the effects of observational selection.
In Section \ref{s:concl} the main findings of the study are summarised.

\section{STABILITY OF MASS LOSS AND COMMON ENVELOPES}
\label{s:ce}

The nature of mass loss by a star after  Roche lobe overflow (RLOF)
is determined by the relationships between three mass-radius exponents
$\zeta_L=\frac {d\ln R_L}{d\ln M},~
\zeta_{th}=\left(\frac{\partial \ln R}{\partial \ln M}\right)_{th},~
\zeta_{ad}=\left(\frac{\partial \ln R}{\partial \ln M}\right)_{ad}$
(Morton, 1960; Paczy{\'n}ski and Sienkiewicz, 1972).
Exponent $\zeta_L$ describes the response of the donor's Roche lobe radius to the mass loss,
$\zeta_{th}$ describes stellar radius response to the mass loss with account for the thermal effects, while
$\zeta_{ad}$ -- restoration of hydrostatic equilibrium on the adiabatic time scale.
If $\zeta_{ad} > \zeta_L > \zeta_{th}$, the star that has lost mass
maintains hydrostatic equilibrium but does not remain in thermal equilibrium, and
mass loss occurs on the  thermal time scale.
If $\zeta_{L} > \zeta_{ad}$, the star cannot maintain hydrostatic equilibrium, and mass loss
proceeds on the dynamical time scale.
Such a mass loss occurs if the donor has an extended convective envelope
or the close binary has a high mass ratio of components.
It is characterized as ``unstable''.
Below, we consider mass loss to be ``unstable'' or ``dynamical'', when the rate of 
$ \dot{M}$ exceeds  $10^{-2}\,M_\odot yr^{-1}$.
At such a $\dot{M}$, models typically cease to converge.

The values of $\zeta_{ad}$ and $\zeta_{th}$ depend on the characteristics of the outer layers
of the star. The boundary of stable mass loss can be found by
calculating \mdot\ in detailed stellar models only (Woods and Ivanova, 2011).
``Mass-exchange stability criteria'' found in the literature, based on
the global characteristics of stars, are only approximate (see, e.g., Temmink et al., 2026).
The limiting rate of the stable mass loss
is determined by the shortest local thermal time scale of the superadiabatic
layers of the stellar envelope (Temmink et al., 2023).

Characteristic time of mass loss by the donor must be compared to the time scale
on which the matter can be added to the accretor.
For an accretor with $M_{2.0} < M_{1.0}$, the thermal time scale is longer
than the thermal time scale of the mass-loss by the donor immediately after RLOF. 
Restoration of the thermal equilibrium is accompanied by an increase in the accretor radius
(Benson, 1970; Yungelson, 1973), which continues until mass ratio of components becomes close to one.

As well, if the accretion rate approaches Eddington limit,
the radius of accretor begins to increase indefinitely due to the low binding energy
of the stellar outer layers (Sch{\"u}rmann and Langer, 2024). 
The increase of radius can lead to the formation of a contact system and then a 
``supercontact'' system with an envelope close to the second
Lagrange point, from the vicinity of which the matter is lost with a significant angular momentum.
As a result, the accretor can be engulfed by the donor envelope.

Deep convective envelopes are already present in some donor stars which overflow their Roche lobes  
in the Hertzsprung gap and in the stars evolving along the Hayashi border.
According to our estimates,
unstable mass-exchange begins when convection engulfs, for example, the outer
45\% by mass in a 4\,\ms\ star, 75\% in an 8\,\ms\ star, and 50\% in a 22\,\ms\ star.
If the mass-exchange is completely non-conservative, mass-loss is unstable for
$q \lesssim 0.4$ (e.g., Henneco et al., 2024).
In this paper, calculations are performed for $q = 0.4(0.1)0.9$.

Common envelopes may be formed if accretion is limited by rotation. 
If the star is rotating like a solid body, its angular velocity attains  
$\omega_{cr}=\sqrt(G M_\star/R^3_{\rm eq})$
already after accretion of (5 - 10)\% of the matter lost by the donor star with specific
angular momentum of accretor (Packet, 1981)\footnote{This estimate assumes that the star does 
does not lose its own angular momentum.}.
Further, accretion is possible only within the limits allowed by the maintenance of $\omega_{cr}$.
As a rule, it is assumed that the excess matter leaves the system with the specific angular momentum
of the accretor due to the radiation pressure of both components of the binary. 
To estimate the maximum rate of the matter removal $\dot{M}_{\rm exc}$ one can assume
that it flows away from the accretor with a mass $M_{\rm 2}$ equal to the radius of its
Roche lobe $R_{\rm L,2}$ (Marchant, 2017).
In this case, if one neglects donor mass,
\begin{equation}
\label{eq:mlossmax}
\dot{M}_{exc} \lesssim  10^{-7.19} \frac{(L_1+L_2)}{\ls} \frac{R_{\rm L,2}}{\rs} 
\frac{M_\odot}{M_2} \quad \frac{M_\odot}{yr}, 
\end{equation}       
where  $L_1$ and $L_2$  are the luminosities of components. 
\begin{figure}[t!]    
\centering   
\includegraphics[width=0.5\textwidth,trim={0 0cm 0 1.5cm},clip]{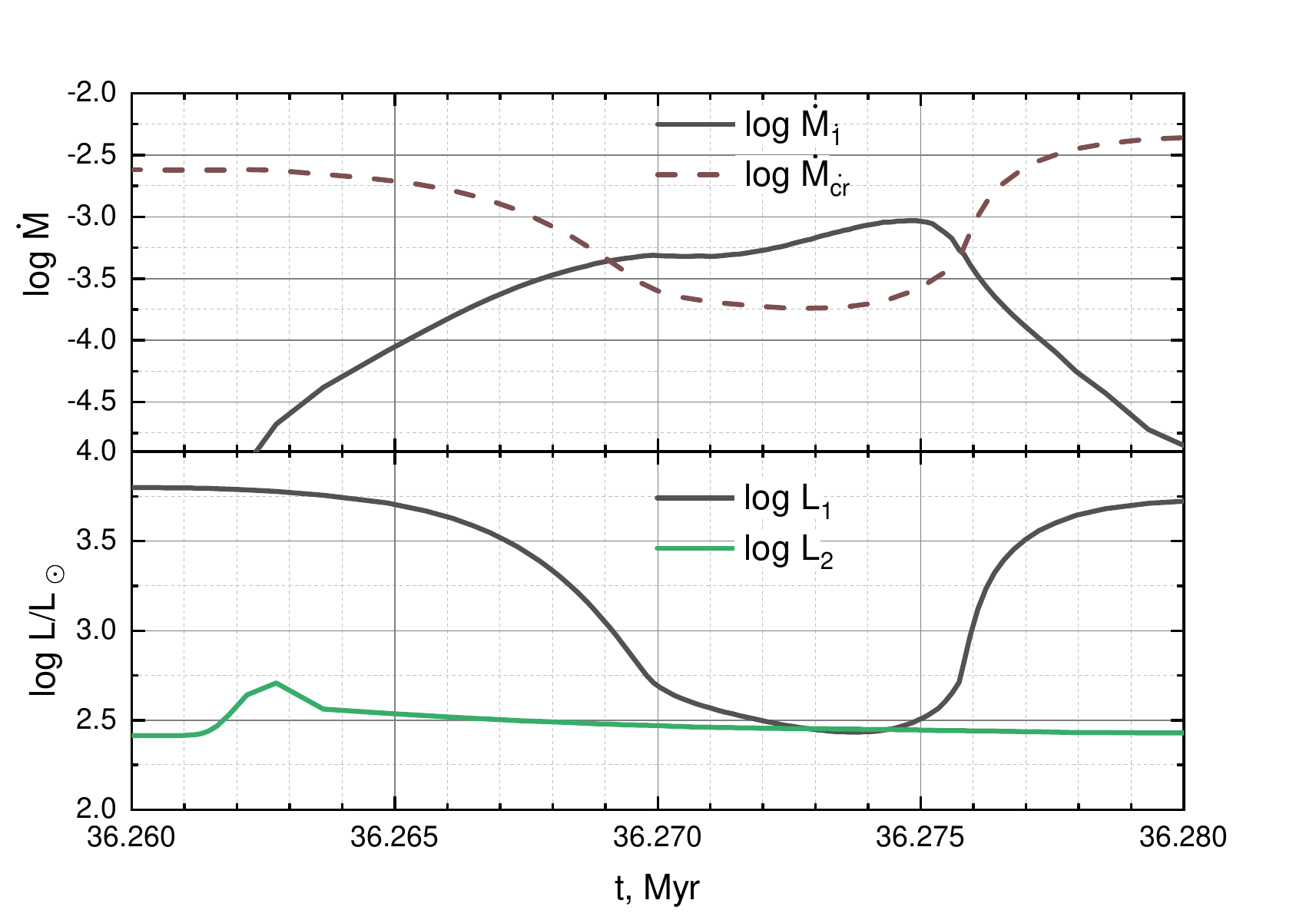}  
\caption{Upper panel panel: donor mass loss rate (solid line)
and maximum possible rate of mass removal from the system (dashed line, Eq.~\ref{eq:mlossmax}). 
Initial  masses of components are 8 and 4 \msun, $P_0$=20\,day.
The rates are  in \myr.
Bottom panel: variation of components luminosity during mass exchange.
}
\label{f:fig_excess}
\end{figure}  

Equation \ref{eq:mlossmax} allows for ambiguous interpretation.
Matter removal rate may depend either on the current parameters of the components or on 
the masses and luminosities at the end of the mass loss stage.
In the later case conditions for mass removal are more favourable. 
Accretion rate may exceed $\dot{M}_{exc}$ not throughout the entire mass exchange stage,
but only while the star is losing a part of its envelope.
Figure~\ref{f:fig_excess} shows an example of evolution during the mass loss stage in which 
${M_{1.0}}$=8\,\ms\ star loses approximately 4\,\ms\ in the regime $\mdot > \dot{M}_{exc}$, while 
the total lost mas is $\simeq 6$\,\ms.
The efficiency of the commonly postulated mass removal mechanism may be actually $<1$, since it is 
unclear what is the distance relative to the accretor from which the matter is removed, what fraction of the 
stellar radiation can  be actually expended on removal of the matter, what is energy of the matter 
that can be considered as removed,
and what is the configuration of the system if $\mdot > \dot{M}_{exc}$ (Henneco et al., 2024).
Formation of a contact system, a circumstellar disk or a disk around the system or  formation 
of a common envelope are possible.
If a contact system forms and the matter is lost from the vicinity of the ${\rm L_2}$ point, 
close binary will likely also plunge into a common envelope, passing before through the
stage of a (super)contact system.
We calculate the evolution assuming that the ``excess'' of the matter is completely lost,
but at the processing stage, consider the possibility that the system could have plunged into a common envelope,
assuming that Eq.~(\ref{eq:mlossmax}) contains the instant parameters of the stars.
The plunge of a close binary into a common envelope, if \mdot$>\dot{M}_{exc}$, is also suggested, 
for example, by Xu et al. (2025).
We consider the loss of matter on the dynamical time scale and the RLOF by accretor as a sufficient 
conditions for formation of common envelopes, like, for example, in the ``universal'' population 
synthesis program POSYDON (Fragos et al., 2023).

No formalism for describing evolution in the case when excess of the matter cannot be removed exists.
We applied an extreme assumption that the system evolves in the same way
as in the case of ``usual'' common envelopes, in accordance with Eq.~(\ref{e:ce}) below.

The outcome of evolution in common envelopes is determined by solution of the equation for the 
balance of the binding energy of the donor envelope and the binding energy of the close binary 
system (Webbink, 1984; de Kool, 1990):
\begin{equation}
 \frac{G M_1 M_e}{\lambda R_{\rm 1,L}} = 
 \alpha_{\rm ce} \left( \frac{G M_c M_2}{2 A_f} - \frac{G M_1 M_2}{2 A_i} \right ),
\label{e:ce}
\end{equation}
where $M_1$ and $M_2$ are initial masses of the donor and accretor,
$M_e$ -- the mass of the donor envelope, 
$M_c$ -- the mass of donor core,
$R_{\rm 1,L}$ -- the radius of the donor Roche lobe,
$A_i$ and  $A_f$ -- initial  and final separations of components of the binary, 
$\alpha_{\rm ce}$ -- the fraction of the binding energy of the system which may be spent on removal of the common envelope (so called ``common envelope efficiency''), 
$\lambda$ -- the parameter which characterizes binding energy of the donor envelope.

Common envelope matter has a significantly higher specific entropy than the 
matter on the surface of the accretor and, therefore, a higher temperature and lower
density than the accretor matter.
Therefore, the mass of the accretor in the common envelope  practically does not increase (Hjellming and Taam, 1991).

Calibration of $\alpha_{\rm ce}$ by observational data is usually limited to the low-mass stars
($\lesssim 1$\,\msun) and did not result in any definite conclusions (e.g.,
Zorotovic et al., 2010; De Marco et al., 2011).
Having in mind this uncertainty, we assumed $\alpha_{\rm ce}=1$.
Considering that in a crude approximation $A_f \propto \alpha_{\rm ce}\lambda A_0$, the decrease  of 
the $\alpha_{\rm ce}$\ values will result in a higher frequency of mergers in the common envelopes.

The parameter $\lambda$ can be calculated based on the equation for the envelope binding energy
\begin{equation}
E_{\lambda, bind} = -\int_{core}^{surf} \left ( -\frac{G m}{R} + \alpha_{\rm th}\epsilon(m)
\right ) dm = \frac{G M_1 M_{1,e}}{\lambda R_1},
\label{e:lambda}
\end{equation}
where  $\epsilon(m)$ is specific internal energy, $\alpha_{\rm th}$ -- a parameter.

An obvious but inevitable drawback of the definition of $\lambda$ above is that it does not take 
into account possible changes in the stellar structure during formation of a common envelope.
We assumed $\alpha_{\rm th}$=1 for our estimates.
The value of $\lambda$ depends on the definition of the boundary between stellar core and 
envelope.
We assumed that the boundary is located in the layer in which the relative hydrogen 
abundance by mass X=0.1 (Dewi and Tauris, 2000).
The dependence of $\lambda$ on stellar mass and radius is shown in Fig.~\ref{f:lambdamr}
(left panel) in the Appendix.
Note that typically $\lambda \ll 1$, and in all cases when, according to our 
calculations, binaries enter common envelopes, the components merge. This
circumstance is illustrated in Fig.~\ref{f:lambdamr} (right panel), which
shows the ratio of the estimated radii of the secondary components after the completion of the common
envelope stage and the radii of their Roche lobes.

\section{EVOLUTIONARY CALCULATIONS AND POPULATION SYNTHESIS }
\label{sec:evol_comp}

For the calculations we applied code MESA (Paxton et al., 2011, 2013, 2015, 2018,
2019; Jermyn et al., 2023) version r24.03.1. 
The physical processes and their parameters are briefly described below.

Mass loss rate by Roche lobe filling stars was calculated taking into account mass loss from both optically thick and thin layers of the star  
(Kolb and Ritter, 1990).
The possibility of ballistic accretion was considered, if stellar radius exceeded the minimum
distance to which the accretion stream approaches the star:
$R_{\rm min} = 0.0425 a (q+q^2)^{0.25}$,
where $a$ is the distance between the components (Lubov and Shu, 1975; Paxton et al., 2015).

Like in Paper I, it was assumed that accretion is limited once the accretor  attains an angular 
velocity
$\omega = 0.95\omega_{cr}$. Later,  accretion is possible
only within the limits necessary to maintain this velocity. Excess of the matter
leaves the system as an enhanced stellar wind with the specific angular momentum of the accretor.

Mixing of the matter was taken into account following B{\"o}hm-Vitense (1958) with the  mixing 
length ${\rm 1.5H_p}$, where ${\rm H_p}$ is the pressure height scale.
The boundary of the convective core was determined using the Ledoux criterion.
Penetration of convective elements beyond the core boundary was described by a step 
function ${\rm f_{ov}H_p}$, where
${\rm f_{ov}}$ varies linearly from 0.1 for M=1.66\,\ms\ to
0.3 for M=20\,\ms\ and then remains constant (Hastings et al., 2021).
The semi-convection parameter $\alpha_{sc}=1$ (Langer et al., 1983).
Thermohaline mixing was taken into account with a parameter $\alpha_{\rm t}$=1,
independent of stellar mass (Cantiello and Langer, 2010).

Rotation of both components and the accompanying instabilities were considered under  assumption 
that the transport of matter and angular momentum is a diffusion process (Heger et al., 2000; 2005).
Eddington-Sweet, Goldreich-Schubert-Fricke, and shear instabilities (dynamical and secular) were 
taken into account, as well as the transport of angular momentum due to the Tayler-Spruit dynamo (Spruit, 2002).
Mixing efficiency coefficient due to the instabilities was $f_c$=1/30 (Chaboyer and Zahn, 1992).
The coefficient describing the sensitivity of instabilities to the chemical composition gradient 
was set to $f_\mu$=0.1 (Yoon et al., 2006).
Initially, components rotate synchronously with $v/v_{\rm crit}$=0.35, where $v$ is the linear rotation velocity
of the star, and $v_{\rm crit}$ is the critical rotation velocity (Hastings et al.,
2020)\footnote{The adopted value of the initial angular rotation velocity has no real impact
on the evolution of the accretor, since the critical rotation velocity
is reached after accretion of only a few percent of the donor mass.}.

Prior to RLOF by stars with $\teff \leq 25000$\,K mass loss due to the stellar wind was 
taken as the larger of the \mdot\ values given by
the formulas of Nieuwenhuijzen and de Jager (1990) and Fink et al. (2001).
For HeS with $M \leq 1.5$\,\msun\, \mdot\  for subdwarfs was used (Krti\v{c}ka et al., 2016).
For more massive stars with $\teff > 25000$\,K and the hydrogen abundance at the stellar surface of $0.4 \leq {\rm X_s} \leq 0.7$, mass loss rate
was determined by interpolating over ${\rm X_s}$ between the values of  \mdot\ of Fink et al.
(2001) and $\dot{M}_{\rm NL}$ for Wolf-Rayet stars (Nugis and Lamers, 2000).
Wind clumpiness was taken into account by multiplying $\dot{M}_{\rm NL}$ by a factor of 1/4 (Cherepashchuk, 2013).
After loss of the hydrogen envelope
\mdot\ was interpolated taking into account chemical composition between
$\dot{M}_{\rm NL}$\ and \mdot\ for WNE and WC subtype stars (Yoon, 2017).

Wind enhancement due to stellar rotation was taken into account (Langer, 1998):
\begin{equation} 
 \dot{M}_{\rm rot}=
  \dot{M}_{\rm nonrot} \times (1/(1 - v/v_{\rm crit}))^{0.43}. 
\label{eq:windenhacement}
\end{equation}

Nuclear reactions grid \texttt{approx21} in MESA was used.
dynamical tides were taken into account using the \verb sync_type=Hut_rad \ scheme in MESA.

In the population synthesis, stellar binarity rate $\mathfrak{B}$ was taken as  a function of 
the primary mass (van Haaften et al., 2013):
\begin{equation}
\label{eq:binfrac}
\mathfrak{B}=\frac{1}{2}+\frac{1}{4} \log \left( \frac{M_{ 1,0}}{\ms} \right), \quad \text{if} \quad 0.08 \leq \frac{M_{1,0}}{\ms} \leq 100.
\end{equation}

For the primary masses, Kroupa (2001) IMF was used:
$dN/dM \propto M^{-\alpha},~\text {where}~\alpha = 1.3
~\text{for}~0.08 \leq M_{1.0}/\ms \leq 0.5~\text{and~} \alpha =
2.3~\text{for~}M_{1.0}/\ms > 0.5\ms$.
Initial distribution of orbital periods for interacting
systems with $M_{1.0} < 15\,\ms$ was assumed to be log-uniform ({\"O}pik, 1924);
for more massive systems, the distribution
$f(\lg(P_{\rm orb})) \propto \lg(P_{\rm orb})^{-0.55}$ was assumed (Sana et al., 2012).
Initial distributions of orbital eccentricities 
$dN/de = \text{const},~e \in [0:1]$, and of component mass ratios
$dN/dq = \text{const},~ q \in [0.1:1]$ were assumed.
Considering the short lifetimes of the relatively massive stars under consideration,
it was assumed that the star formation rate in the Galaxy is constant and equal to
$2\ms yr^{-1}$ (Chomiuk and Povich, 2011).

\begin{figure}[t!]     
\centering
\includegraphics[width=\textwidth, trim={0 1.5cm 0 1.5cm}]{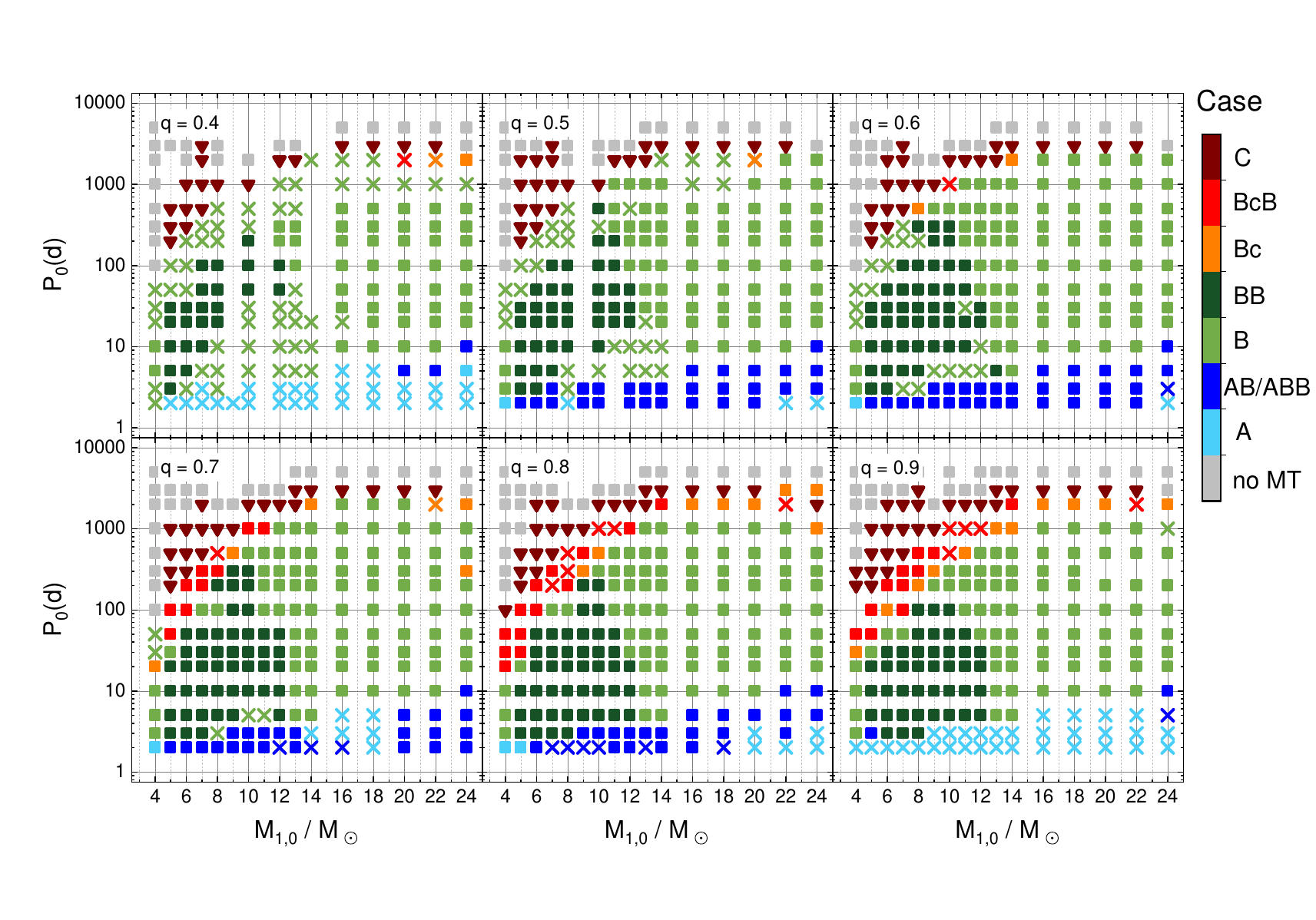}
\caption{The grid of computed models. 
Distribution of models by mass exchange cases is shown.
Wide systems in which mass exchange does not occur are shown in gray.
Crosses mark models in which common envelopes form due to the loss of
matter by the donor on the dynamical time scale.
Downward facing triangles indicate systems evolving in the case C of
mass exchange, in which the donors lose matter almost continuously until formation of a
progenitor of a white dwarf or supernova.
Designations of the ABB, $\mathrm {Bc}$, and $\mathrm {BcB}$ evolution cases are explained in the text.
}
\label{f:case_type}
\end{figure}

\section{Results of the computations}
\label{s:results}
\subsection{The donors.}
\label{ss:donors}
{\it The grid of models.}
The grid of computed  models in the coordinates ($M_{1,0}, P_0$) for various initial
mass ratios of components $q$ is shown in Fig.~\ref{f:case_type}.
Distribution of the models over mass exchange cases is shown.
In addition to the standard notation for cases A, AB, B, BB, and C, ABB denotes close 
binaries in which the donor remnant formed in the cases A or AB
overflows Roche lobe after exhaustion of He in the core. Bc denotes systems in which
the donors overflow Roche lobe
immediately before exhaustion of He in the core and have remnants that remain close 
to the Hayashi border.
If in the latter case the models refill their Roche lobes they are annotated 
as BcB. Examples of the evolution in the cases B, BB, and BcB are discussed below.

\begin{figure}[t!]     
\centering
\includegraphics[width=\textwidth,trim={0 0 0 2cm},clip]{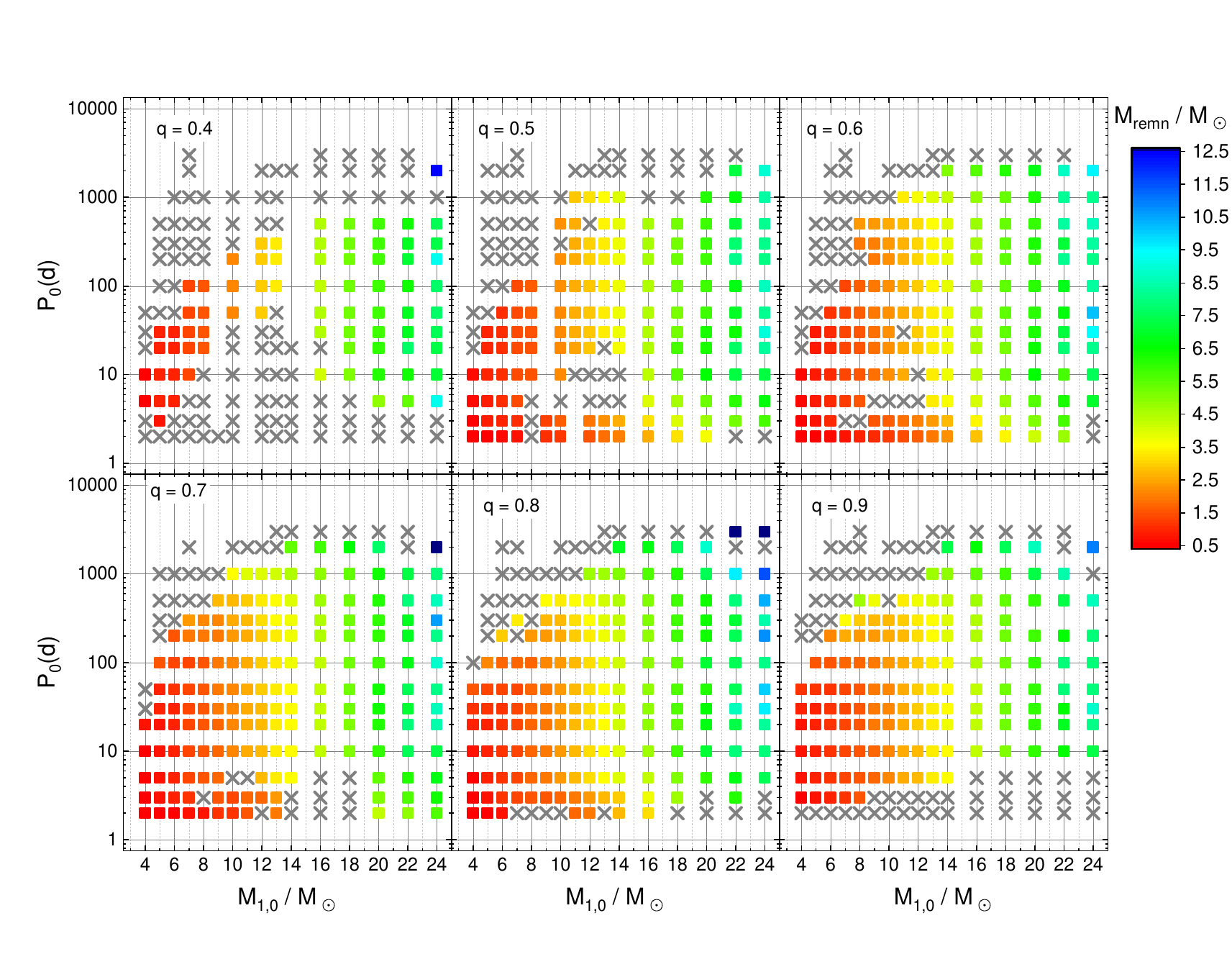}
\vspace{-1cm}
\caption{Dependence of the masses of the 
remnants of primary components of binaries after mass exchange on the initial system parameters 
(color scale). Crosses mark binaries with common envelopes for which the masses of the remnants are  
undefined.}
\label{f:grid_M_HeS_v3}
\end{figure}

The masses of the remnants of the  primary components of close binaries  after mass exchange 
depending on the initial parameters of the systems are shown in Fig.~\ref{f:grid_M_HeS_v3}.

\begin{figure}[t!]   
\begin{minipage}{0.5\textwidth}
\includegraphics[width=\textwidth]{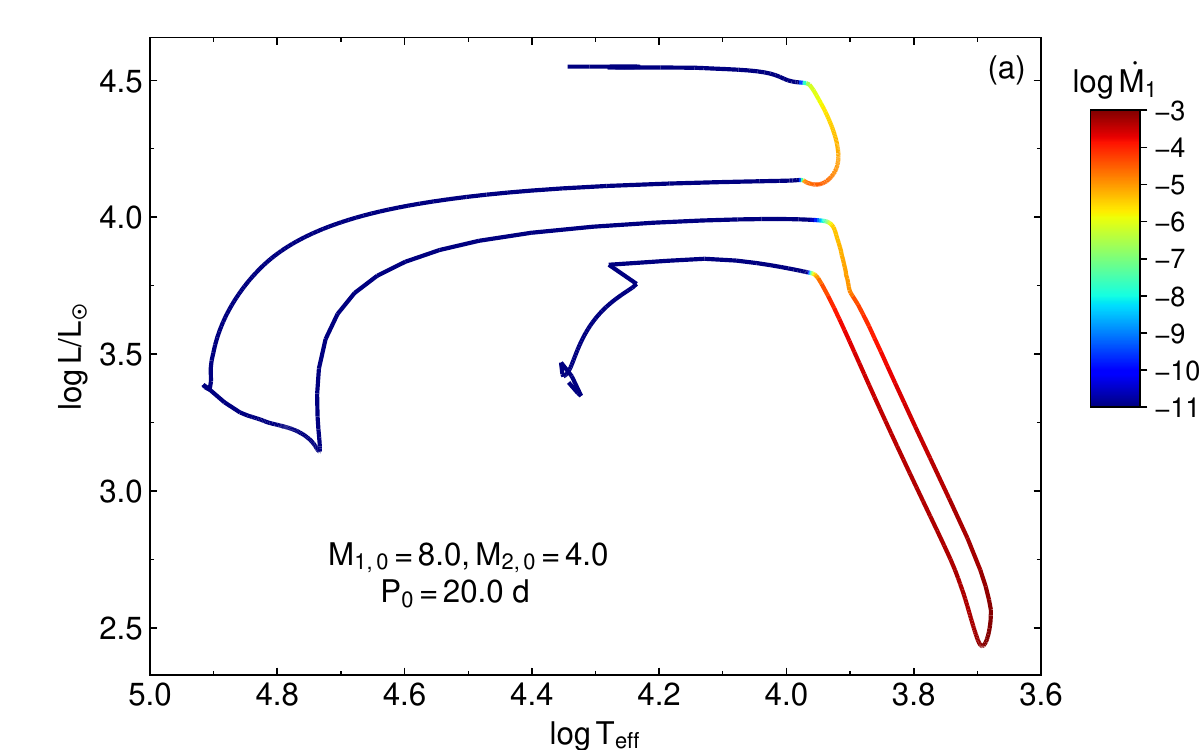} 
\includegraphics[width=\textwidth]{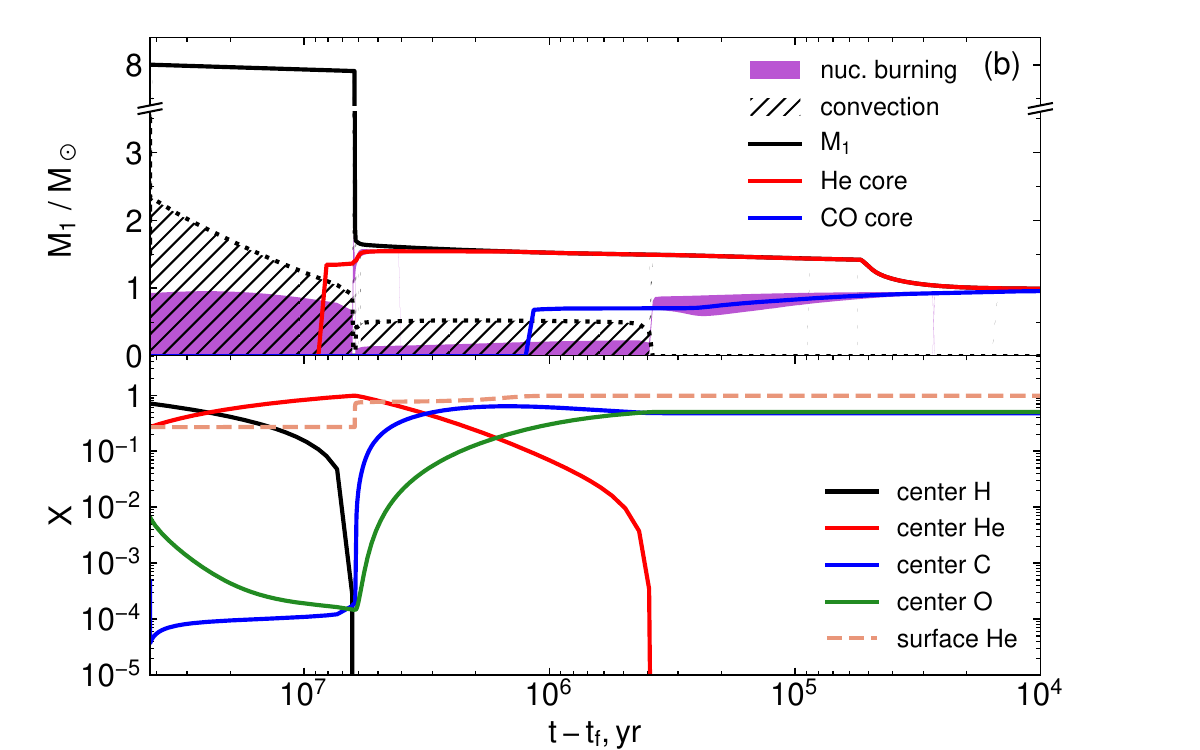} 
\includegraphics[width=\textwidth]{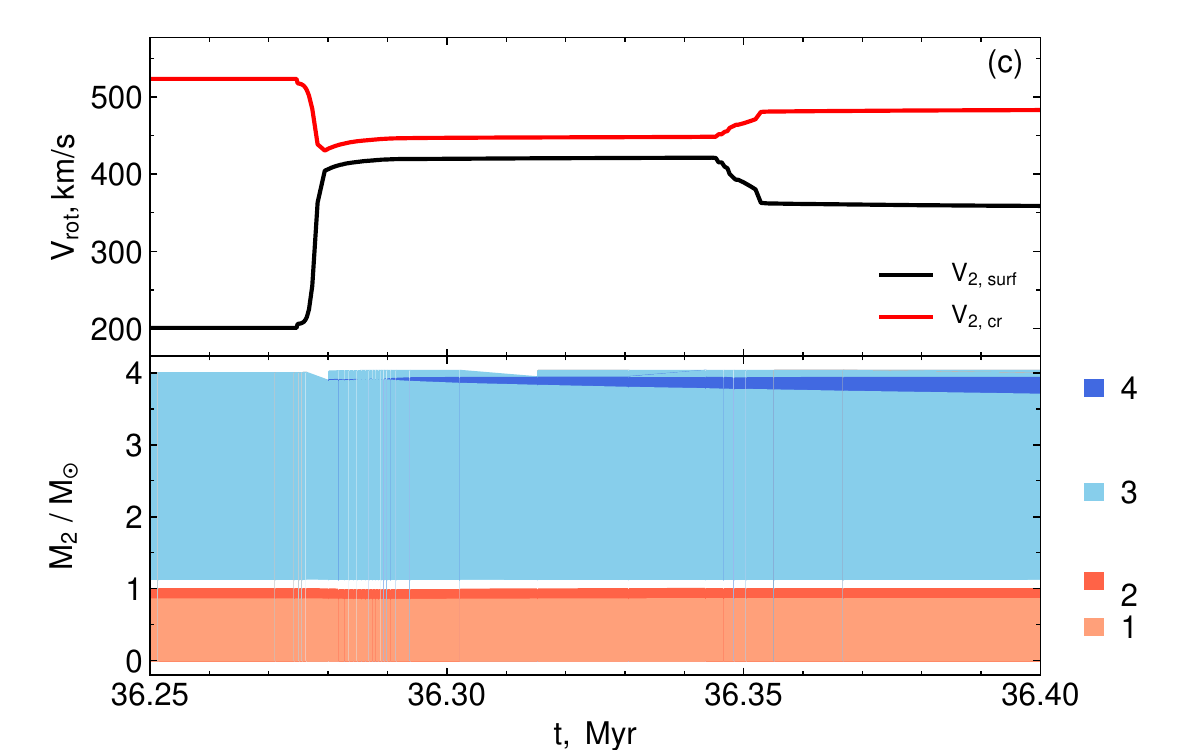} 
\end{minipage}
\begin{minipage}{0.5\textwidth}
\includegraphics[width=\textwidth]{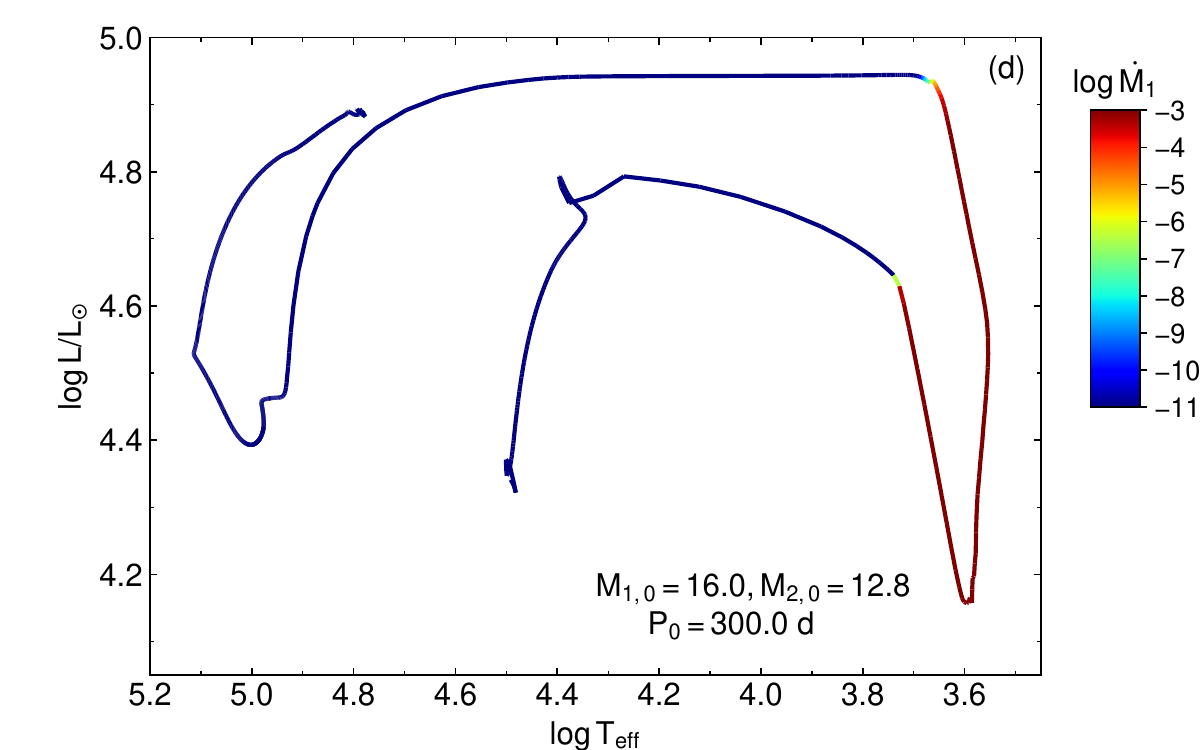} 
\includegraphics[width=\textwidth]{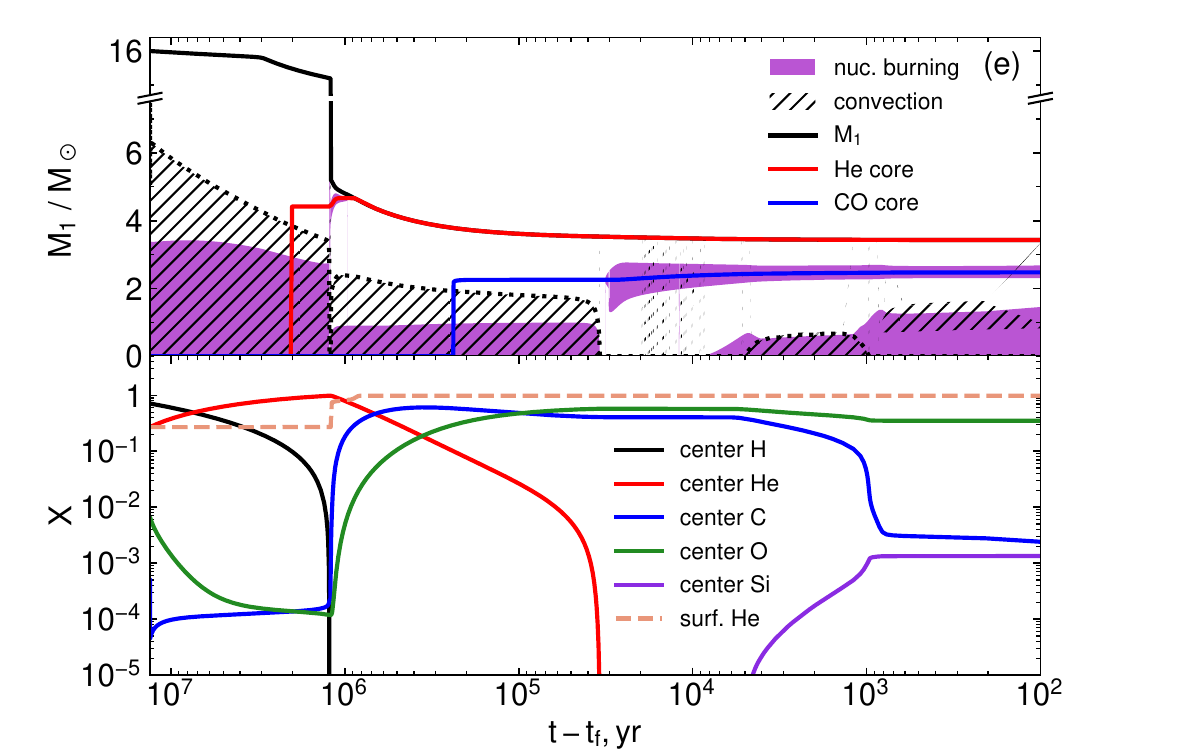} 
\includegraphics[width=\textwidth]{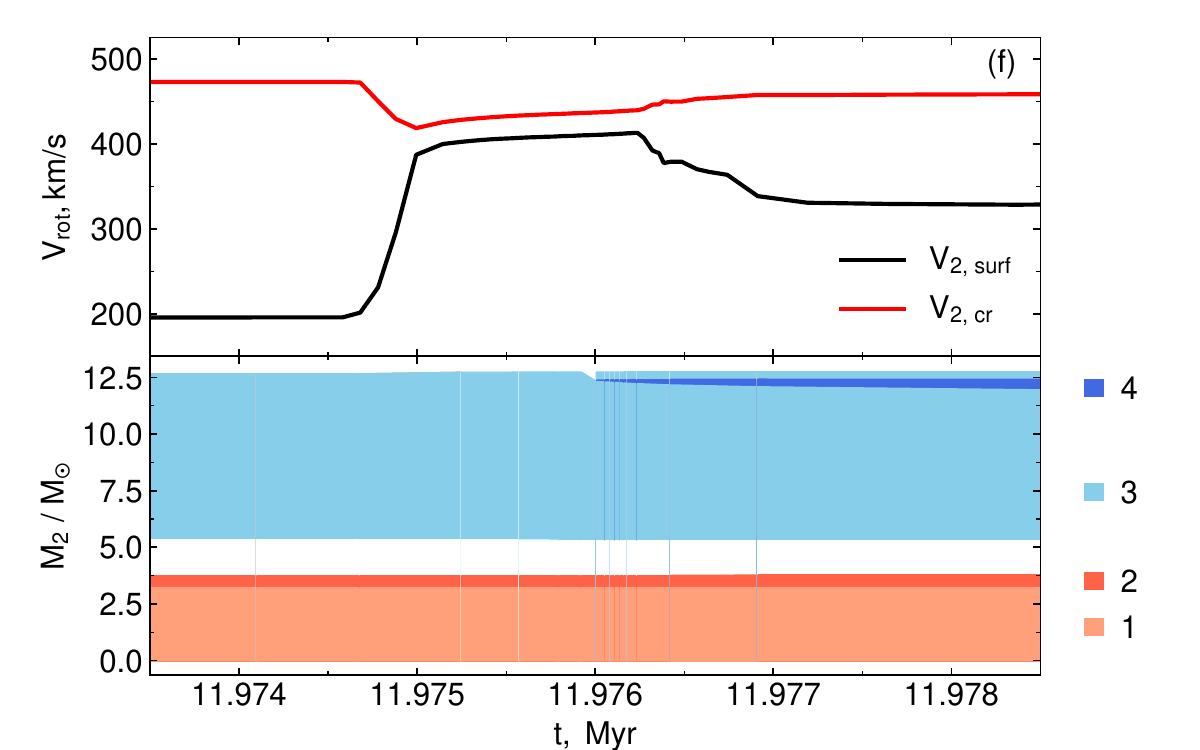} 
\end{minipage}
\caption{Left panel -- variation of the parameters of the system (8+4)\,\ms, 
$P_0=20$\,day in the course of evolution (case BB).
(a) -- evolutionary track of the donor in the Hertzsprung-Russel diagram,
(b) -- Kippenhahn diagram for the donor and variation of central abundances of H, He, C, O 
and of He at the surface,
(с) -- variation of the linear rotation velocity of accretor, critical velocity 
of rotation and location of mixing zones over the first episode of mass exchange:
1 --  convective core, 
2 --  overshooting,
3 --  rotational mixing,
4 --  thermohaline mixing.
Right panel -- the same for the system  (16+12.8)\,\ms, 
$P_0=300$\,day (case B).
} 
\label{f:structure}
\end{figure}
\begin{figure}[t!] 
\includegraphics[width=0.5\textwidth]{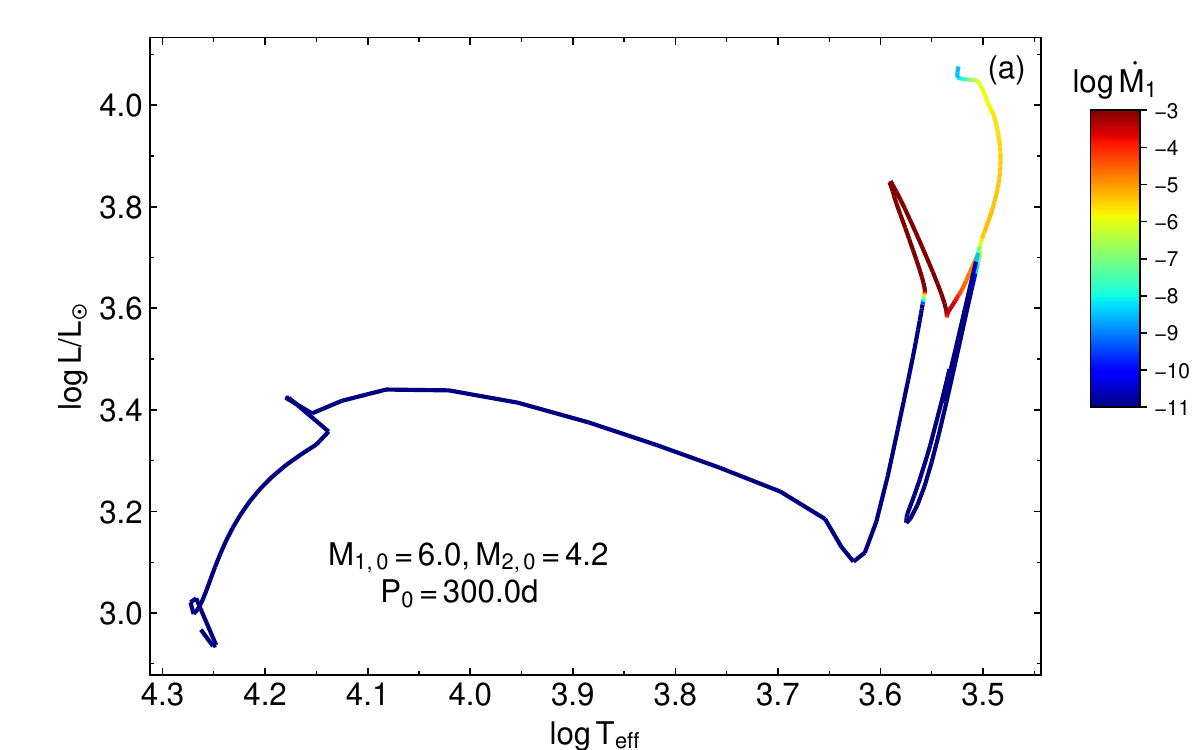} 
\includegraphics[width=0.5\textwidth]{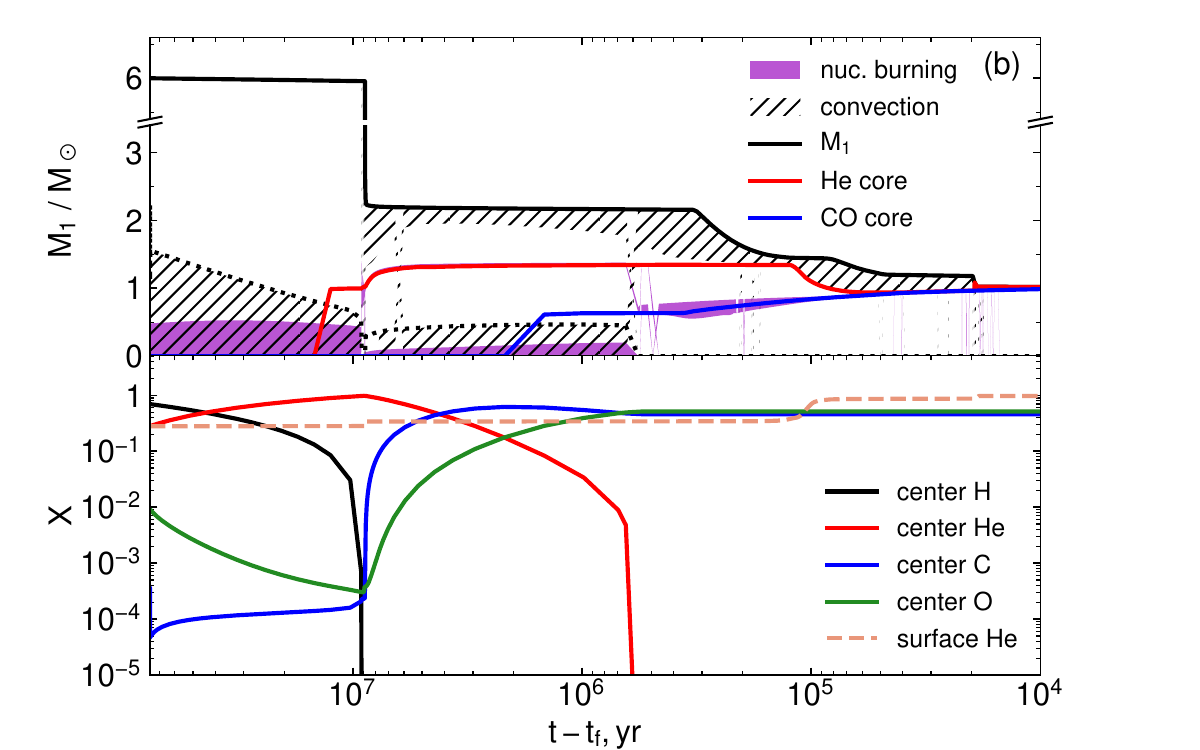} 
\includegraphics[width=0.5\textwidth]{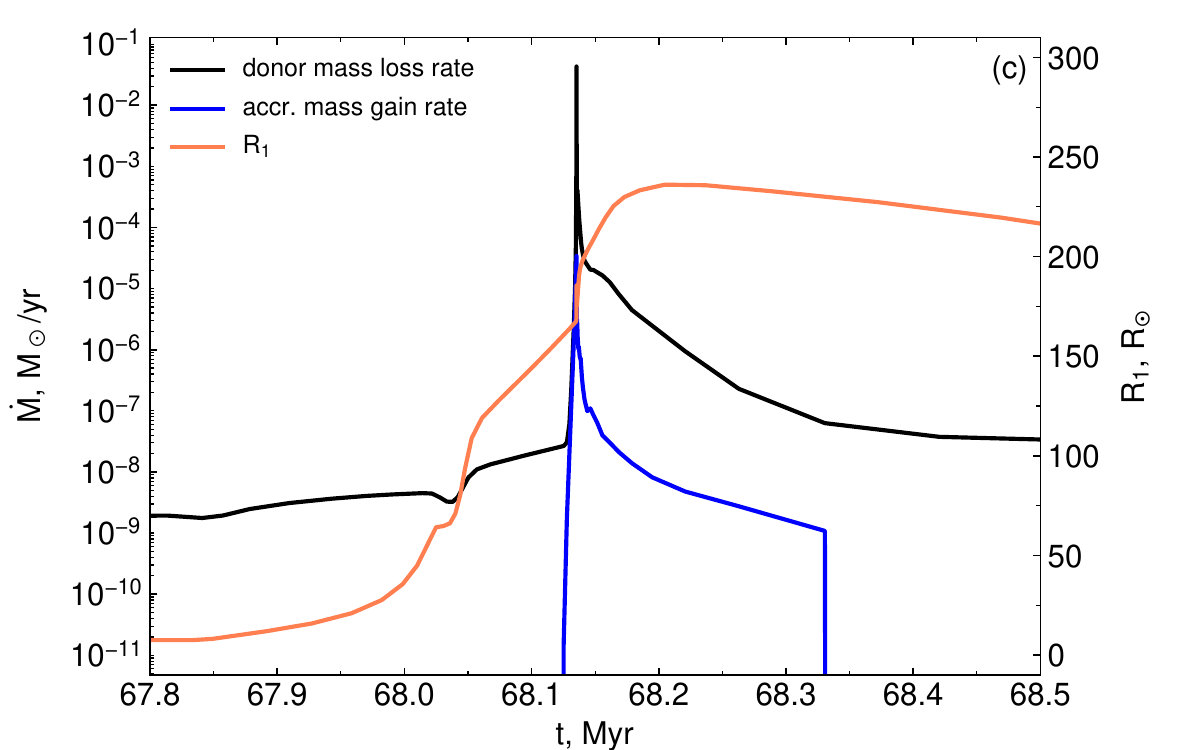} 
\includegraphics[width=0.5\textwidth]{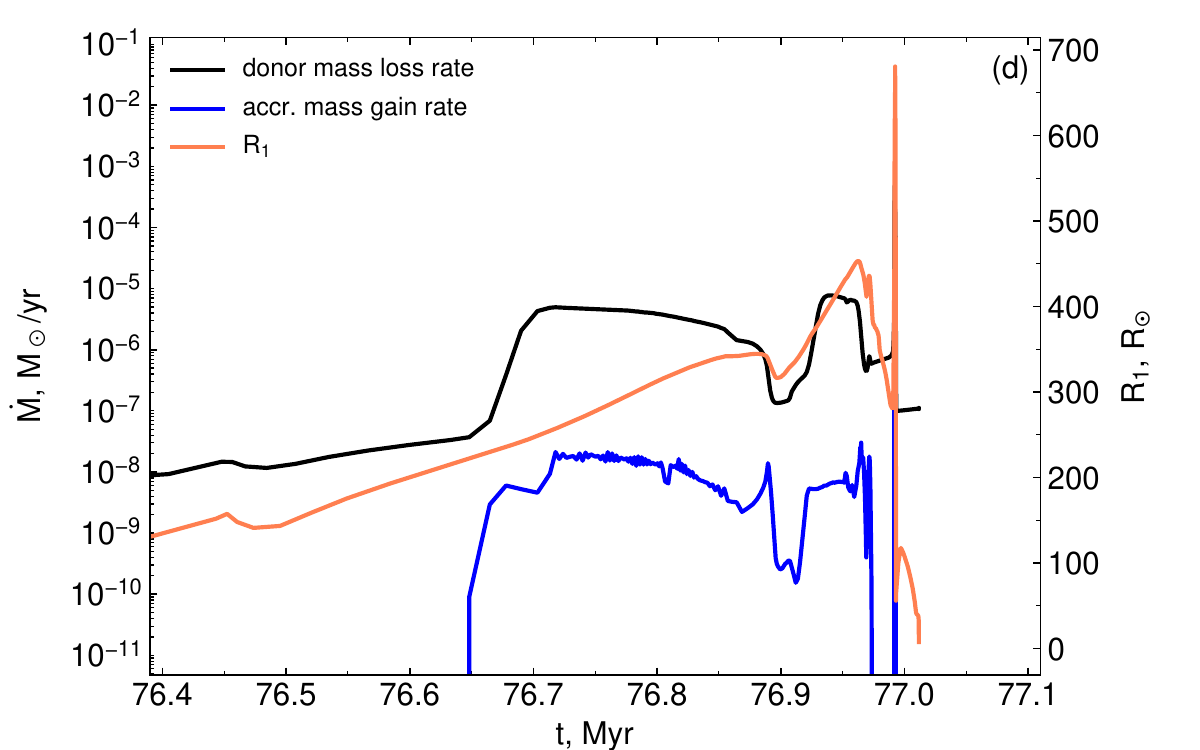} 
\caption{Change of the parameters in the system (6+4.2)\,\ms, $P_0=300$\,day, evolving
in the case Bc.
(a) -- evolutionary track of the star on the Hertzsprung-Russell diagram,
(b) -- Kippenhahn diagram for the donor,
(c) and (d) -- the rate of mass loss and accretion of matter during two episodes of filling the Roche lobe and
change in donor radius at these stages.}
\label{f:Bc_6}
\end{figure}
\begin{figure}[ht!] 
\includegraphics[width=0.5\textwidth]{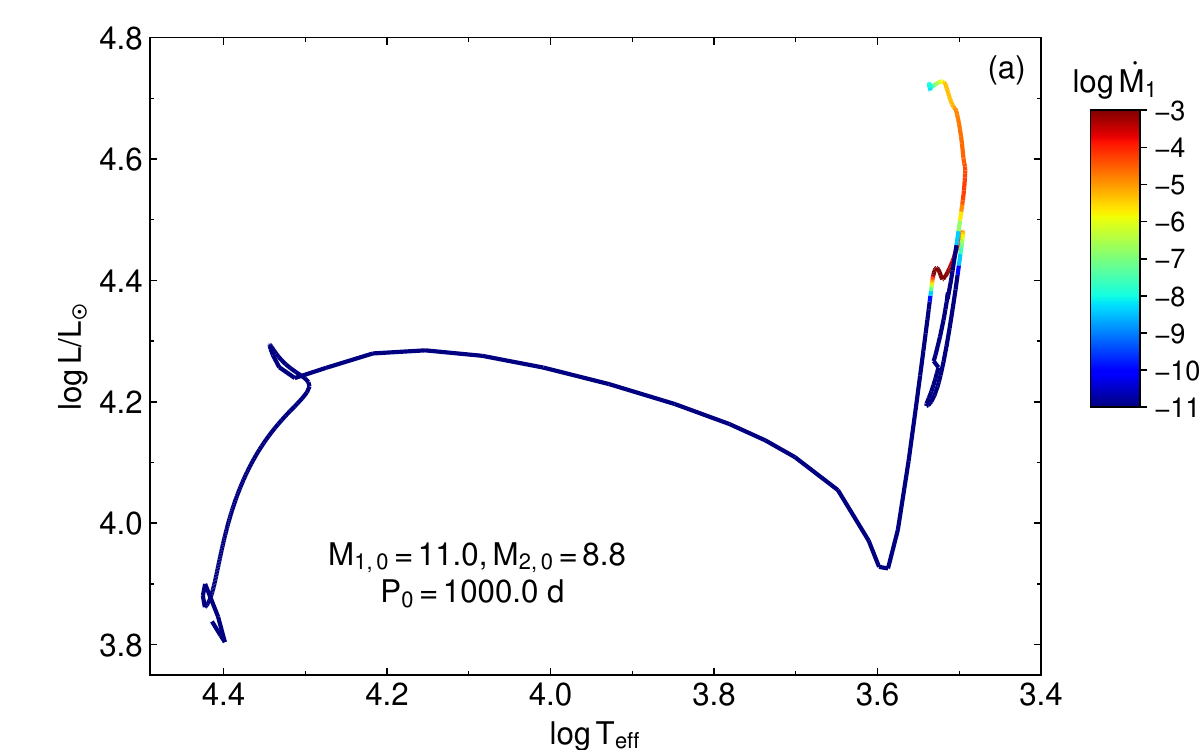} 
\includegraphics[width=0.5\textwidth]{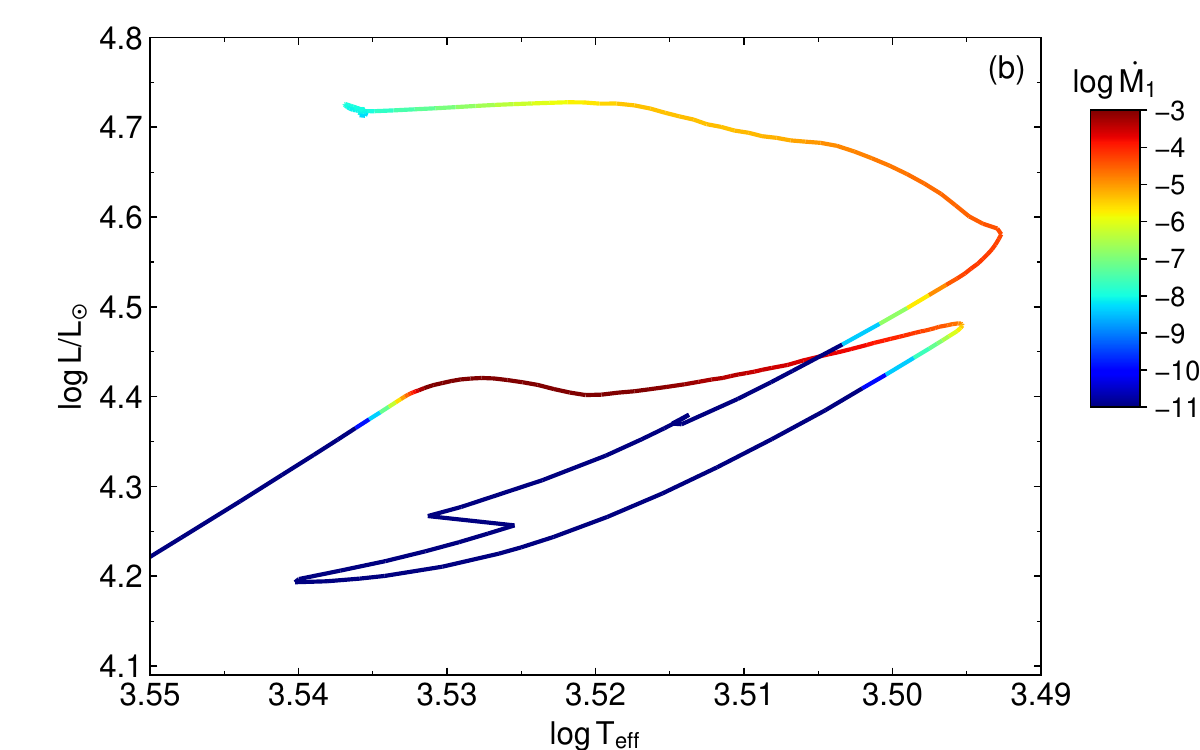} 
\includegraphics[width=0.5\textwidth]{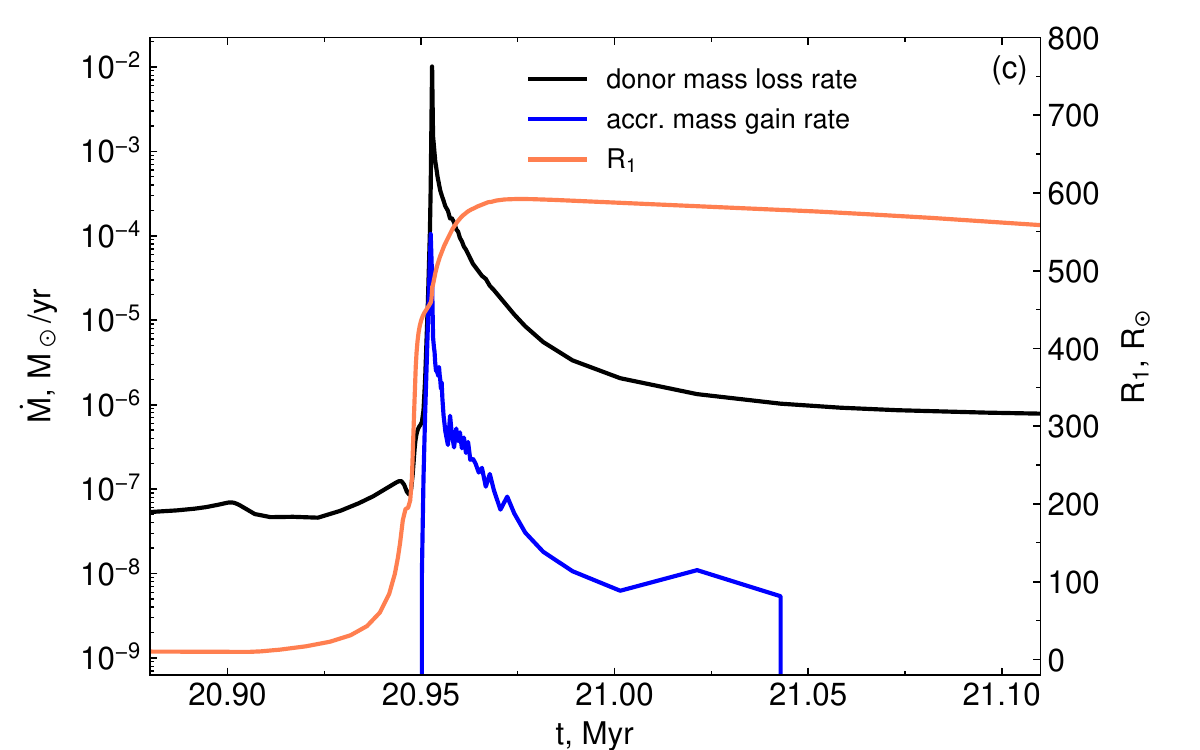} 
\includegraphics[width=0.5\textwidth]{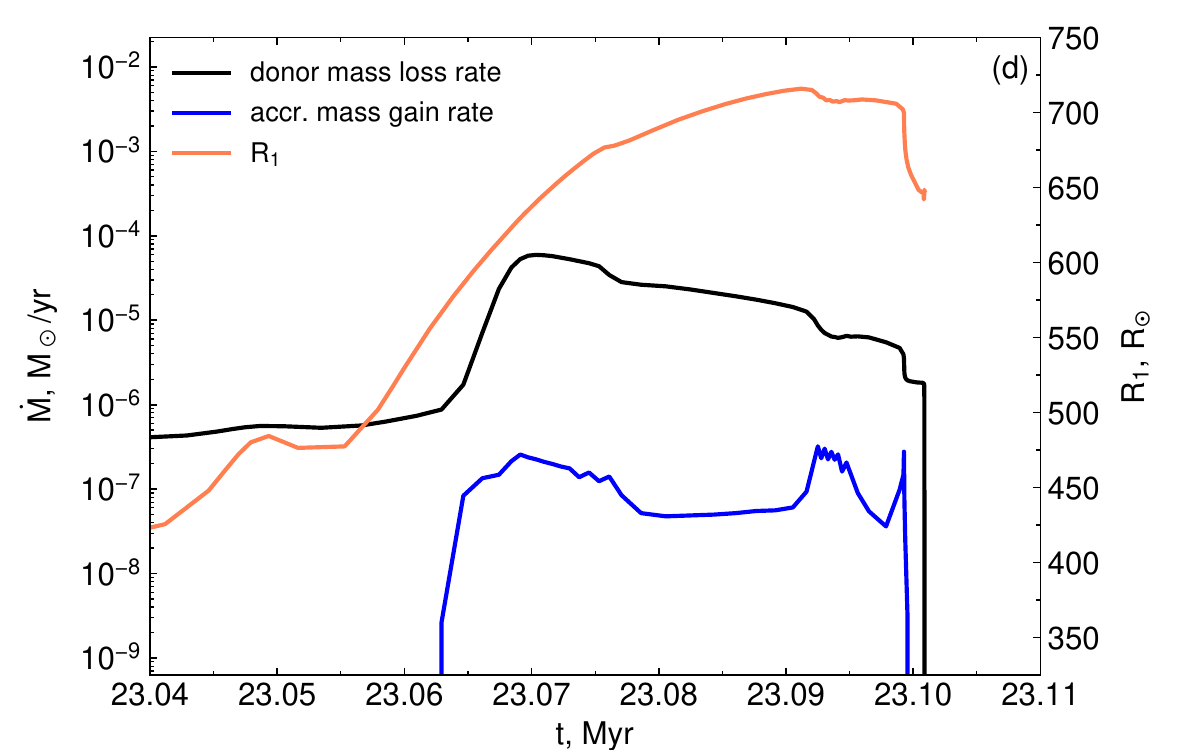} 
\begin{center}
\includegraphics[width=0.5\textwidth]{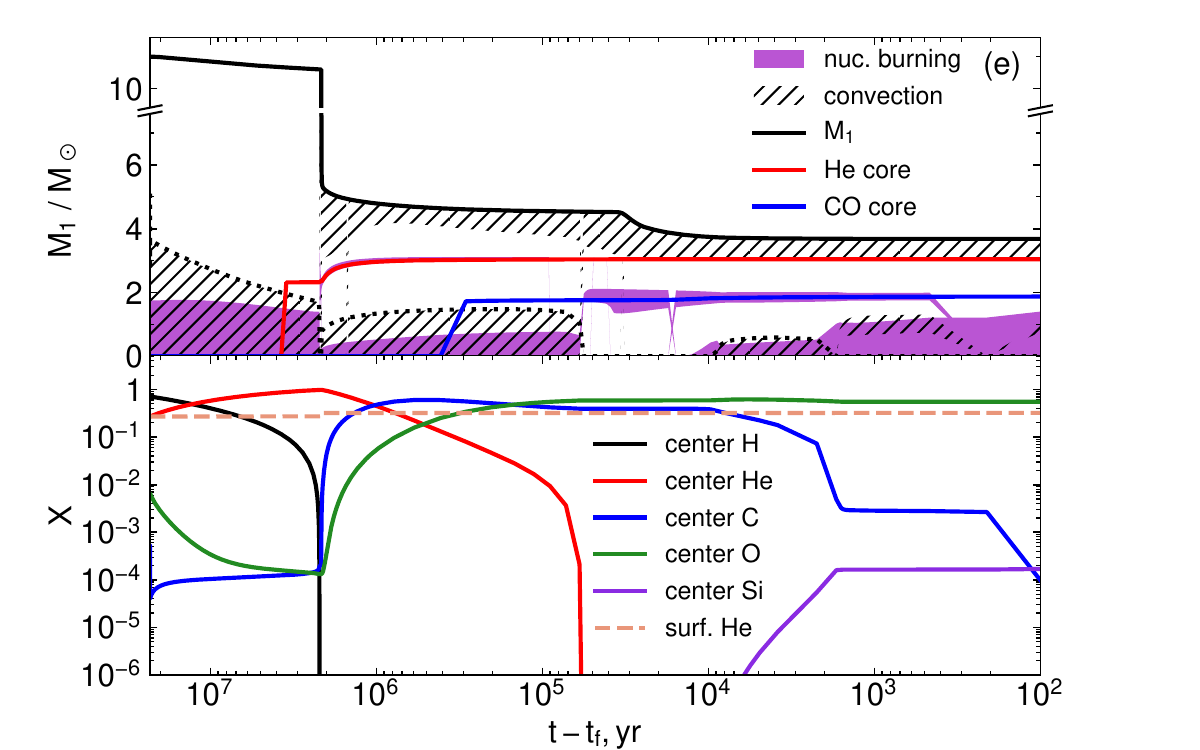} 
\end{center}
\caption{Variation of the model parameters in the system (11+8.8)\,\ms, $P_0=1000$\,day, evolving
in the case Bc.
(a) -- full evolutionary track of the star in the Hertzsprung-Russell diagram,
(b) -- blow-up of the track in the stages with mass exchange,
(c) and (d) -- the rate of mass  loss and accretion during two RLOF episodes
and variations of the donor radius at these stages,
(e) -- Kippenhahn diagram for the donor and the change in the relative abundance of H, He, C,
O, Si in the center of the model and He on its surface.
}
\label{f:Bc_11}
\end{figure}
{\it Cases B and BB of mass exchange.}
Figure~\ref{f:structure} illustrates the changes in the parameters of  
typical close binaries of intermediate mass 
($M_{1,0}=8$\,\ms, $M_{2,0}=4$\,\ms, $ P_0=20$\,day)
and large mass  
($M_{1,0}=16$\,\ms, $M_{2,0}=12.8$\,\ms, $P_0=300$\,day) during evolution.
The difference of the systems is in the different behaviour of the radii of their remnants 
(Paczy{\'n}ski, 1970; Habets, 1986). They experience two (case BB) or one 
(case B) episodes of mass loss, respectively.
In the binary  with $M_{1,0}=8$\,\ms\ the donor first fills the Roche lobe in the hydrogen-shell
burning stage. The second RLOF occurs after He exhaustion in the core\footnote{In MESA, the element 
is considered as ``exhausted'' when its relative abundance by mass decreases to 0.1.
Burning  zones are defined as the regions with
$\epsilon_{\rm nuc} > 10^3 \text{erg/(g} \cdot \text{sec)}$.}.
Evolution of the $M_{1,0}=8$\,\ms\ model was computed almost up to the completion of He-burning 
in the shell and formation of a precursor of a  carbon-oxygen 
white dwarf ($M\simeq\,1$\,\ms) with a low-mass hydrogen envelope. 
The second mass exchange is stable and does not lead to the contact.
This is the result of two factors.
First, the amount of the lost matter is relatively small.
Second, core helium burning time of the accretor is also relatively short, rotation rate of the
accretor does not reduce, it continues to rotate at the velocity rather close to the critical one and 
the amount of the accreted matter is small, since accretion is rotation-limited.  
 
In the system with $M_{1,0}=16$\,\ms\ primary component overflows the Roche lobe
once. Computations were carried out to the beginning of core C-burning. 
The mass of the last computed model is 3.43\,\ms, the mass its CO-core -- 2.48\,\ms, the mass
of the helium shell -- 0.91\,\ms. Hydrogen envelope is completely lost.
A supernova precursor is forming in the system.

In both close binaries, as a result of accretion, secondary components spin up.
Accretion is accompanied by redistribution of the angular momentum in the interiors of the secondary components.
After accretion stops, the star ajustes itself on the thermal time scale, 
rotation velocity drops, but remains higher than before the start of mass exchange and
is about 70\% of the critical value, which is typical for Be stars (Zorec et al., 2016).
A qualitatively similar result with increasing rotation velocity was obtained, for example, by
Wang et al. (2026) in the numerical experiments with a constant rate accretion
onto single rotating stars.

{\it Cases Bc and BcB of mass exchange. }
Of the interest are the relatively rare binaries with $P_0$ from several
tens of day for the systems of intermediate  mass up to $\simeq 1000$ day for massive systems in which
donors overflow Roche lobes shortly before core He-ignition. 
We denote this case as Bc (``B cool''), meaning that stars after mass loss
retain massive hydrogen-helium envelopes and, unlike HeS, do not evolve to
\teff$\apgt$25000K, i.e. remain ``cold''.

In the case Bс, mass loss is interrupted due to the  decrease of the radius accompanying core He-
ignition before the star loses its hydrogen envelope. 
Subsequent evolution occurs near Hayashi border.
Figure~\ref{f:Bc_6} demonstrates the change in the parameters of the system
$M_{1,0}=6$\,\ms, $M_{2,0}=4.2$\,\ms, $P_0=300$\,day.
The mass of the hydrogen-helium envelope remaining after the first RLOF is 
$\simeq 1$\,\ms.
The star refills the Roche lobe in the stage of He-shell ignition, which is 
accompanied by an increase of the radius.
Since the RLOF occurs after core He-exhaustion this case of the evolution is designated
as ВсВ, by analogy with the case ВВ\footnote{Similar case is  
designated as BBB (Tauris et al., 2013), LBB (Yoon et al., 2017), BC (Ercolino et al., 2024).}.
For a significant fraction of computed models, case Bc extends as BcB\footnote{Some of the 
computations were interrupted after core He-exhaustion (at $\mathrm {Y_c=0.01}$).
For several systems, designated as Bc in Fig.~\ref{f:case_type},
computations were interrupted due to the convergence problems.}.
In the example system the star continuously loses matter until H/He envelope
is almost completely lost.
A white dwarf is formed with
the mass of the CO core 1.017\,\ms\ and the mass of the helium mantle 0.003\,\ms.
The mass of WD is slightly larger than the mass of the remnant in the typical case B. 

Figure~\ref{f:Bc_11} shows the changes in the parameters of the binary with 
$M_{1,0}=11\,\ms, M_{2,0}=8.8\,\ms, P_0 = 1000$\,day.
Like in the binaries of lower  mass, two episodes of RLOF occur, at the stages 
of H- and He-burning in the shell sources.
After the end of the first RLOF the star
retains a massive H/He envelope and does not move away from the Hayashi border.
Accretion leads to an increase in the radius of the companion, a main-sequence star,
but by $\simeq$15\% only and no common envelope is formed. 
Computations were carried out to oxygen-ignition in the core. 
The mass of the last computed model is 3.68\,\ms.
The mass of the CO-core of the model is 1.85\,\ms. 
It is surrounded by a 1.19\,\ms\ He-layer.  
At the top there is a H/He shell which retained almost original He abundance
$\rm Y_c=0.28$.
The star is a pre-supernova with the hydrogen in the envelope.
The final configuration is similar to the model Sm11p1000  of Yoon et al. (2017).
In the case of conservative mass exchange, such a binary would have to plunge into a common envelope.
But if accretion is rotation-limited and isotropic reemission occurs, the distance between components increases
more significantly than in the conservative case and the mass loss stops before
the loss of most of the envelope. The star remains ``cold''.

\begin{figure}[ht!] 
\includegraphics[width=0.5\textwidth]{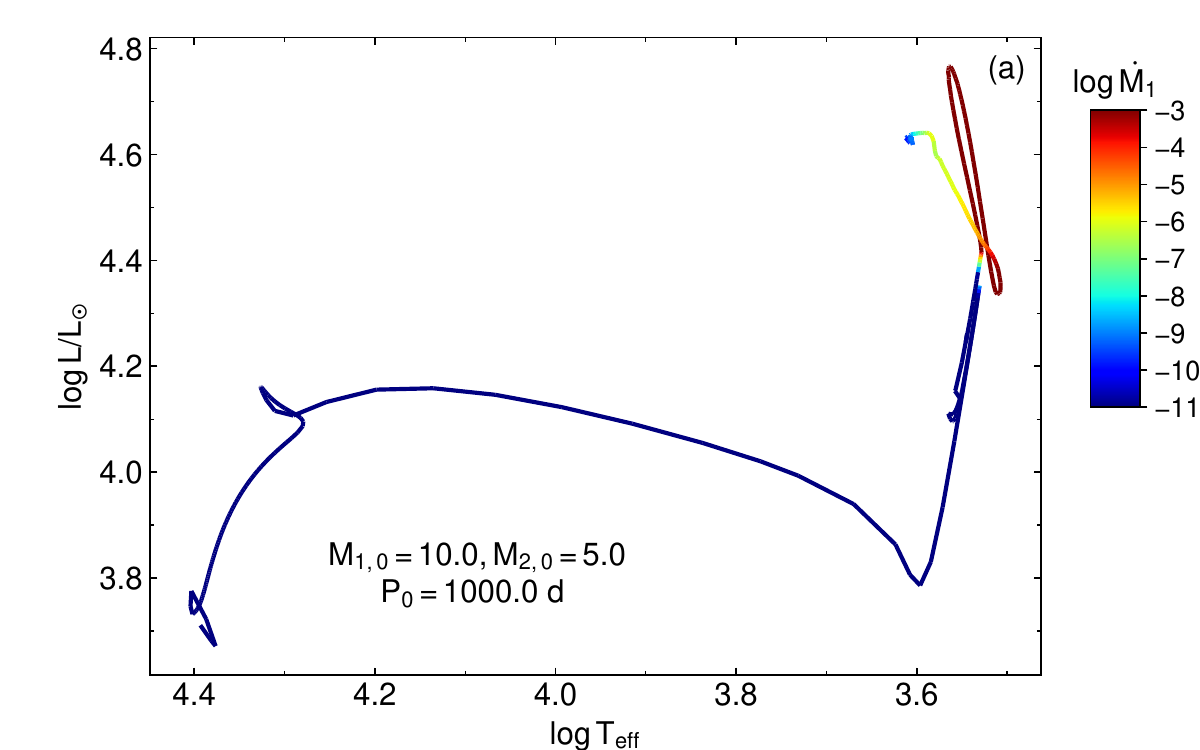} 
\includegraphics[width=0.5\textwidth]{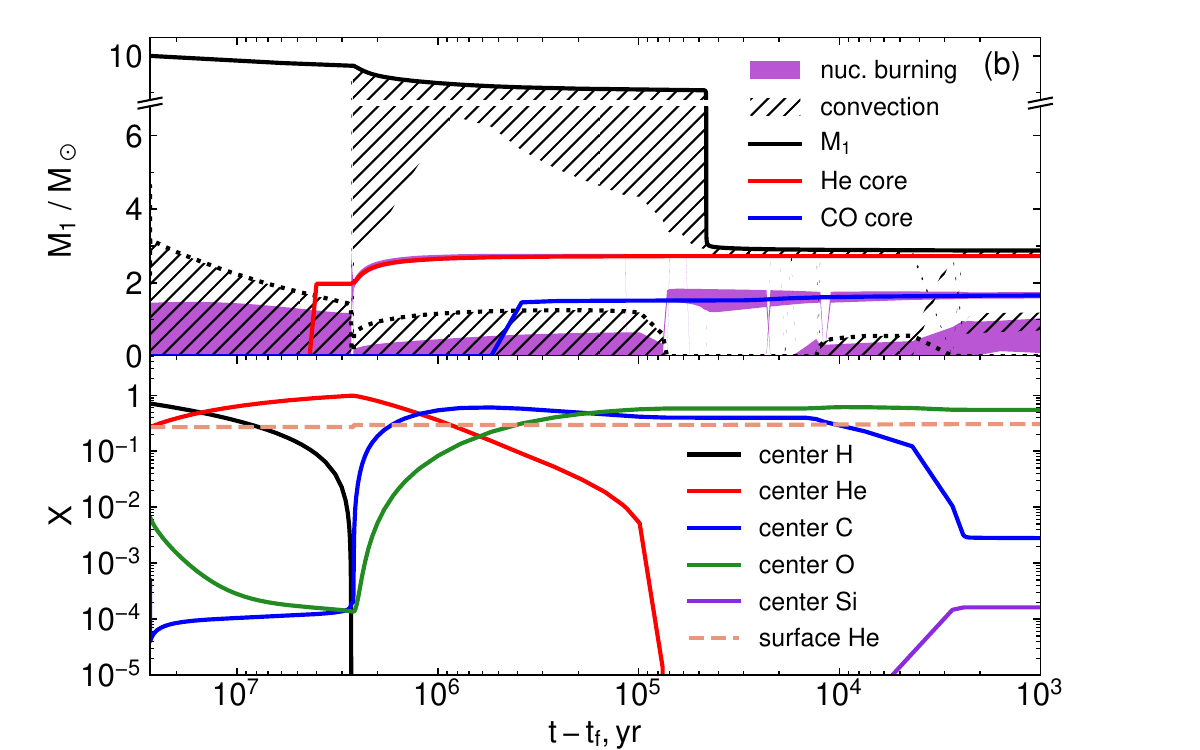} 
\includegraphics[width=0.5\textwidth]{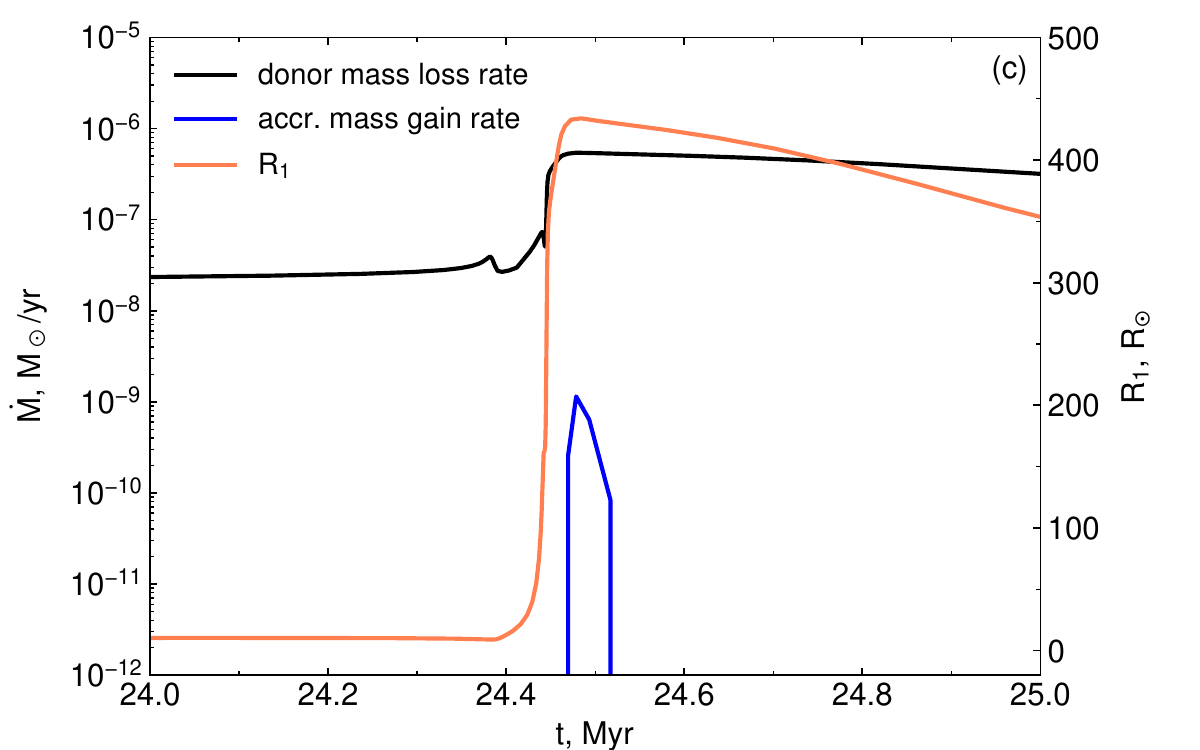} 
\includegraphics[width=0.5\textwidth]{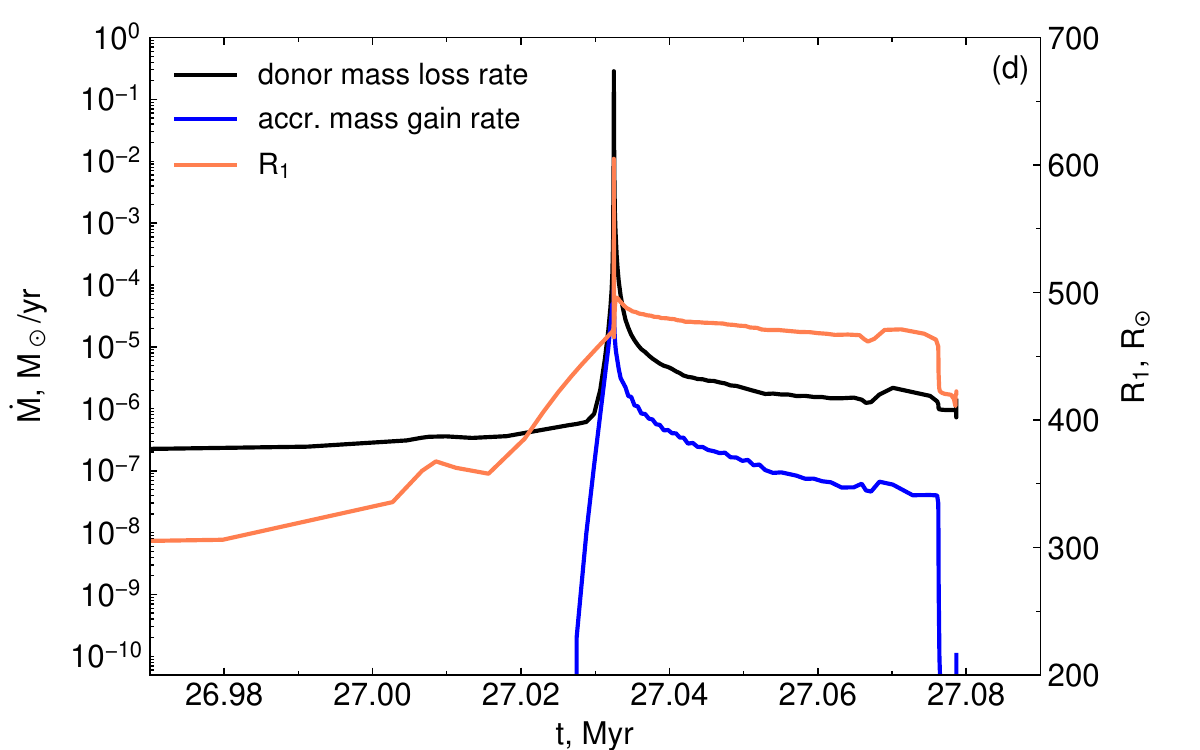} 
\caption{Change of the model parameters in the system (10+5)\ms, $\mathrm {P_0}=1000$\,day,
evolving in the case C.
(a) -- evolutionary track of the star in the Hertzsprung-Russell diagram,
color scale shows mass-loss rate,
{b} -- Kippenhahn diagram and variation of H, He, C, O, Si in the centre of the model
and He on its surface,
(c) and (d) -- changes in the rate of mass loss by the donor, accretion rate 
and donor radius during two episodes of mass loss.
}
\label{f:case-C}
\end{figure}

{\it Case C of mass exchange.}
Evolution in case C is usually associated with the first RLOF occuring 
after He exhaustion in the core (Lauterborn, 1970; Pastetter and Ritter, (1989); Jin et al., 2026; 
Tsai et al., 2026).
Figure \ref{f:case-C} shows the change of the parameters
of a system with $M_{1,0}=10\,\ms, M_{2,0}=5\,\ms, P_0 = 1000$\,day.
The loss of the bulk of the mass by the donor in the He-shell burning stage 
is preceded by the loss of $\simeq$\,2\ms, which is insignificant compared to that in the 
Bc case before He ignition in the core.
RLOF is continuous.
Computations have been carried up to the formation of a supernova precursor.
The total mass of the last model is 2.88\,\ms, it has a CO-core of 1.66\,\ms, and a helium mantle 
of 1.06\,\ms.

In the very final stages of evolution, not considered here,
the stars similar to those 
which evolve in  BcB and C cases
may partially lose their envelopes as a result of the outbursts (e.g., Shiode and Quataert, 2014;
Fuller, 2017; Wu and Fuller, 2021; Ko et al., 2022),
an increase of stellar radius and formation of a common envelope (Wu and Fuller, 2022) or due to 
an enhanced stellar wind (e.g., Ouchi and Maeda, 2017; Matsuoka and Sawada, 2024). However, they 
remain in the red supergiants region and explode as SN\,II of various subtypes.

\begin{figure}[b!] 
\centering
\includegraphics[width=0.55\textwidth]{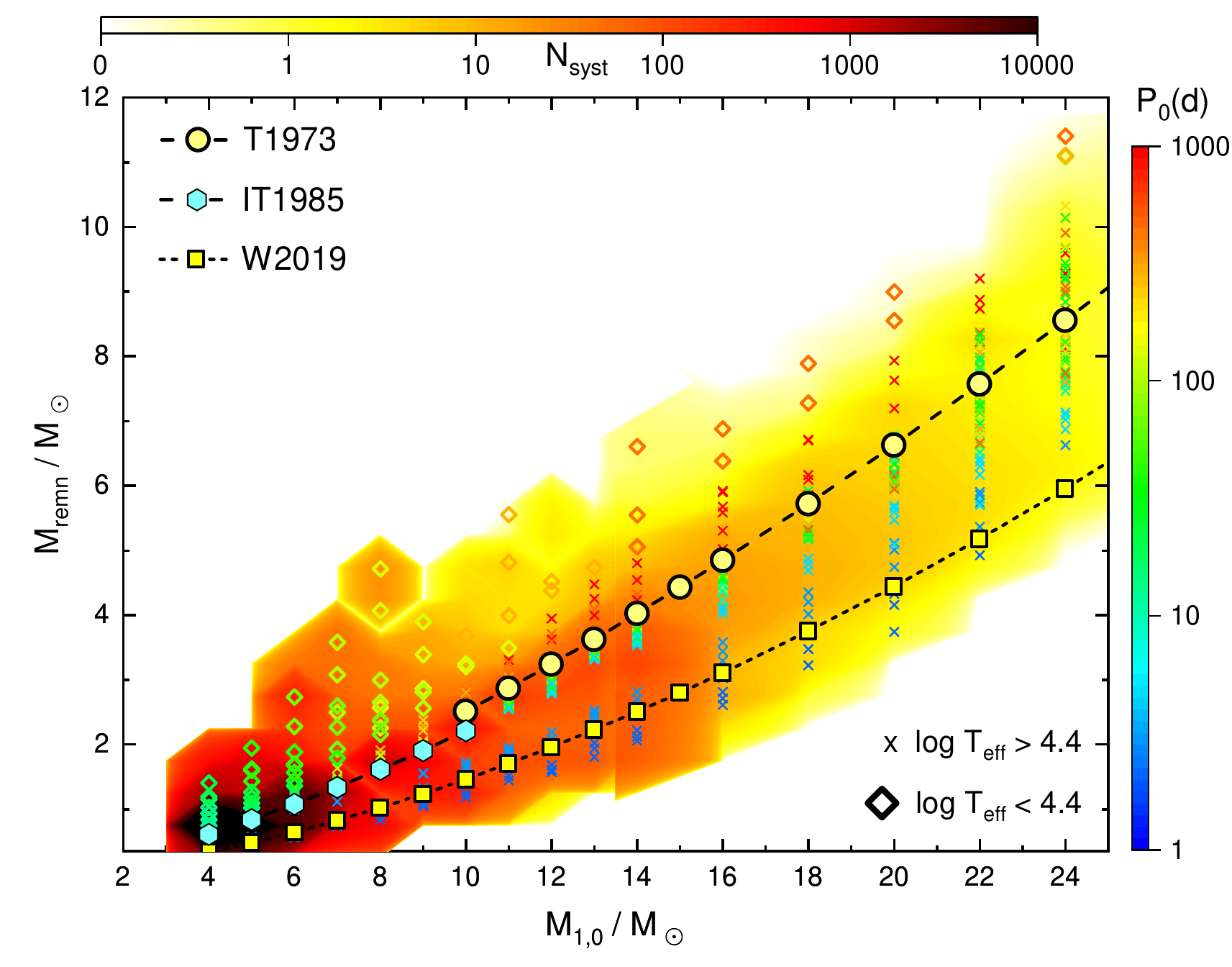}
\caption{Donor remnant masses at the stage of intense core He-burning
(${\mathrm Y_c \approx 0.5}$) for all computed  evolutionary models, depending on their 
ZAMS mass (crosses and diamonds).
Color-coding of symbols shows initial orbital periods of binaries.
The lines show approximations of the remnant mass dependence on the donor ZAMS masses  by
Tutukov and Yungelson (1973, T73), Iben and Tutukov (1985, IT85), and Woosley (2019, W19).
The background shows number distribution of stars.}
\label{f:M_HeS_M0_Y0.5}
\end{figure}

Figure~\ref{f:M_HeS_M0_Y0.5} shows for all
computed models the dependence of masses of donor remnants after mass exchange 
at the stage of intense core He-burning (${\mathrm Y_c \approx 0.5}$) on $M_{1.0}$.
Along evolutionary tracks, at the area of the HRD where 
they have ${\mathrm Y_c \approx 0.5}$ the stars  spend most of their evolutionary 
lifetime  before forming a white 
dwarf, refilling the Roche lobe or exploding as a supernova.
Color scale in the plot shows the dependence of remnants masses $M_{\rm remn}$ on
their initial orbital periods $P_0$.
The masses of remnants  depend significantly on $P_0$.
The remnants are divided into two groups, since we identify HeS with the stars with
${\rm T_{eff}} \geq 25000$K.

For comparison, we show approximations for dependence of $M_{\rm remn}$ on $M_{1,0}$,
based on the calculations of the evolution of primary components of close binaries with masses
$M_{1,0}$=(4 -- 10)\ms: $M_{\rm remn} \approx 0.08 M_{1,0}^{1.4}$ (Iben and Tutukov, 1985) 
and with $M_{1,0} \geq $10\msun:
${\rm M_{\rm remn}} \approx {\rm 0.1M_{1,0}^{1.4}}$ (Tutukov and Yungelson, 1973).
Also shown is an estimate of the masses of cores of single stars with 
$M_{\rm ZAMS}\leq 30\ms$, which is often used as an approximation for the masses of helium stars:
$M_{\rm He} \approx 0.0385M_{\rm ZAMS}^{1.603}$ (Woosley, 2019).
Iben and Tutukov (1985) relation was derived for the models of close binaries  with 
mass exchange in the relatively late case B, but it, however,  slightly  underestimates
the masses of the remnants in the systems with $P_0$ less than a few tens of days.
Tutukov and Yungelson (1973) relation for the models of more massive stars was derived for the 
early case B and it quite satisfactorily reproduces remnant masses for
$P_0$ of the order of tens to several hundred days, due to the rapid evolution of stars.
Woosley (2019) relation significantly underestimates the
masses of HeS, especially massive ones.
For the case Bc masses of remnants of close binary primaries deviate significantly from the general dependence.

Masses of the cores of close binary components, which are close to the remnant masses,
depend on the adopted convective mixing parameters.
Therefore, for example, the mass spread of remnants of stars with $M_{1.0} \leq 6$\,\ms\ is larger 
than that found by Arancibia-Rojas et al. (2024), who also used MESA.

As the background, Fig.~\ref{f:M_HeS_M0_Y0.5} shows number distribution of the models in
the coordinates $\mathrm {M_{1,0} - M_{remn}}$ after taking into account their formation rate and lifetime
from termination of RLOF  to the stage at which $\mathrm {Y_c \approx 0.01}$.
The Figure suggests that the population of HeS should be dominated by the stars with masses
$\aplt$\,2\ms, which observationally will be identified as massive subdwarfs.

In our calculations, the lower limit of the mass of remnants in which iron core collapse with 
subsequent formation of NSs can occur is approximately 2.5\,\ms.
The corresponding $M_{1,0}\in(10 - 12)$\,\msun\ depending on $P_0$.
In the models with this mass, Si-burning  occurs, followed by
the formation of an iron core and collapse.
The value of 2.5\,\ms\ is consistent with the lower mass limit for helium stars
that undergo  Si-burning burst, found by Woosley (2019) for non-rotating helium stars.

In the  HeS  with CO-core masses of approximately (1.37 - 1.43),\ms\ 
after carbon exhaustion electrons are captured in the ONe-core, and these stars can collapse or explode
as peculiar SN~Ia  (Miyai et al., 1980; Isern et al., 1991; Holas et al., 2026).
The range of their progenitor masses depends
on the adopted convective mixing parameters (Chanlaridis et al., 2022).

\begin{figure}[t!]   
\includegraphics[width=0.5\textwidth]{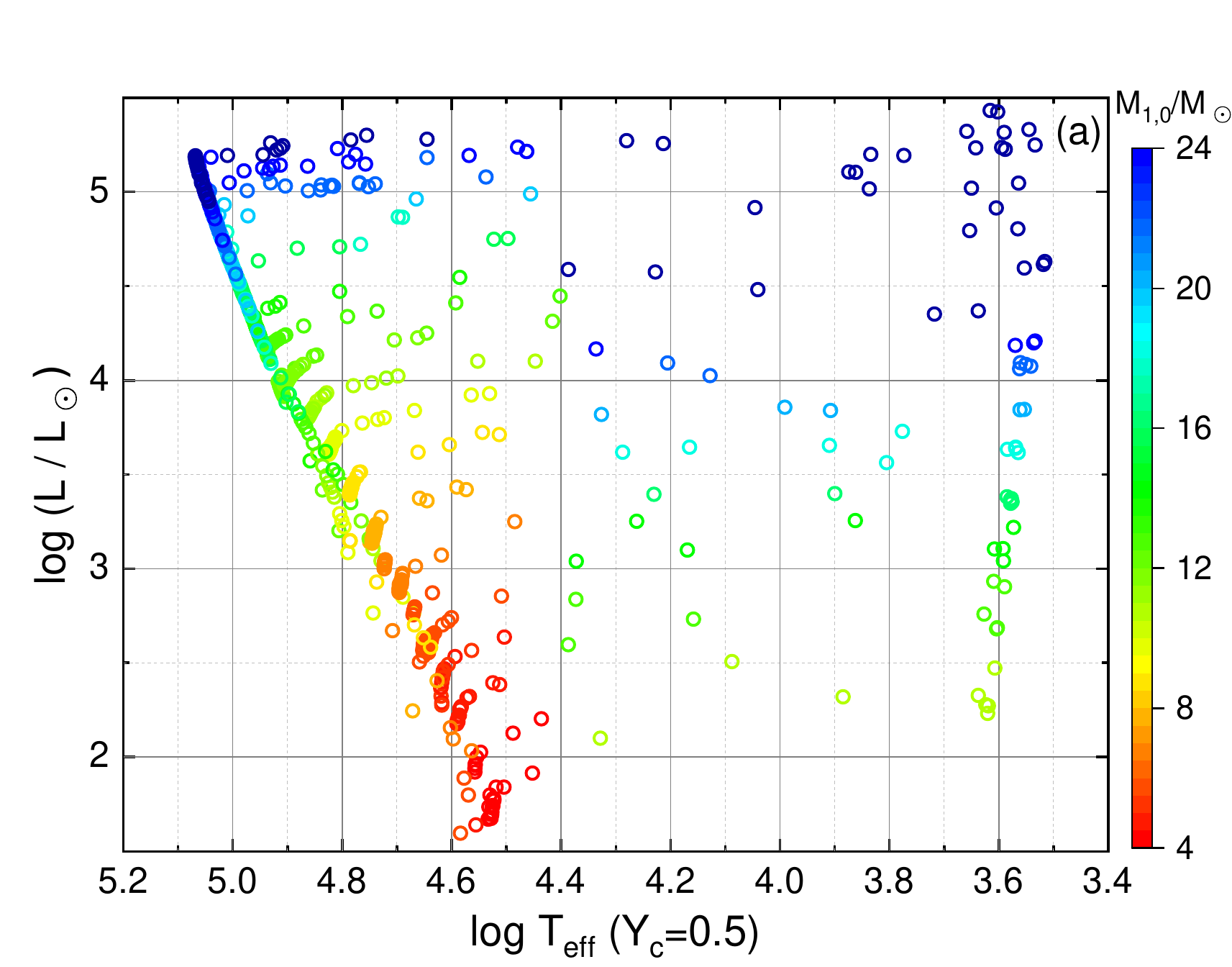}
\includegraphics[width=0.5\textwidth]{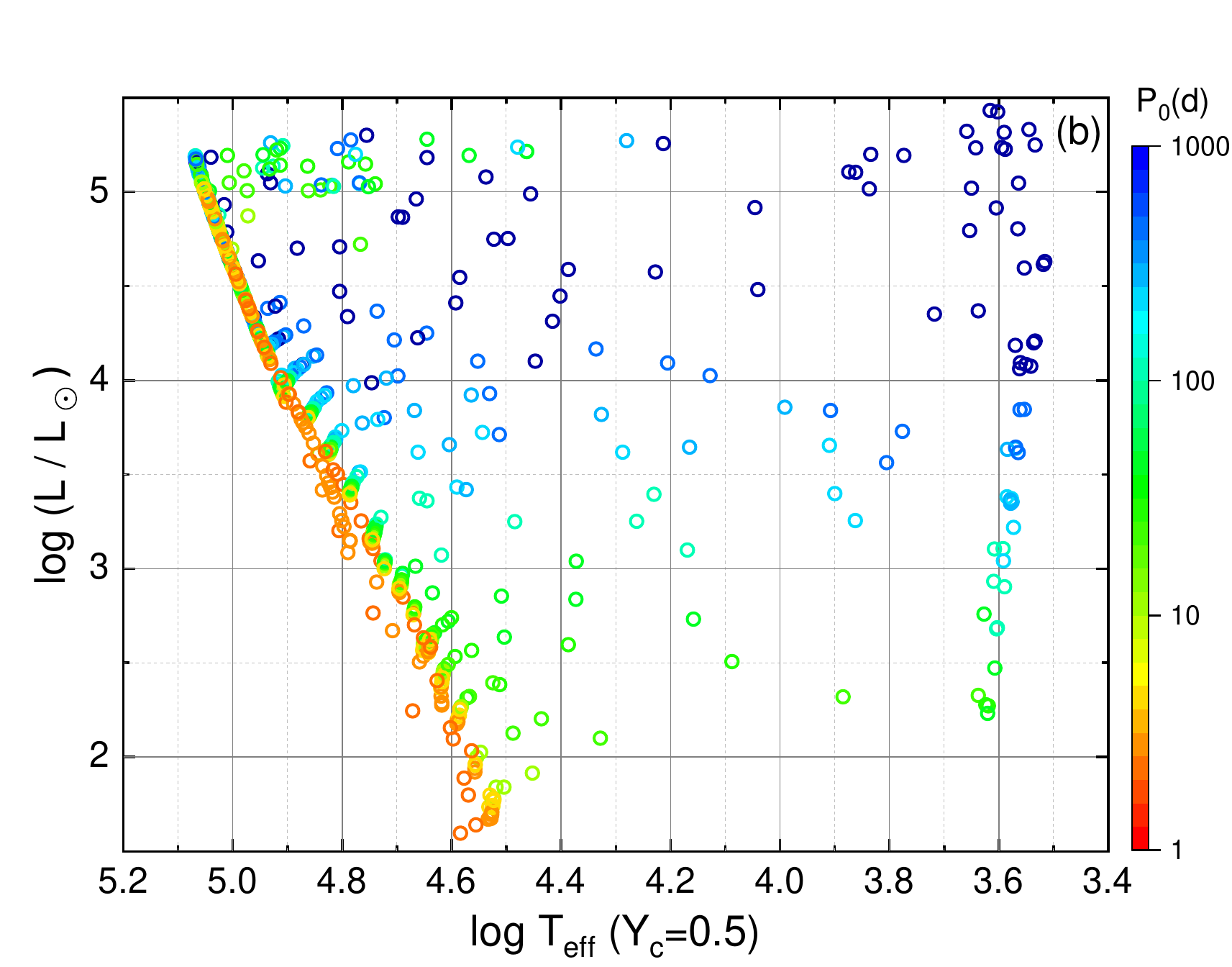}
\includegraphics[width=0.5\textwidth]{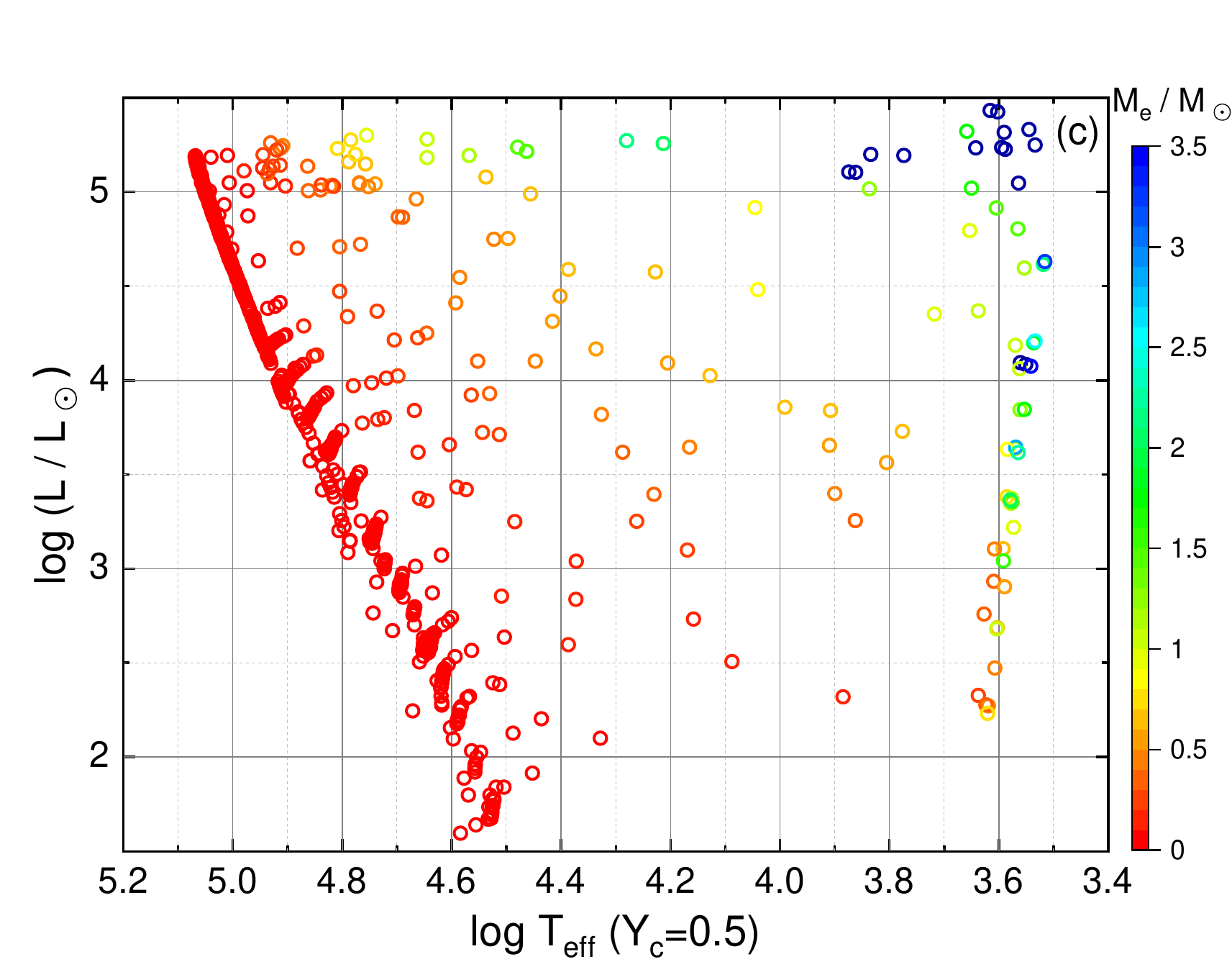}
\includegraphics[width=0.5\textwidth]{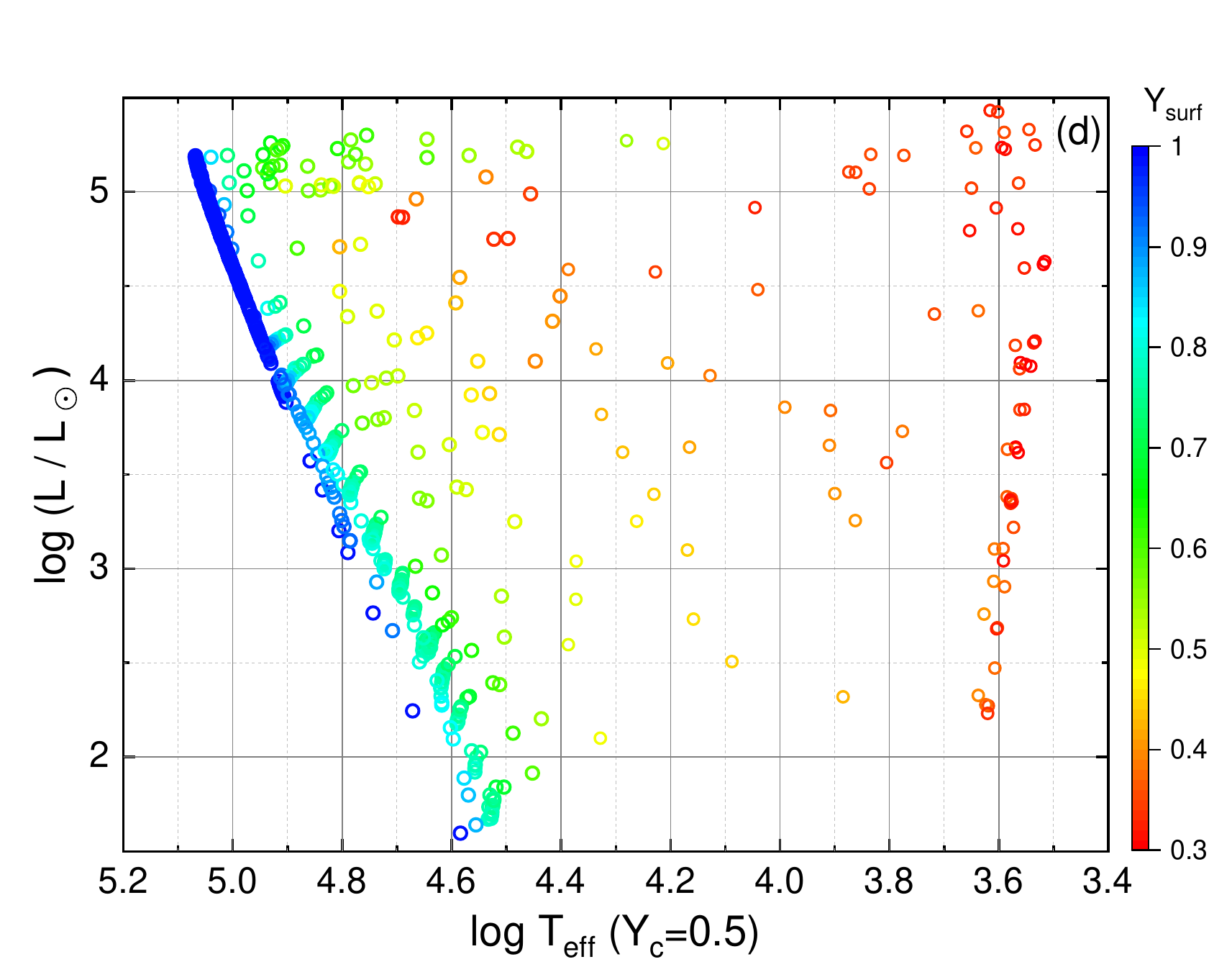}
\caption{Models of donors remnants in the HRD at the time when $\mathrm {{Y_c} \approx 0.5}$.
Color-coded in panel (a) are the masses of the models in HRD,
in panel (b) -- orbital periods ,
in panel (c) -- masses of stellar envelopes,
in panel (d) -- He abundance on the surfaces of the models.
} 
\label{f:M_HeS_M0_Menv}
\end{figure}

Panels (a) and (b) of Fig.~\ref{f:M_HeS_M0_Menv} show positions of the donor remnants in 
the Hertzsprung-Russell diagram as the functions of $M_{1,0}$ and $P_0$.
The models in Fig.~\ref{f:M_HeS_M0_Menv} correspond to the time of intense He burning in the core
($\mathrm {{Y_c} \approx 0.5}$).
A noticeable concentration of low-mass 
models from initially relatively short-period 
binaries is observed in the ``hot'' $(\log{(\teff)} \apgt 4.4)$ part of the diagram. 
Observed objects in this area of the diagram are hot subdwarfs
sdB and sdO $(\log(\teff) \approx (4.4 - 4.7),~\log(L/\ls) \approx (1.0 - 2.5))$ (Heber, 2016).
The most massive models are located in the part of the diagram occupied by nitrogen Wolf-Rayet stars
($\log(\teff) \apgt 4.4,~\log(L/\ls) \apgt 5$, (Shenar et al., 2020b)).

As the initial mass of the donors increases, also grows the number of models in the 
relatively low-temperature area of the diagram.
This is due to an increase in the masses of the retained H/He envelopes (panel (c)).
Over entire range of \teff\ there are remnants of the most massive stars, the components of the 
systems with $P_0$ longer than several tens of days.
Concentration of models corresponding to the case  Bc of mass exchange is noticeable near the Hayashi border.
They are also characterized by the most massive envelopes.
Panel (d) of Fig.~\ref{f:M_HeS_M0_Menv} shows He abundance at the surface of the remnants.
The overwhelming majority of the models have $Y_{surf} \apgt (0.7 - 0.8)$.
The original He abundance is retained by the remnants of the donors that have been formed
in the case Bc of RLOF.
\begin{figure}[t!] 
\centering
\includegraphics[width=\textwidth, trim={0 1cm 0 1cm},clip]{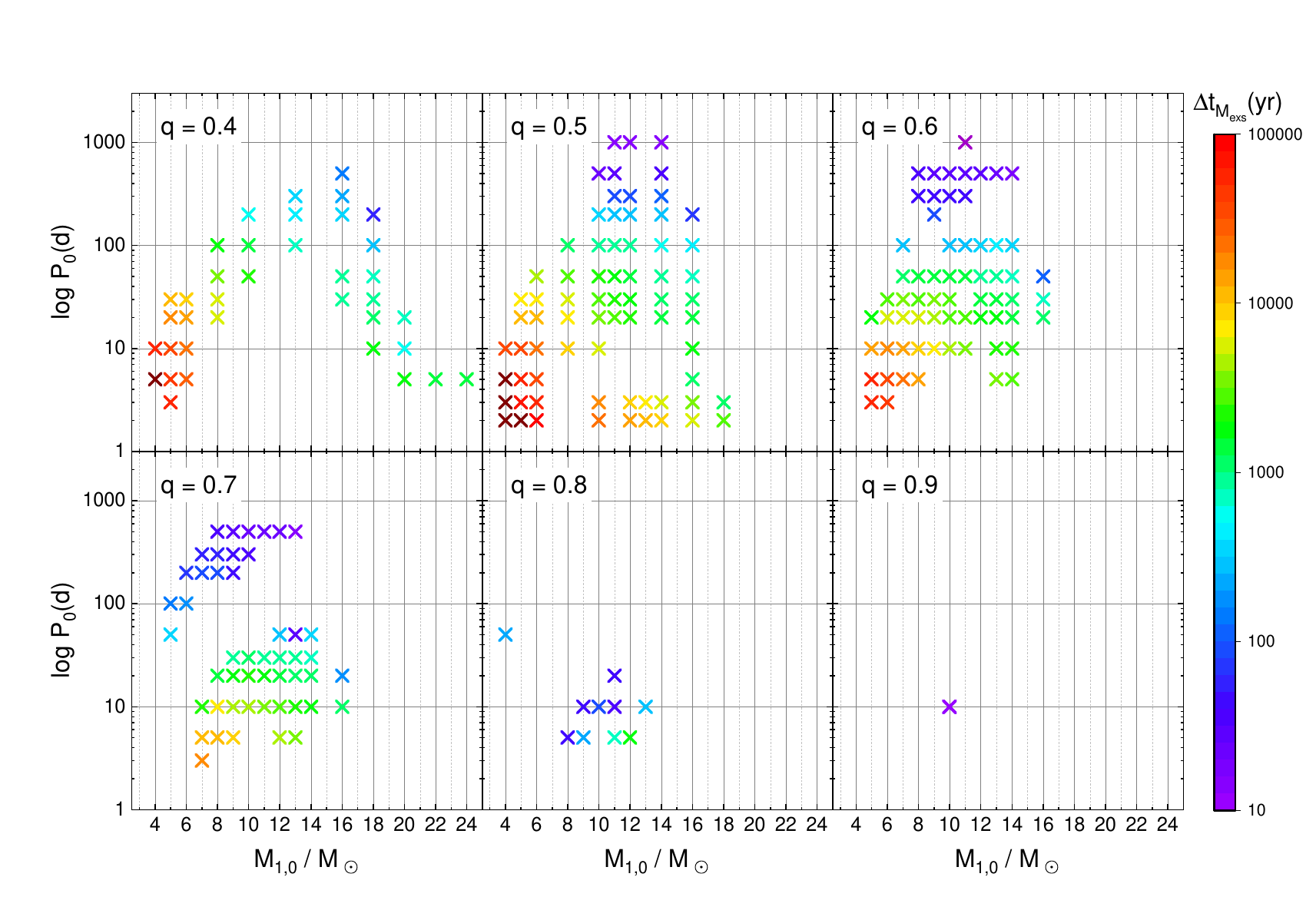}
\includegraphics[width=\textwidth, trim={0 1cm 0 1.5cm},clip]{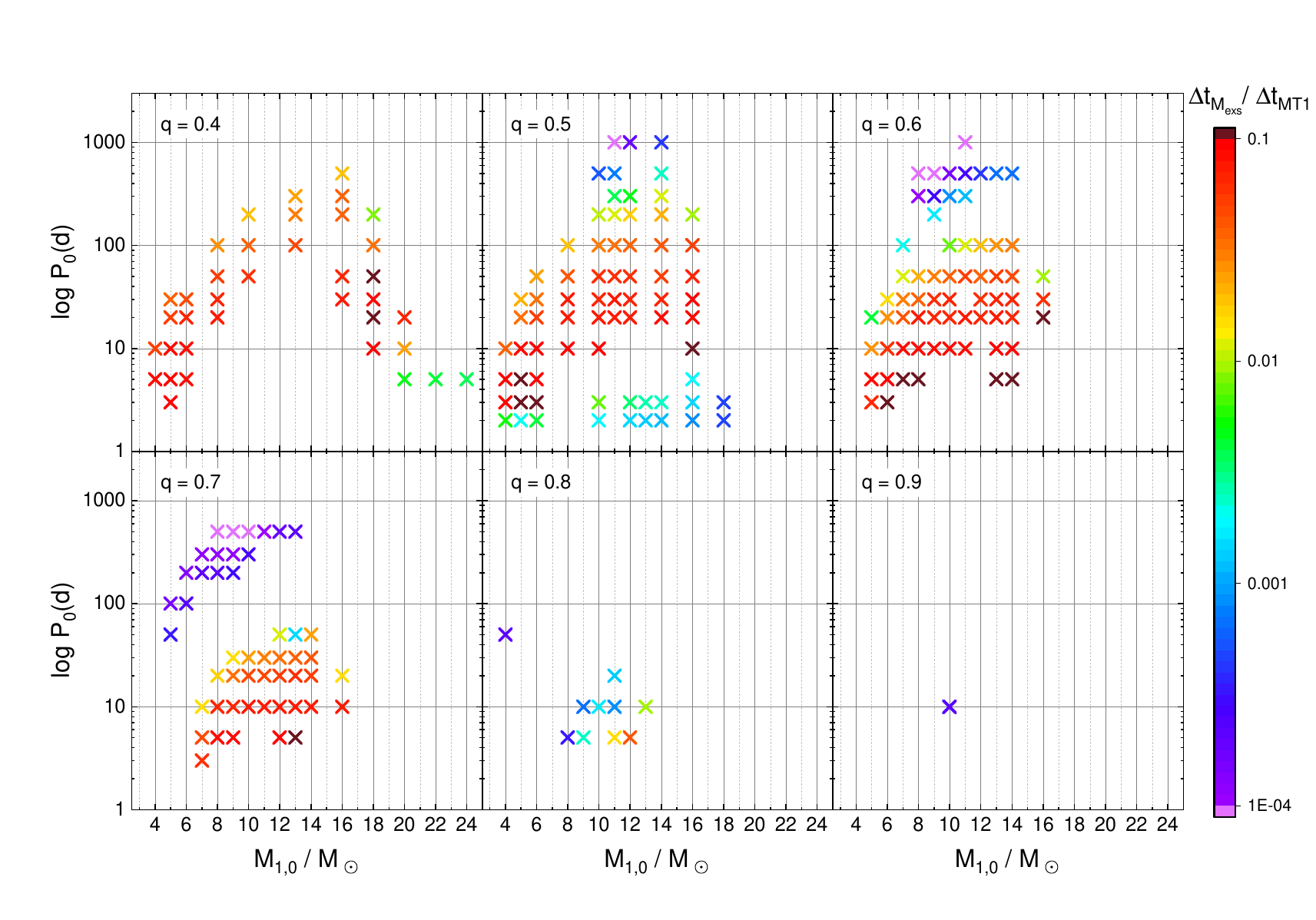}
\caption{Upper panel -- time spent by the donors in the $\dot{M} > \dot{M}_{\rm exc}$ regime.
Bottom panel -- the ratio of the time spent by donors in the regime $\dot{M} > \dot{M}_{\rm exc}$,
and total time of the mass-transfer stage.
}
\label{f:delta-mexc1-2}
\end{figure}
\begin{figure}[ht!] 
\centering
\includegraphics[width=\textwidth, trim={0 1cm 0 1cm},clip]{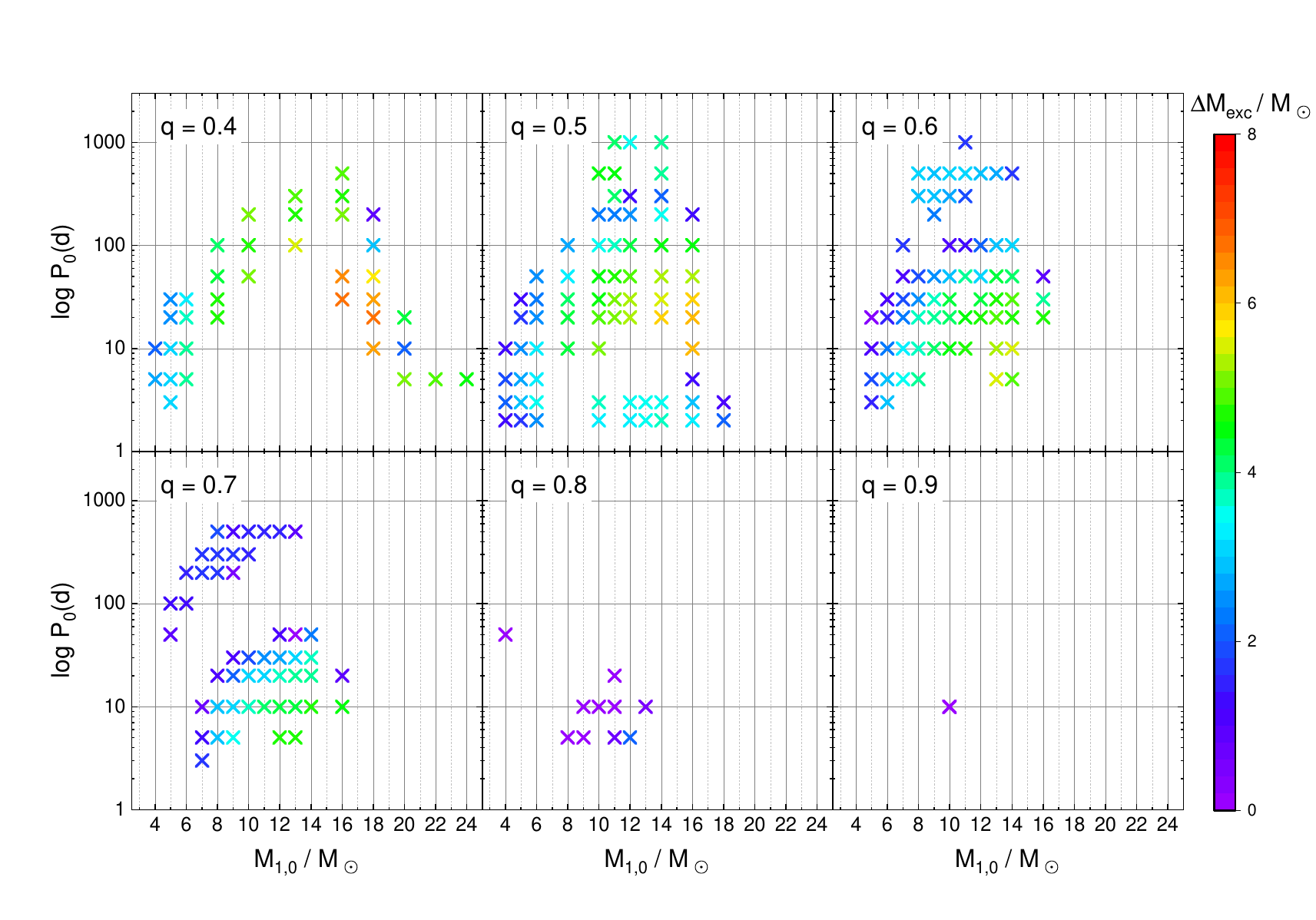}
\includegraphics[width=\textwidth, trim={0 1cm 0 1.5cm},clip]{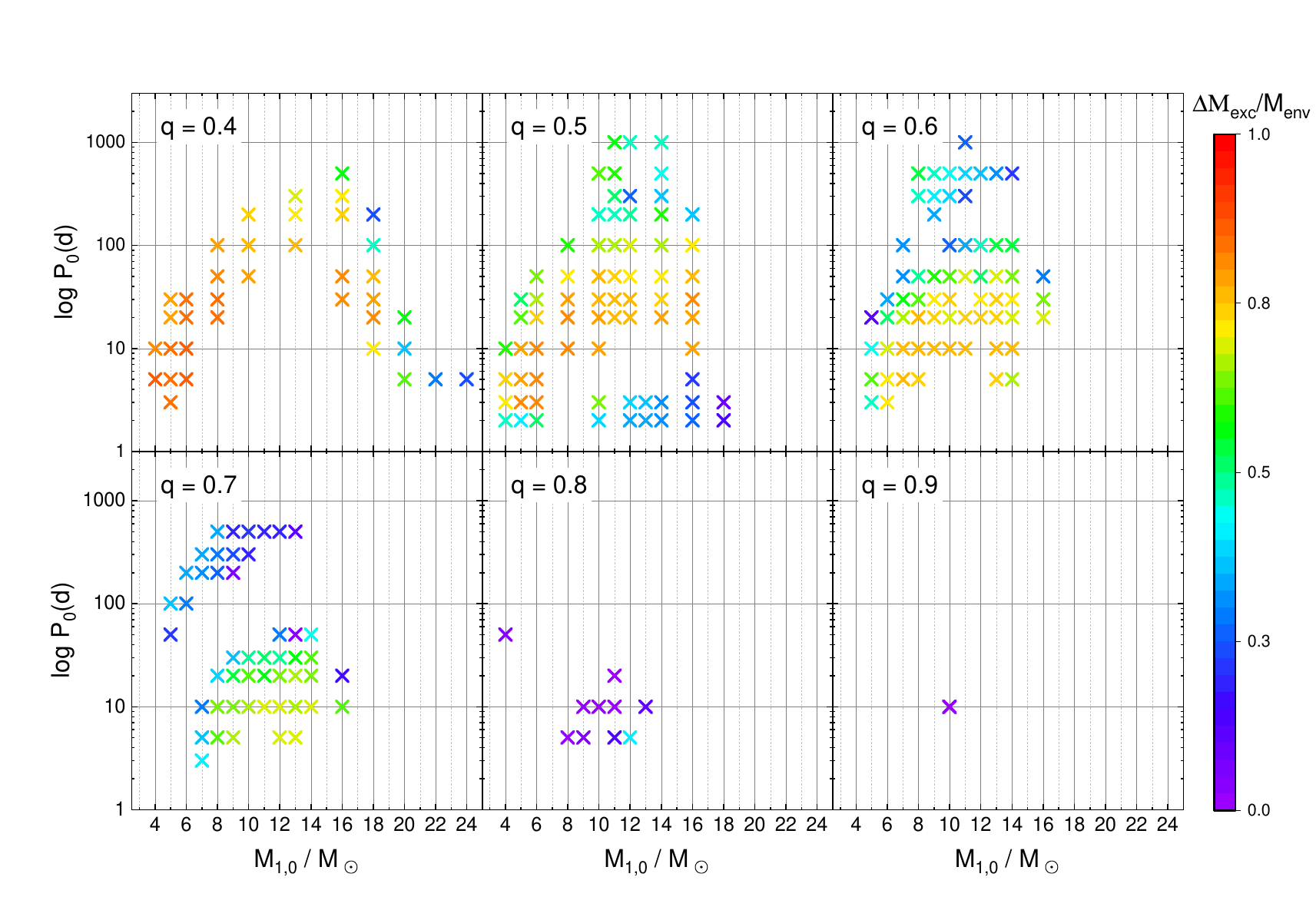}
\caption{Upper panel --  mass lost by the donors while
$\dot{M} > \dot{M}_{\rm exc}$ (color scale).
Bottom panel -- ratio of the mass lost by the donor in the mode
$\dot{M} > \dot{M}_{\rm exc}$ and the total mass lost by the donor.
}
\label{f:delta-mexc3-4}
\end{figure}

Upper panel of Fig.~\ref{f:delta-mexc1-2} shows for various values of $q_0$ the time over which mass-exchange process is dynamically stable, but
$\dot{M} > \dot{M}_{\rm exc}$ (Eq.~(\ref{eq:mlossmax})), while the lower panel shows
the ratio of this time and the total mass-transfer time
$\Delta t_{\rm M,exc} / \Delta T_{\rm MT1}$.
Figure~\ref{f:delta-mexc3-4} shows the mass $\Delta M_{\rm exc}$ that is dynamically stably 
transferred in the regime $\dot{M} > \dot{M}_{\rm exc}$
and the ratio of this mass and the total mass lost by the donor
$\Delta M_{\rm exc} / M_{\rm env}$.
In the regime where radiation pressure is {\it presumably} incapable of removing non-accreted matter 
from the system, the highest ratio
$\Delta M_{\rm exc} / M_{\rm env}$ is typically found for stars
with masses up to (14 - 16)\,\ms\ in the systems with $q \lesssim 0.8$,
producing relatively low-mass stripped helium stars.
For these same close binaries, the ratios $\Delta t_{\rm M,exc} / \Delta T_{\rm MT1} \lesssim 0.1$.
Based on this, it can be concluded that just in these systems formation of the common envelopes by
non-accreted matter is most likely.

\begin{figure}[] 
\vspace{-3.cm}
\includegraphics[width=\textwidth]{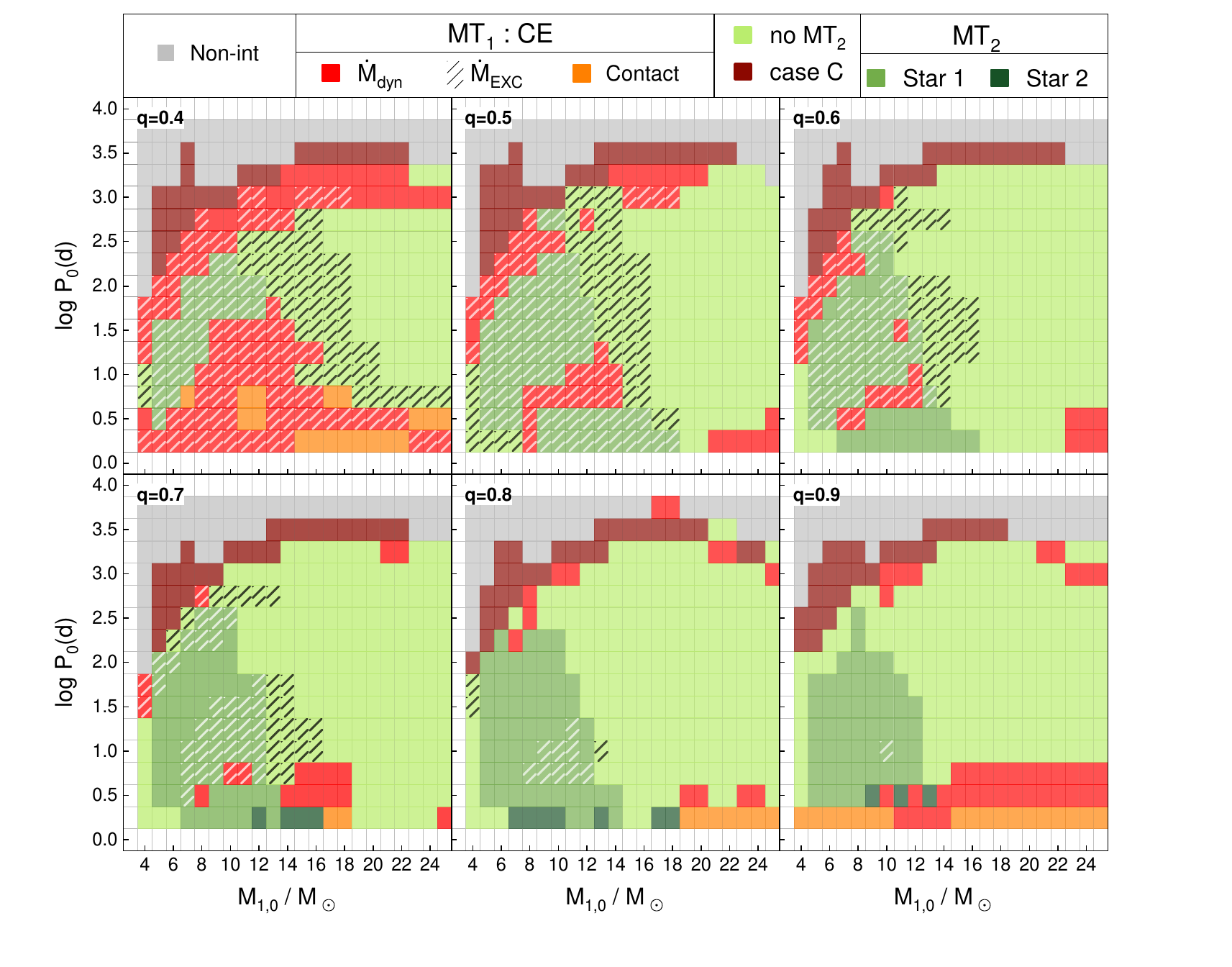}
\caption{Distribution of close binaries over evolutionary channels depending on the initial masses
of primary components, orbital periods, and  mass ratios of components $q$.
Non-interacting systems (Non-int) are marked gray. 
In brown  are the areas of systems in which case C of mass-exchange occurs
and the donor fills the Roche lobe almost continuously.
Areas of systems in which common envelopes ($\mathrm {MT_1:CE}$) may be formed  during the first mass 
exchange as a result of the mass loss  by the donor on the dynamical
time scale ($\mathrm {MT_1:\dot{M}_{dyn}}$) are in red. 
In the orange areas $\mathrm {MT_1:Contact}$ components of close binaries come into contact during the first RLOF.
Hatching  ($\mathrm {\dot{M}_{EXC}}$)
means that in these areas accumulation of non-accreted matter can result in the formation of a common envelope.
Shades of green show the areas of parameters for which the first mass exchange occurs stably.
Light green no~($\mathrm {MT_2}$) indicates the areas of parameters for which stars do not overflow Roche lobes for the second time. 
In green areas (MT$_2$: Star~1) the second RLOF occurs.
By dark green (MT$_2$: Star~2) are highlited parameters areas for which the secondary component of 
the  binary system overflows its Roche lobe before He is exhausted in the core of the primary. 
 }
\label{f:outcomes}
\end{figure}
\begin{table}[t]
\begin{center}
\caption{Formation rate of close binaries with initial components masses (4 - 24)\,\ms\ evolving
over different ecolutionary channels (in yr$^{-1}$).} 
\label{t:outcomes}
 \begin{tabular}{c|c c c c| c c}
 \hline
 \hline
 &   &              &               &            &     \multicolumn{2}{c}{The second } \\
Case of  & \multicolumn{4}{c|}{The first RLOF} 
&\multicolumn{2}{c}{RLOF } \\  
\cline{2-7}
mass exchange     & Stab.         & Exc.          & Dyn.         & Cont.      & S1            & S2 \\
\hline
  (1)       &      (2)     & (3)          &  (4)         &  (5)        &     (6)          & (7)  \\    
\hline  
 A, AB, ABB &     8.1$\times 10^{-4}$  &     2.7$\times 10^{-4}$  &     3.9$	\times 10^{-4}$  &     2.7$\times 10^{-4}$  &    5.1$\times 10^{-4}$  & 1.1$\times 10^{-4}$ \\         
    B,  BB  &     5.7$\times 10^{-3}$  &     3.0$\times 10^{-3}$  &     2.1$\times 10^{-3}$  & 2.5$\times 10^{-5}$ &    1.8$\times 10^{-3}$  &   \\
 Bc, BcB    &     5.9$\times 10^{-4}$  &     3.1$\times 10^{-4}$  &     1.3$\times 10^{-4}$  &              &    7.3$\times 10^{-4}$  &   \\
 \hline
 \hline
 \end{tabular}
\end{center}
\begin{spacing}{1.0}
\footnotesize{{\bf Notes.} 
Column (2) -- stable mass exchange during the
first RLOF and after core-He-exhaustion;
(3) -- a subset of systems with stable mass-exchange, in which common envelopes could have formed 
due to the inability to remove non-accreted matter (a fraction of the systems from the column (2));
(4) -- systems in which during the first RLOF donor loses matter on the dynamical time scale;
(5) -- components come into contact during the first RLOF;
(6) -- second RLOF after core He exhaustion;
(7) -- RLOF by the secondary component before the second RLOF by the remnant of the primary.
The numbers in the ``Stab.'' and ``Exc.'' columns in Table \ref{t:Noutcomes17}
refer to the same evolutionary channels as in this Table.
}
\end{spacing}
\end{table}
\begin{table}{}
\begin{center}
\caption{The number of helium stars formed via different evolutionary channels and of 
their progenitors.}
\label{t:Noutcomes17}
\begin{tabular}{c| c c |c c}
 \hline
 \hline
   & \multicolumn{2}{c|}{Donors} & \multicolumn{2}{c}{Helium stars}\\  
Case of & \multicolumn{2}{c|}{$4\leq {\mathrm {M_{1,0}/M_\odot}} \leq 24$} &  
 \multicolumn{2}{c}{$1\leq {\mathrm {M_{HeS}/M_\odot}} \leq 7 $}  \\ 
mass exchange    &  Stab.         &  Exc.        & Stab.        & Exc.  \\
 \hline
   ~~(1)           & (2)      & (3)          &  (4)          & (5)\\ 
 \hline
A, AB, ABB &     3.2$\times 10^4$    &     1.1$\times 10^4$  &     2.5$\times 10^3$   & 7.8$\times 10^2$  \\
   B, BB   &     9.0$\times 10^4$    &     3.6$\times 10^4$  &     2.6$\times 10^4$   & 1.3$\times 10^4$ \\
 Bc, BcB   &     2.1$\times 10^4$    & 4.4$\times 10^3$  &     1.5$\times 10^4$   & 1.0$\times 10^3$   \\
 \hline
 \hline
 \end{tabular}\\
 \end{center}
\begin{spacing}{1.0}
\footnotesize{{\bf Notes.}
Selected subset of the systems with stable mass exchange where common envelopes may form 
due to impossibility to remove non-accreted matter. 
 (Exc. -- a fraction of systems from cols.  (2) and (4)).}
\end{spacing}
\end{table}

Corresponding results of computations are summarized in Fig.~\ref{f:outcomes} and Tables \ref{t:outcomes} and
\ref{t:Noutcomes17}.
Figure~\ref{f:outcomes} shows combinations of initial parameters $q_0$, $M_{1,0}$ and $P_0$  
for which evolution leads to the merger of components during the first mass-exchange, single RLOF
resulting in formation of a close binary with a white dwarf or a
supernova precursor and a main-sequence star or the second RLOF.
As well are indicated parameters of the models in  which common envelope may form 
due to the {\it presumed} impossibility to remove non-accreted matter. 
Figure~\ref{f:outcomes} shows that mass-exchange becomes more stable
with increase of the  stellar mass, in agreement, for example, with Ge et al. (2020).
The probability of formation of common envelopes decreases significantly with increasing
$M_{1,0}$ and $q$, since the luminosity of stars grows faster than \mdot.
The maximum \mdot\ during the exchange process also decreases with increasing $q$.

Test computations have shown that in the case of the efficiency parameter of common envelopes
$\alpha_{\rm ce}=1$ and stellar envelope binding energy parameter $\lambda$, calculated according to 
Eq.~(\ref{e:lambda}),
regardless of the parameter $\alpha_{th}$, components of binaries  merge. 
Then the data presented in the Tables~\ref{t:outcomes} and \ref{t:Noutcomes17} show
that in the limiting case when possible mergers are taken  into account
the frequency of formation of systems in which RLOF produces detached systems, declines by about 50\%.
This effect is most pronounced in close binaries in which mass-exchange 
cases B and subsequently BB occur.
The estimate of the formation rate of common envelopes is comparable to the estimate obtained by Henneco et al. (2024):
59\% for close binaries with $M_{1,0}=(4.8 - 20.8)$\,\ms, but with $0.1 \leq q \leq 0.9$.

\begin{figure}[t!]   
\centering
\includegraphics[width=0.7\textwidth,trim={0 0 0 2cm},clip]{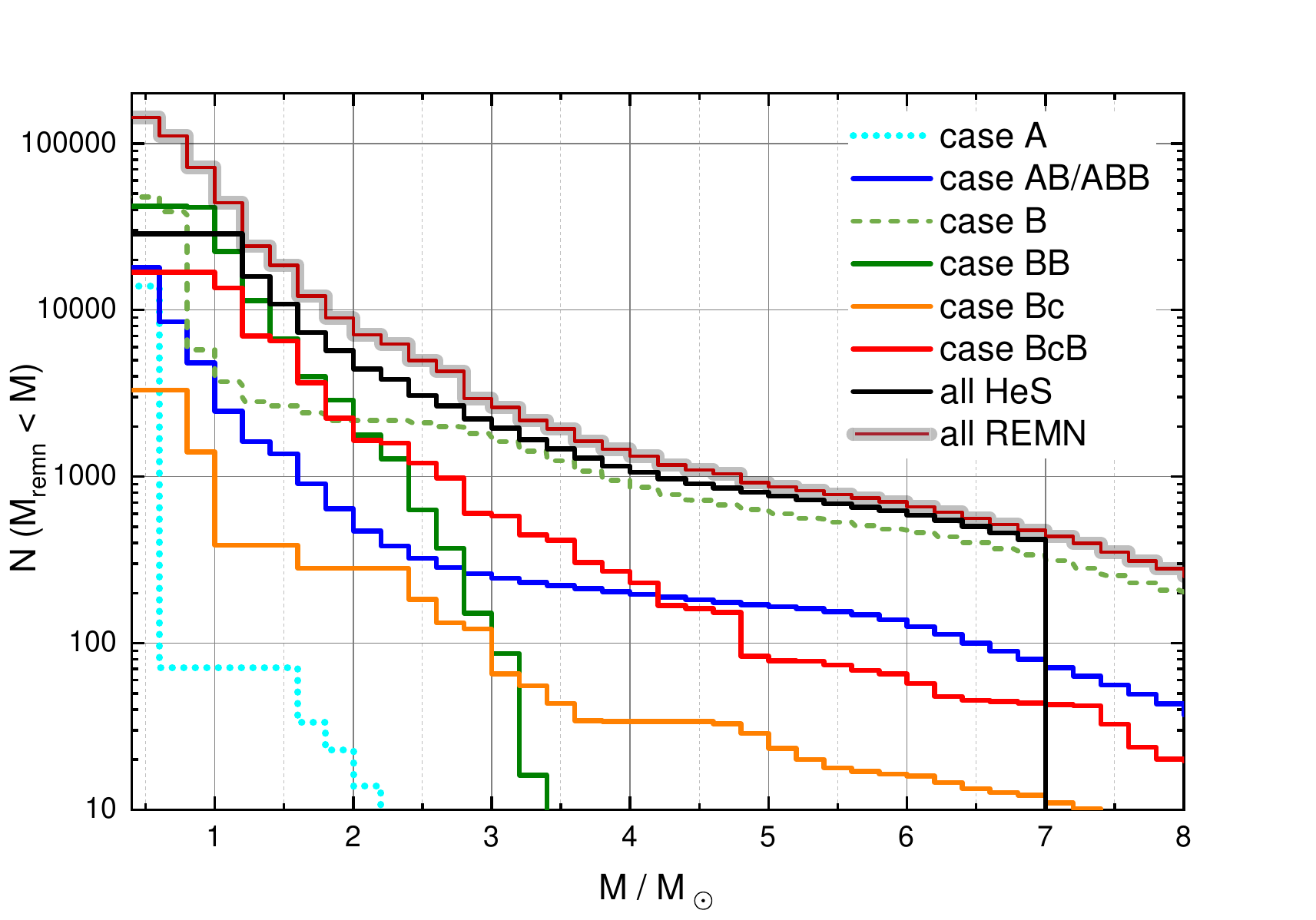}
\caption{Cumulative distribution of the number of stars at the core helium burning stage 
depending on the RLOF case. Helium stars 
($\mathrm{ 1 \leq M_{remn}/\ms \leq 7}$) are singled out by a black line.
}
\label{f:ntotal}
\end{figure} 

\begin{figure}[t!] 
\centering
\includegraphics[width=\textwidth]{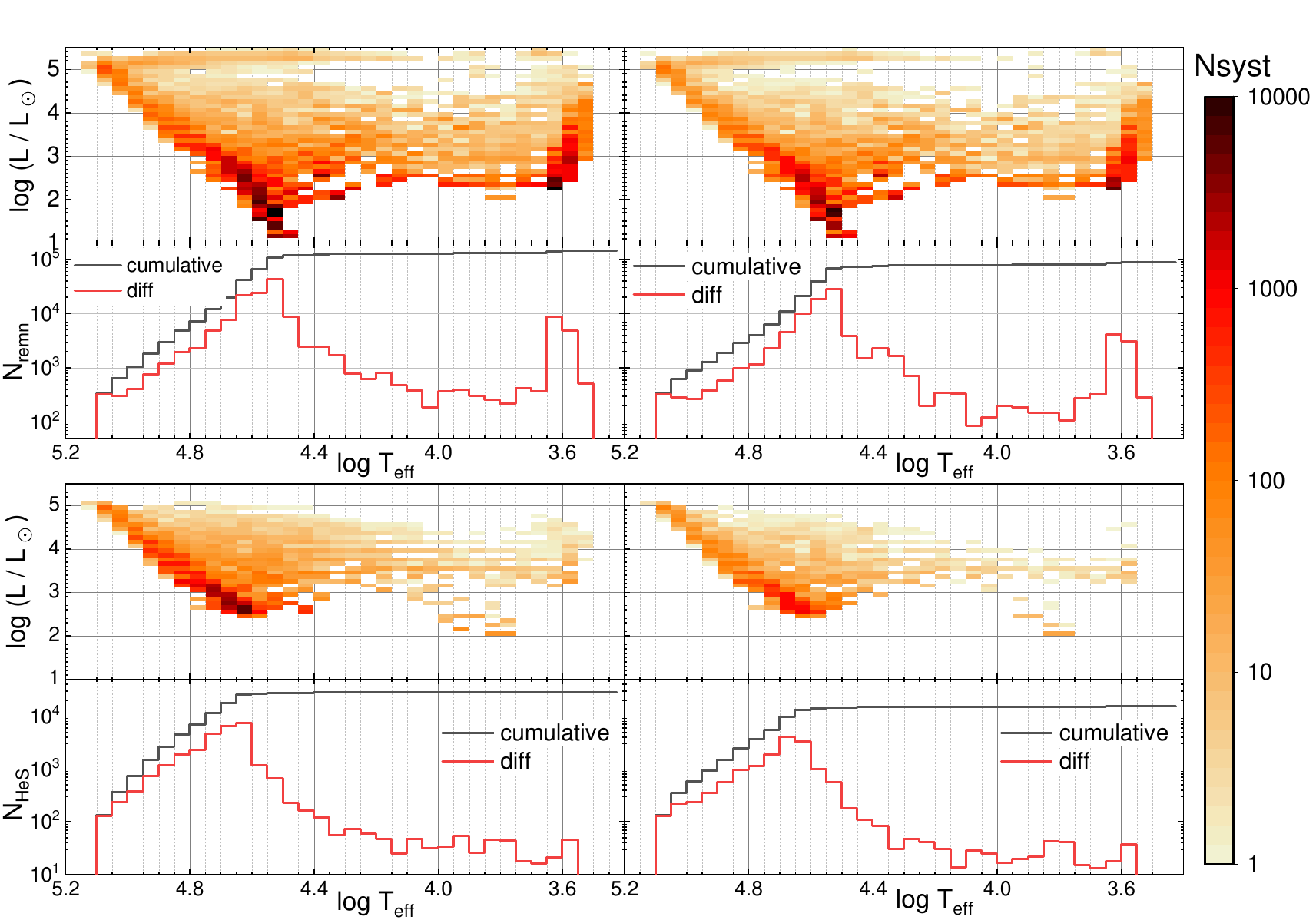}
\caption{Upper panels -- number density distribution of all remnants at the stage from 
RLOF termination to the exhaustion of helium in the core.
Lower -- the same for HeS ($1 \leq M/\ms \leq 7$).
In the panels in the right column the systems where common envelopes may form due to 
the impossibility to remove non-accreted matter are excluded.
}
\label{f:hrremn}
\end{figure} 
\begin{figure}[]   
\centering
\includegraphics[width=0.7\textwidth]{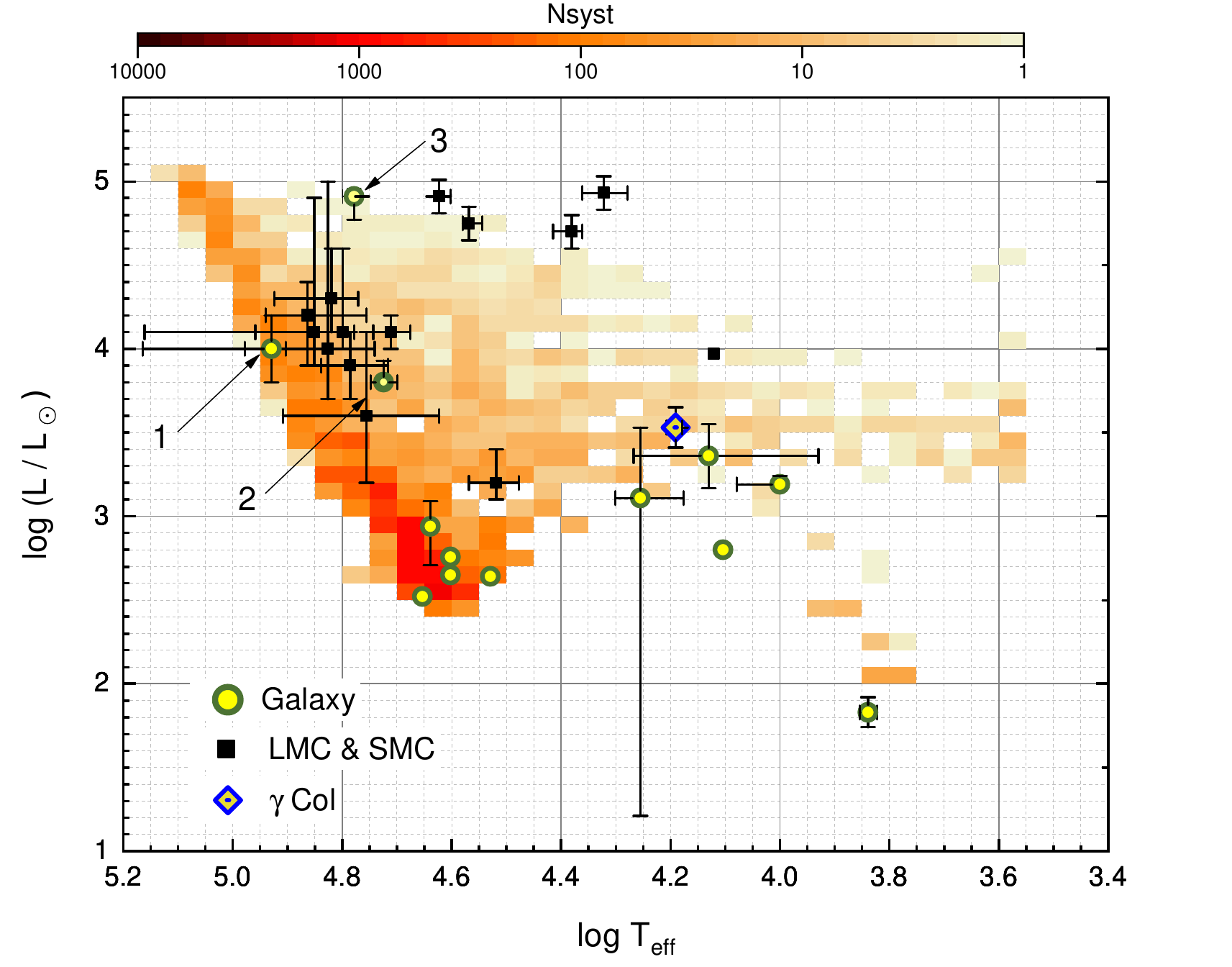}
\caption{Number density of precursors of HeS after termination of RLOF and HeS 
prior to He-exhaustion  in their cores ($\rm Y_c \approx 0.01$). 
Yellow symbols  -- Galactic objects. 
1 -- MWC~656,
2 -- $\phi$~Per,
3 -- WR~2-1.
By a blue diamond is marked position of  $\gamma$~Col.
Black squares are the stars detected in the Magellanic Clouds. }
\label{f:hestobserved}
\end{figure}  

Figure~\ref{f:ntotal} shows the histogram of the mass distribution of remnants of donors 
in close binaries that have mass (0.5 -- 8)\,\ms\ in the stage of helium burning in their cores. 
RLOF cases via which they were formed are indicated.
Helium stars are singled out.  
Total number of the remnants is 42000, while the number of HeS is 28500.
The main input to the total number of remnants comes from the stars evolving 
in the cases B, BB, and BcB, due to the
large range of initial orbital periods of close binaries over which their primary 
components may overflow Roche lobes.
The number of remnants with masses above 7\,\ms\ that can be identified as WR stars 
is close to 400. Given the binarity rate of WR stars, this number is reasonably consistent 
with the lower limit of the observational estimates of their number in the Galaxy.

Number distribution of all remnants and in particular helium stars after the completion of 
the first mass exchange and before  He-ehaustion in their cores is shown in 
Fig. ~\ref{f:hrremn}. Clearly, dominate low-luminosity stars. 

The estimates of the size of the populations of the remnants and HeS presented above
were made under the assumption that common envelopes associated with the impossibility
to remove non-accreted matter do not form.
If, as a limiting case, we assume that common envelopes really form and always lead
to the merger of components, then the systems with initial $q \aplt$(0.6 -- 0.7) and donor masses of
$\aplt$(14 -- 18)\,\ms\ (Fig. \ref{f:outcomes}) become  eliminated from the population of 
progenitors of HeS. 
As it follows from Table~\ref{t:Noutcomes17}, remain $\approx$50\% of HeS,
about 14700 objects.
In the Hertzsprung-Russell diagram (Fig. ~\ref{f:hrremn}) the maximum of number 
distribution shifts toward larger \teff.
Stars with masses of $\aplt$2\,\ms\ disappear. This may be one of the {\it
evolutionary} effects explaining the low number of massive subdwarfs.

Figure~\ref{f:hestobserved} shows the blow-up of the upper part of the lower right panel 
of Fig.~\ref{f:hrremn} with  positions of the only known Galactic HeS star, massive 
Galactic subdwarfs and candidate HeS, detected in the Magellanic Clouds.
We neglect possible slight bias of similar distributions for Galactic and MC stars.
Galactic helium star WR~2-1 has $\mathrm {T_{eff} = 60^{+3}_{-2}\,kK,
\log(L/\ls)=4.91^{0.05}_{-014}}$.
In addition to the subdwarf MWC~656 mentioned in the Introduction
($\mathrm {T_{eff} = 85^{+10}_{-5}\,kK, \log(L/\ls)=4.0\pm 0.2,
M/\ms \approx 1.48^{+0.55}_{-0.46}}$), we show in the Figure the sdO subdwarf
$\phi$\,Per
($\mathrm {T_{eff} = 53 \pm 30\,kK, \log(L/\ls)=3.8\pm 0.13, M/\ms=1.2 \pm 0.2}$,
Murar et al. (2015)). Like for the hot component of MWC~656, its anomalously high 
luminosity suggests that the star is in the He-shell burning stage
(Schootemeijer et al., 2018). Position of $\gamma$~Col is also  shown.
Clearly, most candidate HeS  in the Magellanic Clouds fall within the region of the Hertzsprung-Russell diagram that is well populated by stripped helium stars.

\subsection{Accretors}
\label{ss:accretors}
\begin{figure}[t!]   
\vspace{-1cm}
\centering 
\includegraphics[width=0.8\textwidth]{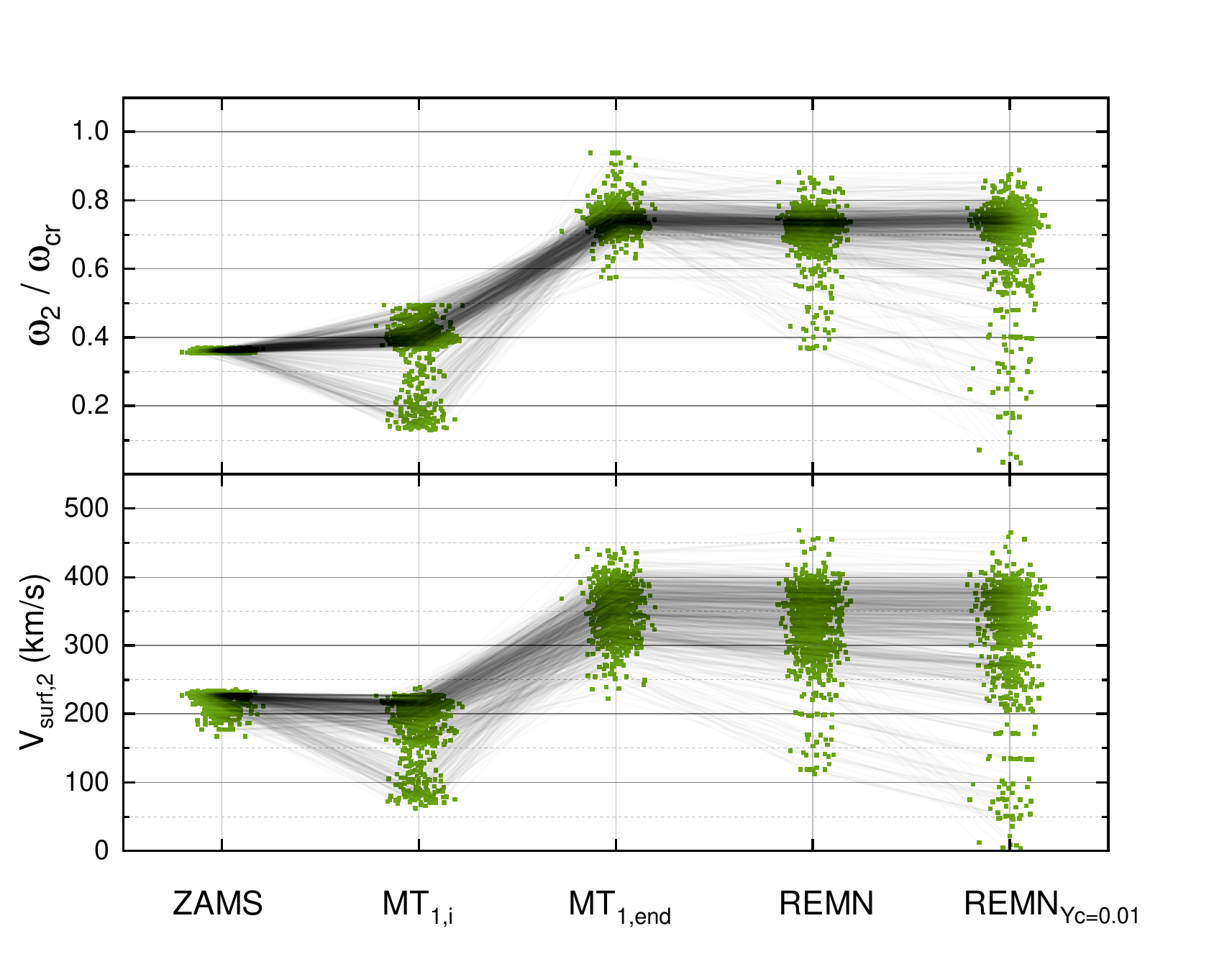}
\caption{
Upper panel -- variation of the ratio of angular velocity of rotation of accretor and critical angular velocity 
$\omega_2 / \omega_{\rm cr}$. 
Lower panel -- variation of the equatorial velocity of rotation $V_{\rm surf,2}$. 
Singled out are certain instants in the evolution of the donor: ZAMS, beginning of 
the first RLOF (${\rm MT_{1,i}}$), 
termination of the first RLOF (${\rm MT_{1,end}}$), 
stage of intense He-burning in the core (REMN) and core He exhaustion (${\rm REMN_{Yc=0.01}}$).  
For convinience the symbols are distributed as ``cloudlets'' around annotated points.
The ``treads'' connect the models of the same binaries. 
See the detailed discussion in the text.
}
\label{f:omegavel}
\end{figure}

Figure~\ref{f:omegavel} shows the change in the ratio of the angular velocity 
of rotation of accretors and their critical angular velocity ($\omega_2/\omega_{\rm cr}$),
as well as linear equatorial velocity of rotation ($V_{\rm surf,2}$)
between the stages of donor evolution for all calculated
models\footnote{In Fig.~\ref{f:omegavel} we show also  accretors in the binaries 
for which the problem of formation of common envelopes 
due to the impossibility of removing of non-accreted matter from the system remains unsolved.}.
The ratio $\omega_2/\omega_{\rm cr}$ for most stars before mass exchange
(${\rm MT_{1,i}}$) varies only slightly.
Slows down rotation of components of the closest systems in which tidal interaction is 
effective.

At the stage of RLOF (between ${\rm MT_{1,i}}$ and ${\rm MT_{1,end}}$),
rotation velocity of accretors first increases, but later, as the accretion rate declines 
at the end of RLOF, rotation velocity for most stars declines below critical and 
becomes about $(0.6-0.8)\,\omega_{\rm cr}$. 
We skip the peak in $\omega_2 / \omega_{\rm cr}$.
Obviously, we exclude from consideration in this stage the systems in which common
envelopes form due to the dynamical mass loss by the donors.  
The stage of evolution after termination of RLOF and before exhaustion of He in the donor 
core (from ${\rm MT_{1,end}}$ to REMN$_{\rm Yc=0.01}$) is short 
 and for the vast majority of former accretors rotation velocity  changes only slightly.
Stars continue to evolve as effectively single objects.
Two competing processes occur: transfer of angular momentum from the contracting core to
the envelope and  its loss due to an increase of radius and previously rotation-enhanced
stellar wind (eq.~(\ref{eq:windenhacement})).
But the wind can significantly slow down only stars with the mass $\gtrsim 20$\,
\msun\ (Langer, 1998; Hastings et al., 2020).
As a result, the distribution of models over $\omega/\omega_{\rm cr}$ retains a peak near
(0.6 -- 0.8). The only exceptions are few systems with initial mass ratios of components close to 1, 
in which during the time of He-burning in the HeS core of the primary component the radius of its companion 
increases substantially (threads going down at the stage REMN - REMN$_{\rm Yc=0.01}$).

Results agree with the model in which Be stars are spun-up accretors of close binaries 
(Waters et al., 1989; Pols et al., 1991). In addition to subdwarfs, Be stars may
have as companions white dwarfs and neutron stars (e.g.,  Naz{\'e}, 2025; Reig, 2026). 

Be stars can also be single objects, former components of close binaries disrupted 
by supernovae explosions. 
Every subsequent supernova explosion destroys $\sim 90\%$ of close binaries in which they 
were located (e.g., Portegies Zwart and Yungelson, 1998).
About 13\% of Be stars in the Galaxy are runaway objects.
This may point to their formation as a result of interactions between components in close binaries (Bubert and Evans, 2018).
Formation of initially rapidly rotating massive stars during star formation is also likely.
There are known Be stars with low-mass companions ($M \lesssim 1$\,\ms) that have not yet 
reached ZAMS (Labadie-Bartz et al., 2025; Naz\'{e} et al., 2025).

\section{Discussion}
\label{s:disc}

Yungelson et al. (2024) estimated the number of HeS with masses (1 - 7)\,\ms\ as 
$\simeq 20000$.
Hovis-Afflerbach et al. (2025, hereafter H2025) also used a grid of tracks computed
by MESA, to estimate the number of HeS. H2025 estimate renormalized to the star formation 
rate SFR=2\,\ms\,yr$^{-1}$ is $\simeq 60000$.
The studies differ in the assumed initial mass functions of the primary components 
(Salpeter and Kroupa, respectively), initial distributions of close binaries over initial  
periods (higher concentration to low periods in H2025), binarity rate of stars
(100\%  in H2025).
In this study, initial mass function of
donors and distribution over periods are similar to that in  H2025.
Binarity rate is mass-dependent and varies from 0.675 for $M_{1,0}=5\,\ms$ to 0.845 for
24\,\ms.
As the result, the number of HeS changed to 28500.
Then, if the possibility of merger of  components in common envelopes formed due to
accumulation of non-accreted matter is neglected, for similar SFR
relative difference to H2025 is about 2, which can be considered tolerable.
However, acceptable at the current level of knowledge differences in the assumed
initial distributions of stars over parameters remain a significant factor influencing the 
estimates.
It should also be noted that, according to H2025, approximately 20\% of
HeS are formed via common envelopes formed due to unstable mass loss by the donors.
In this study it was found that in such common envelopes components  merge.
Under the extreme assumption that evolution in common envelopes formed due to the 
inability to remove non-accreted matter always leads to the merger
of components, the number of HeS decreases to 14700.
The relative difference factor with H2025 becomes  $\simeq 4$.
But the estimate of the number of merging systems depends on the adopted parameters of the 
efficiency of common envelopes $\alpha_{\rm CE}$ and the binding energy of the donor envelope.
Additionally, assumed parameters of mixing in the interiors of stars and stellar wind 
influence dependence of $M_{\rm remn}$ on  $M_{1,0}$.

It is also possible that HeS form in the systems where the secondary experiences RLOF
 after the primary has evolved into a compact object (Zapartas et al., 2017).
Estimate of the number of such HeS  depends on assumptions regarding accretion efficiency, 
common envelopes, and the details of supernova explosion. 
These HeS, if formed, may be disproportionately well  represented in the observed samples, because they are not ``outshined'' by their companions (Blomberg et al., 2026).

Picco et al. (2026) suggested a scenario in which single HeS form by
the merger of components in so-called ``failed'' common  envelopes.
This scenario assumes that the system already contains a HeS and the Roche lobe is
overflown  by its evolved companion.
It occurs only in the closest binaries with initial
$M_{2,0} / M_{1.0} \lesssim 0.75$ and $P_0 \lesssim$30 day.
The probability of occurence of this scenario also depends on the adopted
parameters of the common envelopes and convective mixing that determines stellar radii.

Wang et al. (2026) simulated HeS population in the Galaxy by means of population synthesis code
BSE (Hurley et al. 2002; Keel and Hurley, 2006) with modifications by  Shao and Li (2021).
According to Wang et al., the number of HeS with masses (2 -- 8)\,\ms\ for binarity rate 100\%, 
Kroupa et al. (1993) IMF and SFR=2\,\ms$\rm {yr^{-1}}$, depending on the adopted
parameters of the stellar wind, accretion efficiency and common envelope parameter is
1100 -- 2300. 
After renormalizing to the  Salpeter IMF with a less steep slope for stars with the 
mass $\geq$\,1\ms\, the number of model HeS increases by a factor of 3 and becomes comparable to the 
estimate by Yungelson et al. (2024).
The remaining difference can be attributed to the use by Wang et al. (2026) of the BSE code, which is based on 
approximations  of results of single-star evolution computations for close binaries (Hurley et al., 2000).

Results of computations depend on the assumed algorithm for 
stellar wind, which significantly affects the evolution of stars with the mass
 $\gtrsim$10\,\msun.
For single non-rotating stars with mass
(15 -- 35)\ms\ and $Z=Z_{\odot}$ various combinations of ``recipes'' used in the literature
for the winds of hydrogen-rich stars and Wolf-Rayet stars result in the  difference in masses of
pre-supernovae, reaching 50\% (Renzo et al., 2017).
Extrapolation of the empirical Nugis and Lamers (2000) law beyond the mass interval,
for which it was suggested ($M \apgt 7$\,\msun),
may overestimate the rate of mass loss by HeS, while, for example, the theoretical
Fink’s model (2017), constructed for stars of lower masses, underestimates this rate
(Ramachandran et al., 2024). It should be noted that Fink’s (2017) model was 
elaborated for \teff=50000\,K, lower than the typical \teff\ of HeS.
Helium stars winds we are considering cannot be calibrated empirically, because
there are no relevant observations.
As a numerical experiment, we estimated the possible number of remnants if there hadn't be
stellar wind of HeS. The total number of HeS remained virtually unchanged,
because it is determined by low-mass stars.
But the total number of HeS with masses $\apgt$2\,\ms\ increased by 
a factor 2.

Convective mixing parameters present significant uncertainty,
since a three-dimensional process is described by an one-dimensional model. 
Observational data is insufficient to distinguish and calibrate the contributions of 
individual mixing elements.
There are indications that the commonly accepted parameters need to be revised to increase the masses of the 
He-cores of post-MS stellar models (Johnston, 2021; Johnston et al., 2024).
We have taken into account the dependence of the parameter of convective overshooting on the stellar mass.
However, it remains unclear how other parameters should be modified. Changing the masses 
of the convective cores of models could affect the predicted masses of collapsing stars.
It is possible that the mixing parameters change during evolution, but this issue has not yet been investigated.

The problem of common envelopes is a three-dimensional magnetohydrodynamical one (e.g., 
Webbink, 2008; Ivanova et al., 2013, 2020; R{\"o}pke and De Marco, 2023; Schneider et al., 2026).
Derivation of $\alpha_{\rm ce}$ and $\lambda$ which appear in 
Eq.~(\ref{e:ce}) is still an unsolved problem.
There is an uncertainty regarding the role played by the release of internal energy in the 
common-envelope ejection process (the parameter $\alpha_{\rm th}$ in Eq.~(\ref{e:lambda})).
Additional energy sources for the common envelope ejection are discussed, such as
energy release during accretion onto the companion star, penetration of the companion's 
matter into the nuclear burning shell
of the donor star, leading to explosive energy release, as well as energy loss by 
radiation at the common envelope stage; unclear remains final energy of the lost matter 
(see Ivanova et al., 2013; 2020).
In the case of additional energy sources $\alpha_{\rm ce}$ can be $>$1.
If these factors are not taken into account, $\alpha_{\rm ce}<$1.
The case $\alpha_{\rm ce}=1$ is an idealization that assumes the use of all the orbital
gravitational energy
of the system to dissipate the common envelope, no energy loss to radiation, and zero 
kinetic energy of the envelope matter at infinity.
For $\lambda$, there is an uncertainty associated with the definition of the 
boundary between the core and the envelope. 

We assumed, as in other similar studies, that the masses of the donor remnants after the 
ejection of the common  envelope are the same as the masses of the remnants after 
stable mass exchange, and, based on this, estimated the change in the distance between 
the components in the common envelopes.
However, the masses of remnants are larger than the mass corresponding to 
$\mathrm {X_H=0.1}$ which enters enter the  expression (\ref{e:lambda}). This means that 
the envelope binding energy is underestimated.
Furthermore, we do not take into account the possible increase in the remnant radius 
during the thermal relaxation stage after the dissipation of the common envelope, which can
lead to another stage of mass exchange or to a new common envelope (Ivanova,
2011; Vigna-G{\'o}mez et al., 2022).

Formation of the common envelopes may be also related to the problem of limiting 
mass ratios of components for which donor mass loss can  be formally considered dynamically stable.
In this paper, computations were performed for $q \geq$0.4, and it was assumed that the mass 
loss is unstable if $\mdot \geq 10^{-2}$\ms yr$^{-1}$. Typically, exceeding this limit is 
accompanied by a sharp decrease in the time step between models to fractions of a year and 
divergence of the models.
Figure~\ref{f:outcomes} shows that by these conditions, mass loss by a significant fraction
of low-mass donors can be considered dynamical.
High-mass donors experience relatively more stable mass loss.
In the computations results cited in the literature, the stable mass exchange limit in 
some cases drops to $q$=(0.15 -- 0.35), for example, in Henneco et al. (2024), who adopted
\mdot=10\,\ms\,yr$^{-1}$ as the limit.
However, to ensure numerical stability of the models, Henneco et al. ``stabilized'' the mass accretion rate at 10\% of the rate corresponding to the accretor's thermal time scale.
We considered such systems as merging.

It should be also noted that in the aforementioned study for a 
significant fraction of the models with initial separation of components
$\apgt$500\,\rs\ computations were terminated due to the ``numerical
issues'', which may in fact indicate unstable mass loss and
the associated reduction in the time step between models.
Mass ratio of components $q=0.4$, below which evolution should lead to
the formation of common envelopes or the formation of contact systems followed by a merger,
is in reasonable agreement with the limits found, for example, by Jin et al. (2026) for 
the same mass range of $M_{1.0}$ as in this paper.
It should be noted that Jin et al., unlike to our study,
used the MESA MLT++ calculation scheme, which artificially
limits the efficiency of convection in the stellar envelopes and allows for the calculation of
evolutionary tracks without a catastrophic reduction in the time step (Paxton et al., 2013).

Accretion efficiency $\beta$ is certain  problem.
We adopted the mass and momentum transfer mechanism of Packet
(1981), as built into MESA.
This results in a very low accretion efficiency $\beta \simeq (5 - 10)$\%, which is 
consistent with the values obtained by other authors (e.g., Henneco et al., 2024).
Since stripped helium stars are virtually not observed in the Galaxy, there are no reference points for estimating $\beta$.
Estimates of $\beta$ based on comparisons of observational data for stars of different 
types or individual stars with models are controversial, ranging from
$\simeq$0 (e.g., Hastings et al., 2021) to $\simeq$(0.6 -- 0.8) (e.g., Vinciguerra, 2020;
Sch{\"u}rmann et al., 2022; Xu et al., 2025; Sen et al., 2026).
Parameters of particular stars in a sample of 16
Galactic systems of Be-stars with massive subdwarf companions
($M \approx (1 - 2)$\,\msun), which by our definition are also HeS,
can be reproduced if $\beta$ varies from 0 to 1 (Lechien et al., 2025).
The value of $\beta$ may depend on the evolutionary stage of the donor at the time of RLOF, 
the masses of stars in the close binary (Sch{\"u}rmann et al., 2025), and the specific 
angular momentum of the matter escaping  close binary (Mennekens and Vanbeveren, 2017).
Uncertainty in the estimates is introduced by the dependence of the donor mass on the 
assumptions regarding convective mixing (Sen et al., 2026).
From a theoretical standpoint, the question of the efficiency of accretion onto rotating 
stars also remains open.
Colpi et al. (1991), Paczy{\'n}ski (1991), Popham and Narayan (1991), and Bisnovatyi-Kogan 
(1993), using polytropic models, showed that accretion onto critically rotating stars can occur, if the  
excess momentum is lost through interactions between the star and accretion disk. 
However, these studies were not continued further, with the exception of an attempt to parameterize the results of
Popham and Narayan (Xing et al., 2026), as solving the problem requires three-dimensional gas-dynamical calculations, which are currently unfeasible.
Virtual absence of observed HeS in the Galaxy may indicate inefficiency of the 
mechanism for removing non-accreted matter by radiation pressure
of the components and a small fraction of the close binary orbital energy that can be 
used to expel common envelopes.

Possible reasons for the paucity of HeS stars in the Galaxy discussed above are related to 
the {\it evolution} of stars.
However, there are also {\it selection effects} that hinder their detection.

\begin{figure}[t!] 
\centering
\includegraphics[width=0.5\textwidth,angle=-90]{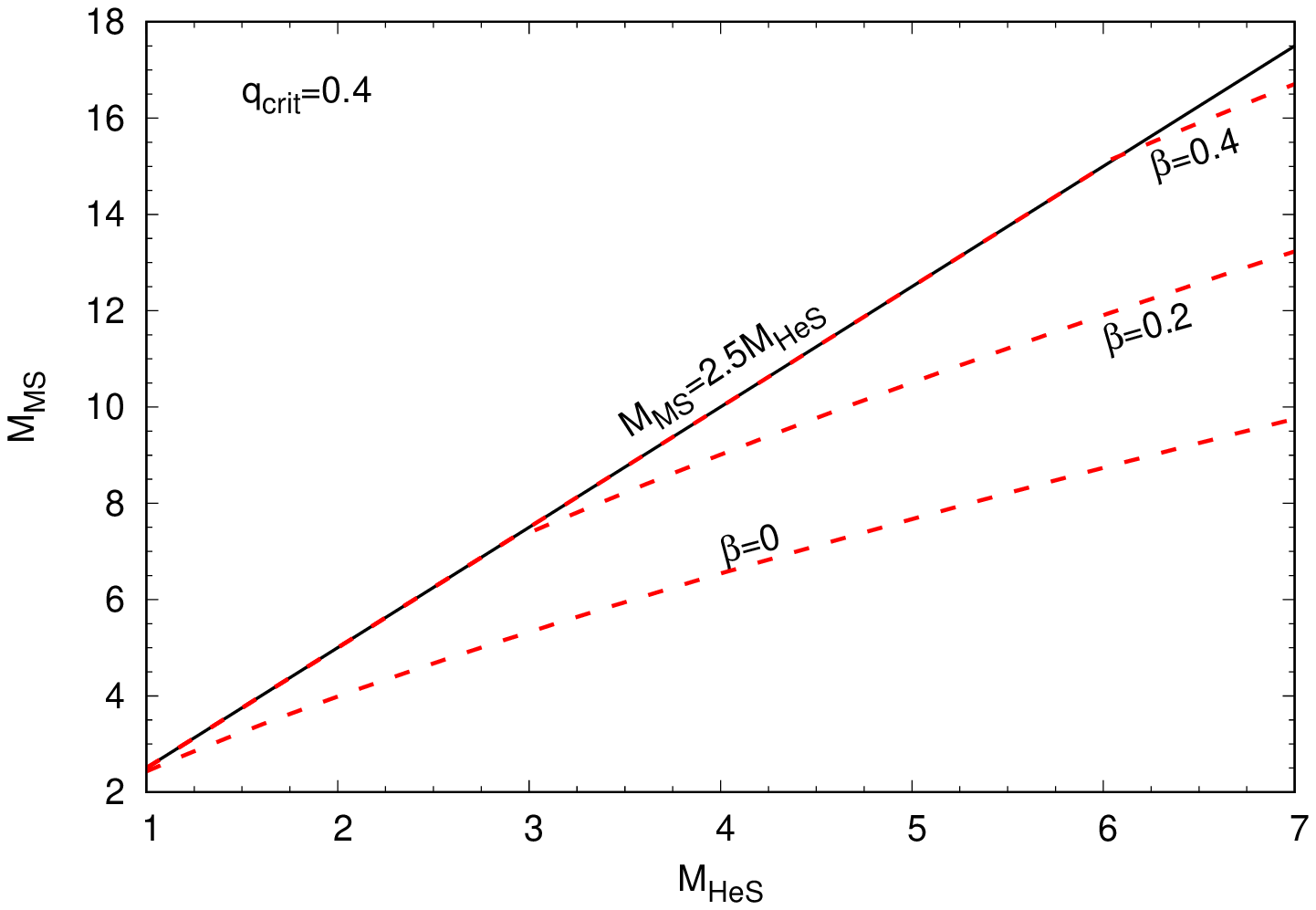}
\caption{Relation  between the masses of components in close binaries of stripped 
helium stars and their main-sequence companions.
If $\mathrm {M_{MS} \apgt 2.5M_{HeS}}$,
detection of HeS by the UV excess in the spectral energy distribution of HeS+MS  pair is 
hardly possible.
Red dashed lines are lower limits of areas where
HeS can be observed for the different values of the accretion efficiency $\beta$.}
\label{f:qlim}
\end{figure}

Drout et al. (2023) and Ludwig et al. (2026) identified candidate HeS in the
Magellanic Clouds based on their positions in the UV color-luminosity diagrams.
The excess UV radiation  due to HeS in 
the combined spectral energy distributions of 
``HeS + MS'' pairs, can be observed if,
roughly, $\mathrm {M_{HeS} \apgt 0.4\,M_{MS}}$\footnote {Of course, this ratio
also depends on the extent of evolution  of the HeS companion.}.
For a rough estimate, we assume that the same condition
should be  satisfied for the Galactic stars. 
To prevent a close binary from plunging into a common envelope
during its evolution resulting in the formation of a HeS paired with a MS-star, 
it is necessary for the initial mass ratio of components to exceed certain critical 
value $q_{crit}$, which can be taken as 0.4.
Defining the accretion efficiency as
$\beta = (M_{\rm MS} - M_{2.0}) / (M_{1.0} - M_{\rm HeS})$ and using
relation $M_{\rm HeS} = 0.08 M_{1.0}^{1.4}$, we obtain the ``detection condition''
\begin{equation}
\beta \left [\left ( M_{\rm HeS}/0.08 \right )^\alpha - M_{\rm HeS} \right ]
+ q_{\rm crit} \left ( M_{\rm HeS}/0.08 \right )^\alpha \aplt 2.5 M_{\rm HeS},
\label{eq:hes}
\end{equation}
where $\alpha = 1/1.4$.
In the $\mathrm {M_{HeS} - M_{MS}}$ diagram (Fig.~\ref{f:qlim}), one can identify
the areas of ratios between $\mathrm {M_{HeS}}$ and $\mathrm {M_{MS}}$ that allow the 
existence of HeS components potentially detectable by UV excess for different $\beta$ values.
A slight difference
between the initial-to-final mass ratios for moderate- and high-mass stars
has virtually no effect on this rough estimate.

As the accretion efficiency increases, HeS companions become more massive,
limiting the number of HeS potentially detectable by their UV color excess.
For $\beta \apgt$0.4, HeS virtually disappear.
In the most favourable case of almost completely non-conservative accretion, for example, 
limited by the critical rotation velocity as assumed in this paper,
the number of HeS+MS pairs within the ``observable'' area is $\approx 3000$.
Out of this number the stars with masses $\apgt 2$\,\msun\ comprise 
$N_{\rm max}\approx 1000$.
This is the maximum estimate, taking into account the existence of interstellar absorption and the limited capabilities of the instruments.

According to Fig.~\ref{f:qlim}, HeS with
$M_{\rm HeS} \lesssim$ (4 -- 5)\,\ms\ should be detectable, since the initial donor mass 
function
favours the detection of low-mass objects and prevents the detection of massive HeS.
On the other hand, $q_{\rm crit}$ and the degree of non-conservativeness of the mass 
exchange may depend on the masses of close binary components.
Nevertheless, the ``outshining'' of HeS by their MS companions in UV  may be the main
{\it observational selection effect} that prevents HeS detection
by photometry. 
Note, HeS in the Magellanic Clouds and the Galaxy with known
parameters have been detected using spectroscopy in the optical and ultraviolet
spectra.

Thus, evolutionary effects prevent formation of HeS stars
in a significant fraction of the close binaries in which they could form  and 
observational selection effects prevent detection of the stars that form.
The limit set by evolution was also noted by G{\"o}tberg et al. (2018), who suggested that
the mass ratio of components in close binaries, the progenitors of the systems with 
potentially detectable HeS stars, should be $\mathrm {M_{2.0}/M_{1.0}} > 0.25$,
based on the fact that at smaller $\mathrm {M_{2.0}/M_{1.0}}$ common envelopes form always.

\section{Conclusion}
\label{s:concl}

We investigated formation of the population of helium remnants of donors in close binaries 
with masses (4 -- 24)\,\ms.
The mass of the least massive remnants, which fall within the region of the Hertzsprung-
Russell diagram with $\log(\teff)\apgt$4.4 in the stage of core helium burning  varies 
from 0.36\ms\ to $\simeq 1$\ms\, depending on the initial period of the close binary system.
The most massive remnants, which fall into the same region of  $\log(\teff)$, have masses from 6.5\ms\ to 10.5\ms.
Remnants with masses of (1 -- 7)\ms\ can be identified with the most massive hot subdwarfs
sdB/O and the unique stripped helium star discovered in the Galaxy.

About 25\% of all remnants are formed by mass transfer in  cases A and AB, while the rest are formed by mass transfer in case B.
For HeS, these proportions are 12\% and 88\%, respectively.
The total number of the remnants and HeS ranges from 75,000 to 120,000 and from 14,700 to 28,500, respectively,
depending on the assumptions concerning common  envelopes.
If we restrict ourselves to the stars with $M \geq 2\,\ms$, the  number of HeS ranges from 3,200 to 5,500.

The main factor that influences the model estimate of the number of objects and reduces it 
is possible immersion of components of the binaries in the common envelopes and their merger.
Formation of the common envelopes may be caused by the unstable mass loss by the donors or the 
impossibility  to remove non-accreted matter from the system if accretion efficiency is 
limited by rotation. The common envelope factor may be one of the {\it evolutionary} 
reasons why only one HeS star with a mass greater than 2\,\ms\ has been detected in the 
Galaxy so far.

Significant reasons for the discrepancies between the estimates of HeS number obtained by 
different authors are differences in the assumptions regarding common envelopes and 
initial mass functions of primaries of close binaries and distributions of stars over 
initial orbital periods adopted in the population synthesis. 

Observational selection effects severely limit the possibility of detecting HeS photometrically, since
in the spectra of the systems harbouring HeS stars and MS stars, the latter
 ``outshine'' HeS if
$\mathrm {M_{\rm HeS} \lesssim 0.4\,M_{\rm MS}}$. Systems in which detectable HeS stars 
could potentially form must have small initial mass ratios of components, making 
them unstable with respect to formation of common envelopes and possible mergers.

Due to significant interstellar absorption, detection efficiency of HeS stars in the 
Galaxy should be lower than that in the Magellanic Clouds (10\%), and we can expect 
detection of much less than 1000 objects.

Besides HeS stars, the Galaxy may contain several hundred ``cold'' stars 
($\log(\teff) \aplt 4.4$) which were primaries 
of close binaries which have lost only part of their envelopes and are in the  helium-shell burning stage.

Stripped helium stars are accompanied by rapidly rotating stars. Estimates indicate that their 
rotation velocities should be (0.6 -- 0.8)$\omega_{\rm cr}$.

\vspace{5mm}
The authors acknowledge A.M. Cherepashchuk, K.A. Postnov, L. Piersanti, and
I.A. Shaposhnikov for discussion of various aspects of this study and the
referees for helpful comments.
This research was supported by the Russian Science Foundation grant No. 25-22-00295,
https://rscf.ru/project/25-22-00295/.
NASA ADS bibliographic database have been used in this work.

\pagebreak
{\it APPENDIX}
\counterwithin*{figure}{part}
\stepcounter{part}
\renewcommand{\thefigure}{A\arabic{figure}}

\begin{figure}[h!]
\includegraphics[width=0.5\textwidth,trim={0 0 0 0},clip]{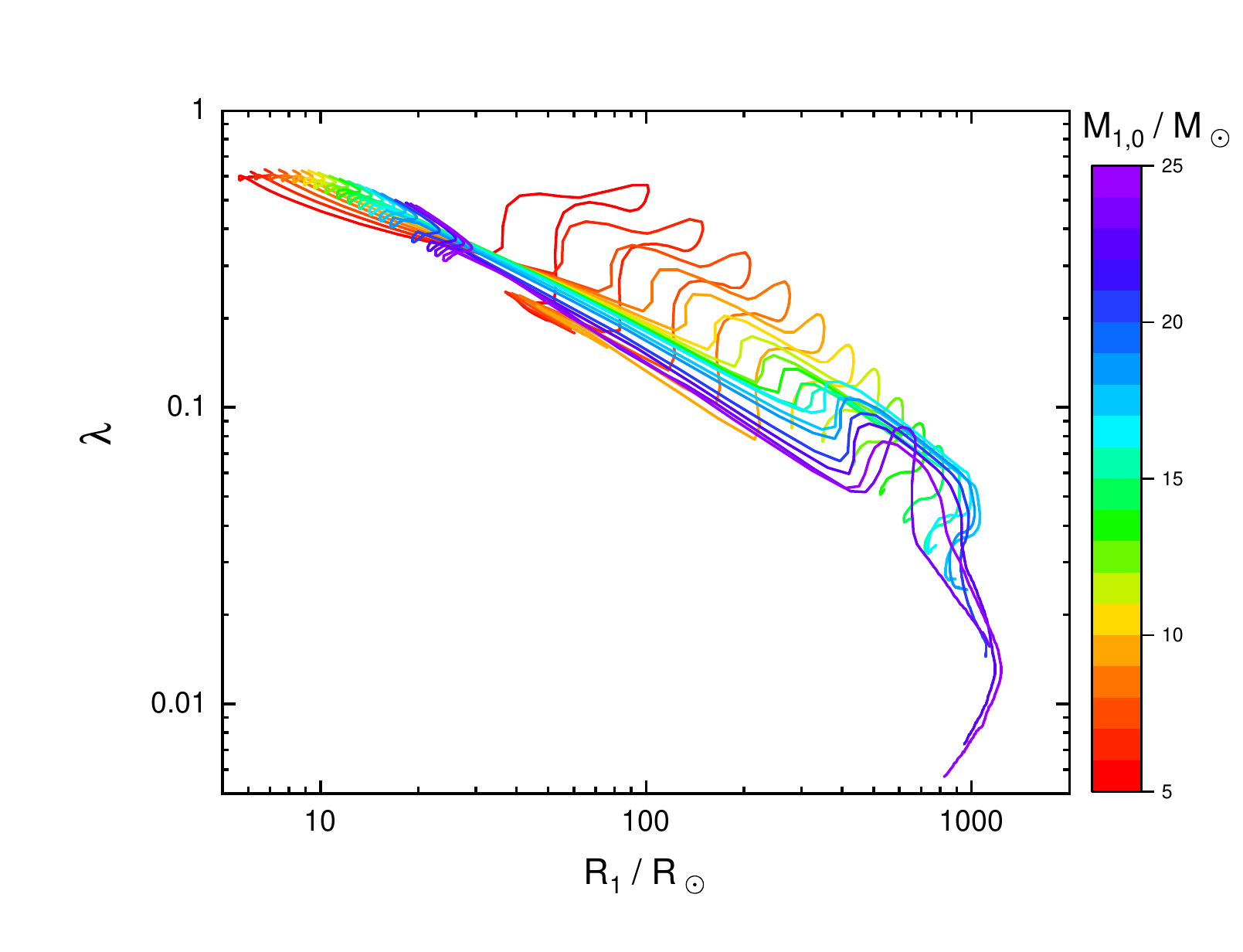} 
\includegraphics[width=0.5\textwidth,trim={0 0 0 0},clip]{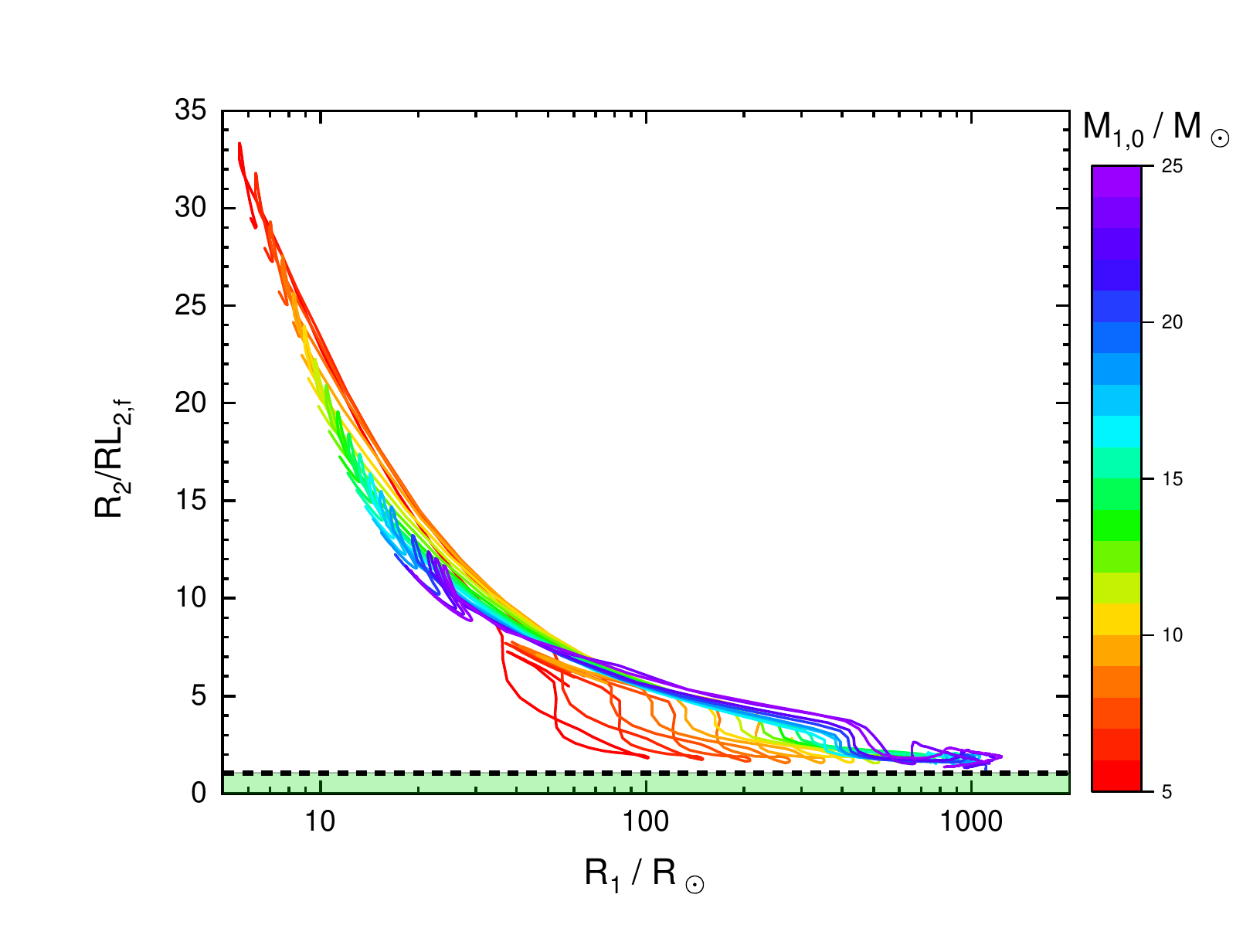} 
\caption{Left panel -- dependence of the stellar envelope binding energy  parameter 
$\lambda$ on the masses and radii of the stars.
Right panel -- ratio of the radii of the remnants of the secondary components and the radii of their Roche lobes
after the completion of common envelope stage. Components can avoid merger
only if $\mathrm {R_2/R_{L2,f}} \leq 1$. Color-coded are the masses of stars.}
\label{f:lambdamr}
\end{figure}
\begin{figure}[h!]
\centering
\includegraphics[width=0.8\textwidth,trim={0 0 0 1.5cm},clip]{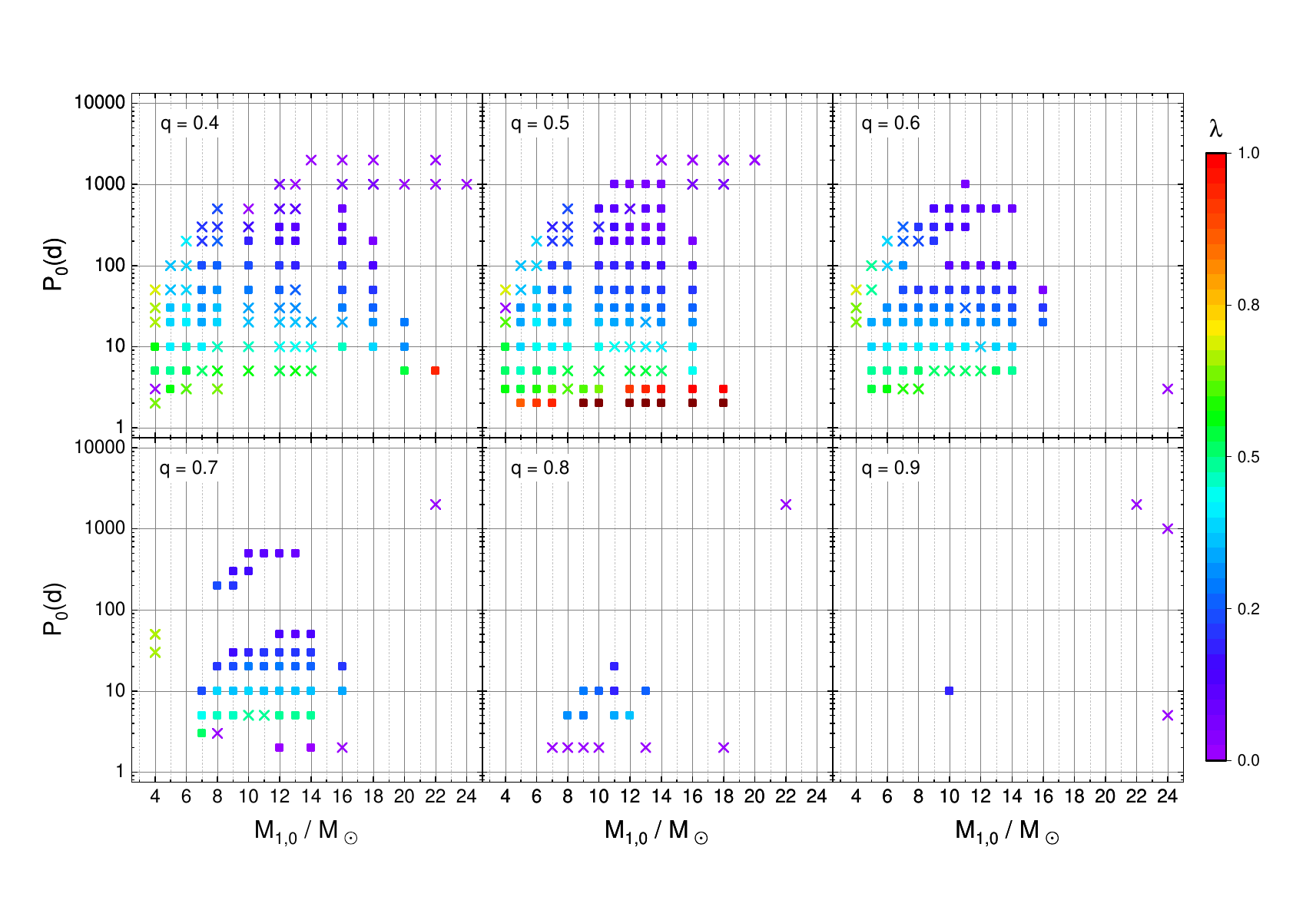} 
\caption{Envelope binding energy parameter $\lambda$ for the stars that plunge into common envelopes
due to dynamical mass loss (crosses) and due to impossibility to remove non-accreted matter from the system (squares).
}
\label{f:l_grid}
\end{figure}
\clearpage
\nopagebreak
   
\section*{\it {References}}
{\footnotesize
\begin{spacing}{1.0}
\begin{enumerate}[label=\arabic*.]

\item E.~Arancibia-Rojas, M.~Zorotovic, M.~Vu\v{c}kovi\'{c},
A.~Bobrick, J.~Vos, F.~Piraino-Cerda, ``The mass range of hot subdwarf B stars from MESA simulations'', \mnras~\textbf{527}, 11184 (2024).

\item R.S.~Benson, Ph.D. Thesis, U. California, Berkeley (1970).

\item E.~B\"{o}hm-Vitense, ``\"{U}ber die Wasserstoffkonvektionszone in Sternen verschiedener Effektivtemperaturen und Leuchtkr\"{a}fte. Mit 5 Textabbildungen'', \zap~\textbf{46}, 108 (1958).

\item G.S.~Bisnovatyi-Kogan, ``A self-consistent solution for an accretion disc structure around a rapidly rotating non-magnetized star'', \aap~\textbf{274}, 796 (1993).

\item L.~Blomberg, K.~El-Badry, B.~Ludwig, M.R.~Drout,
Y.~G\"{o}tberg, ``Intermediate-mass Stripped Stars in the Magellanic Clouds: Forward Modeling the Observed Population Discovered Via UV Excess'', \pasp~\textbf{138}, 024202 (2026).

\item D.~Boubert, N.W.~Evans, ``On the kinematics of a runaway Be star population'', \mnras~\textbf{477}, 5261 (2018).

\item L.~Wang, D.R.~Gies, G.J.~Peters, Z.~Han, ``The Orbital and Physical Properties of Five Southern Be+sdO Binary Systems'', Astron. J. \textbf{165}, 203 (2023).

\item (G.-Y.~Wang, Y.~Shao, J.-G.~He, Y.-D.~Nie, X.-J.~Xu, X.-D.~Li, ``A Rare Population of Intermediate-mass Helium Stars between Hot Subdwarfs and Wolf-Rayet Stars'', \apj~\textbf{1004}, 34 (2026).

\item L.M.~van Haaften, G.~Nelemans, R.~Voss, S.~Toonen, S.F.~Portegies Zwart, L.R.~Yungelson, M.V.~van der Sluys, ``Population synthesis of ultracompact X-ray binaries in the Galactic bulge'', Astron. Astrophys. \textbf{552}, A69 (2013).

\item L.B.F.M.~Waters, O.R.~Pols, S.J.~Hogeveen, J. Cote, E.P.J. van den Heuvel,
``The formation and detectability of Be + white dwarf systems'', \aap~\textbf{220}, L1 (1989).  

\item R.F.~Webbink, ``Double white dwarfs as progenitors of R Coronae Borealis stars and type I supernovae'', \apj~\textbf{277}, 355 (1984).

\item (R.F.~Webbink, ``Common Envelope Evolution Redux'', Astrophys. Space Sci. Library \textbf{352}, 233 (2008).

\item J.I.~Villase\~{n}or, D.J.~Lennon, A.~Picco, T.~Shenar, P.~Marchant, N.~Langer, P.L.~Dufton, F.~Nardini, et al., ``The B-type Binaries Characterisation Programme --- II. VFTS 291: a stripped star from a recent mass transfer phase'', MNRAS \textbf{525}, 5121 (2023).

\item J.I.~Villase\~{n}or, H.~Sana, L.~Mahy, T.~Shenar, J.~Bodensteiner, N.~Britavskiy, D.J.~Lennon, M.~Moe, et al., ``Binarity at LOw Metallicity (BLOeM): Enhanced multiplicity of early B-type dwarfs and giants at Z = 0.2 Z$_\odot$'', Astron. Astrophys. \textbf{698}, A41 (2025).

\item S.~Vinciguerra, C.J.~Neijssel, A.~Vigna-G\'{o}mez, I.~Mandel,
Ph.~Podsiadlowski, T.J.~Maccarone, M.~Nicholl, S.~Kingdon, et al., ``Be X-ray binaries in the SMC as indicators of mass-transfer efficiency'', \mnras~\textbf{498}, 4705 (2020).

\item A.~Vigna-G\'{o}mez, M.~Wassink, J.~Klencki, A.~Istrate,
G.~Nelemans, I.~Mandel, ``Stellar response after stripping as a model for common-envelope outcomes'', \mnras~\textbf{511}, 2326 (2022).

\item S.C.~Wu, J.~Fuller, ``A Diversity of Wave-driven Presupernova Outbursts'', \apj~\textbf{906}, 3 (2021). 
\item S.C.~Wu, J.~Fuller, ``Extreme Mass Loss in Low-mass Type Ib/c Supernova Progenitors'', \apj~\textbf{940}, L27 (2022).

\item Т.E.~Woods, N.~Ivanova, ``Can We Trust Models for Adiabatic Mass Loss?'', \apj~\textbf{739}, L48 (2011).

\item S.E.~Woosley, ``The Evolution of Massive Helium Stars, Including Mass Loss'', \apj~\textbf{878}, 49 (2019).

\item I.A.~Gabitova, A.C.~Carciofi, T.H.~de Amorim, M.~Suffak, A.S.~Miroshnichenko, S.V.~Zharikov, A.C.~Rubio, S.~Danford, ``dynamical Masses and Radiative Transfer Modeling of HD 698: A Be Binary in Evolutionary Transition'', Astrophys. J. \textbf{995}, 180 (2025).


\item Y.~G\"{o}tberg, S.E.~de Mink, J.H.~Groh, T.~Kupfer, P.A.~Crowther, E.~Zapartas, M.~Renzo, ``Spectral models for binary products: Unifying subdwarfs and Wolf-Rayet stars as a sequence of stripped-envelope stars'', Astron. Astrophys. \textbf{615}, A78 (2018).

\item Y.~G\"{o}tberg, M.R.~Drout, A.P.~Ji, J.H.~Groh, B.A.~Ludwig, P.A.~Crowther,
   N.~Smith, A.~de Koter, et al., ``Stellar Properties of Observed Stars Stripped in Binaries in the Magellanic Clouds'', Astrophys. J. \textbf{959}, 125 (2023).

\item H.~Ge, R.F.~Webbink, X.~Chen, Z.~Han, ``VizieR Online Data Catalog: Adiabatic Mass Loss in Binary Stars. III'', Astrophys. J. \textbf{899}, 132
(2020).

\item J.D.M.~Dewi, T.M.~Tauris, ``On the energy equation and efficiency parameter of the common envelope evolution'', \aap~\textbf{360}, 1043 (2000).

\item M.~de Kool, ``Common Envelope Evolution and Double Cores of Planetary Nebulae'', \apj~\textbf{358}, 189 (1990).

\item O.~De Marco, J.-C.~Passy, M.~Moe, F.~Herwig, M.~Mac Low, B.~Paxton, ``On the $\alpha$ formalism for the common envelope interaction'', \mnras~\textbf{411}, 2277 (2011).

\item M.S.~Hjellming, R.E.~Taam, ``The response of main-sequence stars within a common envelope'', \apj~\textbf{370}, 709 (1991).

\item A.S.~Jermyn, E.B.~Bauer, J.~Schwab, R.~Farmer, W.H.~Ball, E.P.~Bellinger, A.~Dotter, M.~Joyce, et al., ``Modules for Experiments in Stellar Astrophysics (MESA): Time-dependent Convection, Energy Conservation, Automatic Differentiation, and Infrastructure'', Astrophys. J. Suppl. Ser. \textbf{265}, 15 (2023).

\item H.~Jin, N.~Langer ``Chemical fingerprints of binary mass 
transfer in massive stars'', arxiv:2606.11940v1 (2026).

\item H.~Jin, N.~Langer, A.~Ercolino, S.E.~de Mink, ``A comprehensive grid of massive binary evolution models for the Galaxy: Surface properties of post-mass-transfer stars'', Astron. Astrophys. 	\textbf{707}, A56 (2026).

\item C.~Johnston, ``One size does not fit all: Evidence for a range of mixing efficiencies in stellar evolution calculations'', \aap~\textbf{655}, A29 (2021).

\item C.~Johnston, M.~Michielsen, E.H.~Anders, M.~Renzo, M.~Cantiello, P.~Marchant, A.~Goldberg, R.H.~Townsend, ``Modelling Time-dependent Convective Penetration in 1D Stellar Evolution'', Astrophys. J. \textbf{964}, 170 (2024).

\item M.R.~Drout, Y.~G\"{o}tberg, B.A.~Ludwig, J.H.~Groh, S.E.~de Mink, A.J.G.~O'Grady, N.~Smith, ``An observed population of intermediate-mass helium stars that have been stripped in binaries'', Science \textbf{382}, 1287 (2023).

\item K.~Dsilva, T.~Shenar, H.~Sana, P.~Marchant, {\it Massive Stars Near and Far, IAU Symp. 361} (Ed. J.~Mackey, J.S.~Vink and N.~St-Louis: Cambridge U. Press, 2024), p.~496.

\item D.~Dutta, J.~Klencki, ``Evolutionary nature of puffed-up stripped star binaries and their occurrence in stellar populations'', Astron. Astrophys. \textbf{687}, A215 (2024).

\item S.-C.~Yoon, ``Towards a better understanding of the evolution of Wolf-Rayet stars and Type Ib/Ic supernova progenitors'', \mnras~\textbf{470}, 3970 (2017).

\item S.-C.~Yoon, N.~Langer, C.~Norman, ``Single star progenitors of long gamma-ray bursts. I. Model grids and redshift dependent GRB rate'', \aap~\textbf{460}, 199 (2006).

\item S.-C.~Yoon, L.~Dessart, A.~Clocchiatti, ``Type Ib and IIb Supernova Progenitors in Interacting Binary Systems'', \apj~\textbf{840}, 10 (2017).

\item Е.~Zapartas, S.E.~de Mink, S.D.~Van Dyk, O.D.~Fox, N.~Smith, K.A.~Bostroem,
A.~de Koter, A.V.~Filippenko,  ``Predicting the Presence of Companions for Stripped-envelope Supernovae: The Case of the Broad-lined Type Ic SN 2002ap'', \apj~\textbf{842}, 125 (2017). 

\item J.~Zorec, Y.~Fr\'{e}mat, A.~Domiciano de Souza, F.~Royer,
L.~Cidale, A.-M.~Hubert, T.~Semaan, C.~Martayan, et al., ``Critical study of the distribution of rotational velocities of Be stars. I: Deconvolution methods, effects due to gravity darkening, 
macro-turbulence and binarity'', \aap~\textbf{595}, A132 (2016).

\item M.~Zorotovic, M.-R.~Schreiber, B.T.~G\"{a}nsicke,
A.~Nebot G\'{o}mez-Mor\'{a}n, ``Post-common-envelope binaries from SDSS. IX: Constraining the common-envelope efficiency'', \aap~\textbf{520}, A86 (2010).

\item I.~Iben~Jr., A.V.~Tutukov, ``On the evolution of close binaries with components of initial mass between 3 M and 12 M'', Astrophys. J. Suppl. Ser. \textbf{58}, 661 (1985).

\item I.~Iben~Jr., A.V.~Tutukov, ``Evolutionary Scenarios for Intermediate-Mass Stars in Close Binaries'', Astrophys. J. \textbf{313}, 727 (1987).

\item N.~Ivanov), ``Common Envelope: On the Mass and the Fate of the Remnant'', \apj~\textbf{730}, 76 (2011).

\item N.~Ivanova, S.~Justham, S.~Chen, O.~De~Marco, C.L.~Fryer, E.~Gaburov, H.~Ge, E.~Glebbeek, et al., ``Common envelope evolution: where we stand and how we can move forward'', Astron. Astrophys. Rev. \textbf{21}, 59 (2013).

\item N.~Ivanova, S.~Justham, P.~Ricker, Common Envelope Evolution, IOP Publishing: AAS-IOP Astronomy Book Series (2020).

\item J.~Isern, R.~Canal, R.J.~Labay, ``The Outcome of Explosive Ignition of ONeMg Cores: Supernovae, Neutron Stars, or ``Iron'' White Dwarfs?'', \apj~\textbf{372}, L83 (1991).

\item А.~Irrgang, N.~Przybilla, G.~Meynet, ``$\gamma$ Columbae as a recently stripped pulsating core of a massive star'',  
Nat. Astronomy~\textbf{6}, 1414 (2022). 

\item M.~Cantiello, N.~Langer, ``Thermohaline mixing in evolved low-mass stars'', \aap~\textbf{521}, A9 (2010).

\item P.D.~Kiel, J.R.~Hurley, ``Populating the Galaxy with low-mass X-ray binaries'', \mnras~\textbf{369}, 1152 (2006).

\item R.~Kippenhahn, A.~Weigert, ``Entwicklung in engen Doppelsternsystemen I. Massenaustausch vor und nach Beendigung des zentralen Wasserstoff-Brennens'', \zap~\textbf{65}, 251 (1967).

\item R.~Klement, T.~Rivinius, D.R.~Gies, D.~Baade, A.~M\'erand, J.D.~Monnier, G.H.~Schaefer,
   C.~Lanthermann, et al., ``The CHARA Array Interferometric Program on the Multiplicity of Classical Be Stars: New Detections and Orbits of Stripped Subdwarf Companions'', Astrophys. J \textbf{962}, 70 (2024).

\item J.~Klencki, G.~Nelemans, A.G.~Istrate, O.~Pols, ``Massive donors in interacting binaries: effect of metallicity'', A\&A \textbf{638}, A55 (2020).

\item T.~Ko, D.~Tsuna, Y.~Takei, T.~Shigeyama, ``Eruption of the Envelope of Massive Stars by Energy Injection with Finite Duration'', \apj~\textbf{930}, 168 (2022).

\item U.~Kolb, H.~Ritter, ``A comparative study of the evolution of a close binary using a standard and an improved technique for computing mass transfer'', \aap~\textbf{236}, 385 (1990).

\item M.~Colpi, M.~Nannurelli, M.~Calvani, ``An analytical model for a Keplerian disc and a boundary layer around a rapidly rotating star'', \mnras~\textbf{253} (1), 55 (1991).

\item P.~Kroupa, C.A.~Tout, G.~Gilmore, ``The Distribution of Low-Mass Stars in the Galactic Disc'', \mnras~\textbf{262}, 545 (1993).

\item P.~Kroupa, ``On the variation of the initial mass function'', \mnras~\textbf{322}, 231 (2001).

\item J.~Krti\v{c}ka, J.~Kub\'{a}t, I.~Krti\v{c}kov\'{a}, ``Stellar wind models of subluminous hot stars'', \aap~\textbf{593}, A101 (2016).

\item J.~Labadie-Bartz, M.~Suffak, C.~Jones,
Y.~Naz\'{e}, K.~Gayley, G.~Peters, R.~Rast, A.~Ravikumar, et al., ``Exploring the binary origin of B and Be rapid rotators'', \apss~\textbf{370} (12), 134 (2025).

\item N.~Langer, K.J.~Fricke, D.~Sugimoto, ``Semiconvective diffusion and energy transport'', \aap~\textbf{126} (1), 207 (1983).

 \item N.~Langer, ``Coupled mass and angular momentum loss of massive main sequence stars'', \aap~\textbf{329}, 551 (1998).

\item D.~Lauterborn, ``Evolution with mass exchange of case C for a binary system of total mass 7 M sun'', Astron. Astrophys. \textbf{7}, 150 (1970).

 \item T.~Lechien, S.E.~de Mink, R.~Valli, A.C.~Rubio, L.A.C.~van Son, R.~Klement, H.~Jin, O.~Pols, ``Binary Stars Take What They Get: Evidence for Efficient Mass Transfer from Stripped Stars with Rapidly Rotating Companions'', Astrophys. J. Lett. 	\textbf{990}, L51 (2025).

\item Z.~Lei, R.~He, P.~N\'{e}meth, X.~Zou, H.~Xiao, Y.~Yang, J.~Zhao, ``Mass Distribution for Single-lined Hot Subdwarf Stars in LAMOST'', \apj~\textbf{953}, 122 (2023).

\item D.J.~Lennon, J.~Ma\'{i}z Apell\'{a}niz, A.~Irrgang, R.~Bohlin, S.~Deustua, P.L.~Dufton,
   S.~Sim\'{o}n-D\'{i}az, A.~Herrero, et al., ``Hubble spectroscopy of LB-1: Comparison with B+black-hole and Be+stripped-star models'', Astron. Astrophys. \textbf{649}, A167 (2021).

\item Lubow, F.H.~Shu, ``Gas dynamic of semidetached binaries'', Astrophys. J. \textbf{198}, 383 (1975).

\item B.~Ludwig, M.R.~Drout, Y.~G\"{o}tberg, et al., ``The Stripped-star Ultraviolet Magellanic Cloud Survey (SUMS): The Ultraviolet Photometric Catalog and Stripped-star Candidate Selection'', \apj~\textbf{999}, 73 (2026).

\item P.~Marchant, Ph.D. Thesis, Rheinische Friedrich Wilhelms University of Bonn, (2017).

\item T.~Matsuoka, R.~Sawada, ``Binary Interaction Can Yield a Diversity of Circumstellar Media around Type II Supernova Progenitors'', \apj~\textbf{963}, 105 (2024). 

\item J.G.~Mengel, J.~Norris, P.G.~Gross, ``Binary Hypothesis for the Subdwarf B Stars'', \apj~\textbf{204}, 488 (1976).

\item N.~Mennekens, D.~Vanbeveren, ``A comparison between observed Algol-type double stars in the solar neighborhood and evolutionary computations of galactic case A binaries with a B-type primary at birth'', \aap~\textbf{599}, A84 (2017).

\item S.~Miyaji, K.~Nomoto, K.~Yokoi, D.~Sugimoto, ``Supernova Triggered by Electron Captures'', \pasj~\textbf{32}, 303 (1980).

\item D.C.~Morton, ``Evolutionary Mass Exchange in Close Binary Systems'', \apj \ \textbf{132}, 146 (1960).

\item D.~Mourard, J.D.~Monnier, A.~Meilland, D.~Gies, F.~Millour,
M.~Benisty, X.~Che, E.D.~Grundstrom, et al., ``Spectral and spatial imaging of the Be+sdO binary $\phi$ Persei'', Astron. Astrophys. \textbf{577}, A51 (2015).

\item J.~M\"{u}ller-Horn, K.~El-Badry, A.A.C.~Sander, et al., ``A Galactic intermediate-mass stripped star with a Wolf-Rayet-like wind '', arXiv:2608.05276 (2026a).

\item J.~M\"{u}ller-Horn, V.~Ramachandran, K.~El-Badry, A.A.C.~Sander, J.~Bodensteiner, D.R.~Gies, Y.~G\"{o}tberg, Th.~Rivinius, ``Ultraviolet spectroscopy reveals a hot and luminous companion to the Be star+black hole candidate MWC 656'', Astron. Astrophys. \textbf{708},
A187 (2026b).

\item  Y.~Naz\'{e} ``Going Forward to Unveil the Nature of $\gamma$~Cas Analogs'',
Galaxies~\textbf{13}, 8 (2025).

\item Y.~Naz\'{e}, G.~Rauw, P.A.~Ko\l{l}aczek-Szyma\'{n}ski, N.~Britavskiy,
J.~Labadie-Bartz, ``A family of binaries with an extreme mass ratio'', \aap~\textbf{703}, A239 (2025).

\item T.~Nugis, H.J.G.L.M.~Lamers, ``Mass-loss rates of Wolf-Rayet stars as a function of stellar parameters'', \aap~\textbf{360}, 227 (2000).

\item H.~Nieuwenhuijzen and C.~de~Jager, ``Parametrization of stellar rates of mass loss as functions of the fundamental stellar parameters M, L, and R'', \aap~\textbf{231}, 134 (1990).

\item S.~Nedhath, S.~Rani, A.~Subramaniam, E.~Pancino, ``AstroSat/UVIT Study of NGC 663: First detection of Be+sdOB systems in a young star cluster'', \aap~\textbf{699}, L1 (2025).

\item R.~Ouchi, K.~Maeda, ``Radii and Mass-loss Rates of Type IIb Supernova Progenitors'', \apj~\textbf{840}, 90 (2017).

\item W.~Packet, ``On the spin-up of the mass accreting component in a close binary system'', Astron. Astrophys. \textbf{102}, 17 (1981).

\item B.~Paxton, L.~Bildsten, A.~Dotter, F.~Herwig, P.~Lesaffre, F.~Timmes, ``Modules for Experiments in Stellar Astrophysics (MESA)'',
  \apjs~\textbf{192}, 3 (2011).

\item B.~Paxton, M.~Cantiello, P.~Arras, L.~Bildsten, E.F.~Brown, A.~Dotter, C.~Mankovich, M.H.~Montgomery, et al., ``Modules for Experiments in Stellar Astrophysics (MESA): Planets, Oscillations, Rotation, and Massive Stars'', \apjs~\textbf{208}, 4 (2013).

\item B.~Paxton, P.~Marchant, J.~Schwab,
E.B.~Bauer, L.~Bildsten, M.~Cantiello, L.~Dessart, R.~Farmer, et al., ``Modules for Experiments in Stellar Astrophysics (MESA): Binaries, Pulsations, and Explosions'', \apjs~\textbf{220}, 15 (2015).

\item B.~Paxton, J.~Schwab, E.B.~Bauer, L.~Bildsten, S.~Blinnikov, P.~Duffell, R.~Farmer, J.A.~Goldberg, et al., ``Modules for Experiments in Stellar Astrophysics (MESA): Convective Boundaries, Element Diffusion, and Massive Star Explosions'', \apjs~\textbf{234}, 34 (2018).

\item B.~Paxton, R.~Smolec, J.~Schwab, A.~Gautschy, L.~Bildsten, M.~Cantiello, A.~Dotter, R.~Farmer, et al., ``Modules for Experiments in Stellar Astrophysics (MESA): Pulsating Variable Stars, Rotation, Convective Boundaries, and Energy Conservation'', \apjs~\textbf{243}, 10 (2019).

\item L.~Pastetter and H.~Ritter, ``Evolution of close binary systems that undergo a dynamicaly stable late case C mass transfer'', \aap~\textbf{214}, 186 (1989).

\item B.~Paczy\'{n}ski, ``Evolution of Close Binaries. V. The Evolution of Massive Binaries and the Formation of the Wolf-Rayet Stars'', Acta Astron. \textbf{17}, 355 (1967).

\item B.~Paczy\'{n}ski, ``Evolution of Single Stars. II. Core Helium Burning in Population I Stars'', \actaa~\textbf{20}, 195 (1970).

\item B.~Paczy\'{n}ski, {\it Structure and Evolution of Close Binary Systems,
IAU Symp. 73} (Ed. P.~Eggleton, S.~Mitton, and J.~Whelan, Dordrecht: D. Reidel Publ. Co., 1976), p.~75.

\item B.~Paczy\'{n}ski, ``A Polytropic Model of an Accretion Disk, a Boundary Layer, and a Star'', \apj~\textbf{370}, 597 (1991).

\item B.~Paczy\'{n}ski and R.~Sienkiewicz, ``Evolution of Close Binaries VIII. Mass Exchange on the dynamic Time Scale'', Acta Astron. \textbf{22}, 73 (1972).

 \item A.~Picco P.~Marchant, D.~Pauli, H.~Sana, ``Mergers via failed common envelope as a route toward intermediate-mass stripped stars'', \aap~\textbf{710}, L12 (2026).

\item O.R.~Pols, J.~Cote, L.B.F.M.~Waters, J.~Heise, ``The formation of Be stars through close binary evolution'', \aap~\textbf{241}, 419 (1991).

\item R.~Popham, R.~Narayan, ``Does Accretion Cease When a Star Approaches Breakup?'', \apj~\textbf{370}, 604 (1991).

\item S.F.~Portegies Zwart, L.R.~Yungelson, ``Formation and evolution of binary neutron stars'', \aap~\textbf{332}, 173 (1998).

\item V.~Ramachandran, J.~Klencki, A.A.C.~Sander, D.~Pauli, T.~Shenar,
   L.M.~Oskinova, W.-R.~Hamann, ``A partially stripped massive star in a Be binary at low metallicity. A missing link towards Be X-ray binaries and double neutron star mergers'', Astron. Astrophys. \textbf{674}, L12 (2023).

\item V.~Ramachandran, A.A.C.~Sander, D.~Pauli, J.~Klencki, F.~Backs, F.~Tramper,
   M.~Bernini-Peron, P.~Crowther, et al., ``X-Shooting ULLYSES: Massive stars at low metallicity: VIII. Stellar and wind parameters of newly revealed stripped stars in Be binaries'', Astron. Astrophys. \textbf{692}, A90 (2024).

\item M.~Renzo, C.D.~Ott, S.N.~Shore, S.E.~de~Mink, ``Systematic survey of the effects of wind mass loss algorithms on the evolution of single massive stars'', Astron. Astrophys.
\textbf{603}, A118 (2017).

\item P.~Reig, ``Be/X-Ray Binaries: Phenomenology, Variability, and Accretion 
dynamic'', Universe \textbf{12}, 7, 201 (2026). 

\item F.K.~R\"{o}pke, O.~De~Marco, ``Simulations of common-envelope evolution in binary
stellar systems: physical models and numerical techniques'', Liv. Rev. Comp. Astrophys.
\textbf{9}, 2 (2023).

\item G.~Rate, P.A.~Crowther, ``Unlocking Galactic Wolf-Rayet stars with Gaia DR2 --- I. Distances and absolute magnitudes'', \mnras~\textbf{493}, 1512 (2020).

\item H.~Sana, S.E.~de~Mink, A.~de Koter, N.~Langer, C.J.~Evans, M.~Gieles, E.~Gosset, R.G.~Izzard, et al., ``Binary Interaction Dominates the Evolution of Massive Stars'', Science \textbf{337}, 444 (2012).

\item P.~W.~A. Roming, T.~E. Kennedy,  K.~O. Mason, J.~ A. Nousek, L. Ahr, R.~E. Bingham, P.~S. Broos, 
M. Carter, et al., ``The Swift Ultra-Violet/Optical Telescope'' Space Sci. Rev. \textbf{120}, 95 (2005).

\item H.~Sana, T.~Shenar, J.~Bodensteiner, N.~Britavskiy, N.~Langer, D.J.~Lennon, L.~Mahy, I.~Mandel, et al., ``A high fraction of close massive binary stars at low metallicity'', Nature Astronomy \textbf{9}, 1337 (2025).

\item K.~Sen, M.~Renzo, H.~Jin, N.~Langer, A.~Schootemeijer, J.I.~Villaseñor, L.~Mahy, A.~Grichener, ``Interacting Binaries on the Main Sequence as In Situ Tracers of Mass-transfer Efficiency and Stability'', Astrophys. J. \textbf{1000}, 2 (2026).

\item Z.~Xing, T.~Fragos, V.~Kalogera, S.~Gossage, K.A.~Rocha, E.~Zapartas), ``Disk-regulated Mass Transfer between Rotating Nondegenerate Stars: Insights from Be and sdOB Binaries'',
Astrophys. J. 	\textbf{1004}, 94 (2026).

\item A.~Schootemeijer, Y.~G\"{o}tberg, S.E.~de~Mink, D.~Gies, E.~Zapartas, ``Clues about the scarcity of stripped-envelope stars from the evolutionary state of the sdO+Be binary system $\phi$ Persei'', Astron. Astrophys. \textbf{615}, A30 (2018).

\item H.C.~Spruit, ``Dynamo action by differential rotation in a stably stratified stellar interior'', \aap~\textbf{381}, 923 (2002).

\item X.-T.~Xu, С.~Sch\"{u}rmann, N.~Langer, C.~Wang, A.~Schootemeijer, T.~Shenar, A.~Ercolino, F.~Haberl, ``Populations of evolved massive binary stars in the Small Magellanic Cloud: I. Predictions from detailed evolution models'', \aap~\textbf{704}, A218 (2025).

\item Т.M.~Tauris, N.~Langer, T.J.~Moriya, Ph.~Podsiadlowski, S.-C.~Yoon,
S.I.~Blinnikov, ``Ultra-stripped Type Ic Supernovae from Close Binary Evolution'', \apj \ \textbf{778}, L23 (2013).

\item K.D.~Temmink, O.~Pols, S.~Justham, A.G.~Istrate, S.S.~Toonen, ``Coping with loss. Stability of mass transfer from post-main-sequence donor stars'', \aap \ \textbf{669}, A45 (2023).

\item K.D.~Temmink, S.~Justham, O.~Pols, ``One's loss is (not) another's gain: Isotropic re-emission destabilises mass transfer from radiative donor stars'', \aap~\textbf{709}, L6 (2026).

\item S.H.~Tsai, K.-J.~Chen, K.~Maeda, P.-S.~Ou, F.K.~R\"{o}pke, ``Interacting Binary Stars as Progenitors for Interacting Supernovae'', ApJL~\textbf{1005}, L36 (2026).

\item A.~Tutukov, L.~Yungelson, ``Evolution of massive close binaries'', Nauchnye Informatsii \textbf{27}, 70 (1973), in Russian.

\item А.~Tutukov, L.~Iungelson, ``On the origin of faint blue stars'', {\it Second Conference on Faint Blue Stars, IAU Colloquium 95} (Schenectady, NY: L. Davis Press, Inc., 1987), p.~435 (1987)  

\item Yu.~A. Fadeyev et al. ``Elusive Helium Stars in the Gap between Subdwarfs and Wolf-Rayet Stars II. Nonlinear Pulsations of Stripped Helium Stars'', 
Astron. Lett.~\textbf{51}, 25 (2025).

\item J.-S.~Vink, A.~de~Koter, H.J.G.L.M.~Lamers, ``Mass-loss predictions for O and B stars as a function of metallicity'', \aap~\textbf{369}, 574 (2001).

\item J.S.~Vink, ``Winds from stripped low-mass helium stars and Wolf-Rayet stars'', Astron. Astrophys. \textbf{607}, L8 (2017).

\item T.~Fragos, J.J.~Andrews, S.S.~Bavera, C.P.L.~Berry, S.~Coughlin, A.~Dotter, P.~Giri, V.~Kalogera, et al., ``POSYDON: A General-purpose Population Synthesis Code with Detailed Binary-evolution Simulations'', \apjs~\textbf{264}, 45 (2023).

\item J.~Fuller, ``Pre-supernova outbursts via wave heating in massive stars --- I. Red supergiants'', \mnras~\textbf{470}, 1642 (2017).

\item G.M.H.J.~Habets, ``The evolution of a single and a binary helium star of 2.5 solar mass up to neon ignition'', Astron. Astrophys. \textbf{165}, 95 (1986).

\item S.~Chanlaridis, J.~Antoniadis, D.R.~Aguilera-Dena, G.~Gräfener, N.~Langer, N.~Stergioulas, ``Thermonuclear and electron-capture supernovae from stripped-envelope stars'', \aap~\textbf{668}, A106 (2022).

\item J.R.~Hurley, O.R.~Pols, C.A.~Tout, ``Comprehensive analytic formulae for stellar evolution as a function of mass and metallicity'', \mnras~\textbf{315}, 543 (2000).

\item J.R.~Hurley, C.A.~Tout, O.R.~Pols, ``Evolution of binary stars and the effect of tides on binary populations'', \mnras~\textbf{329}, 897 (2002).

\item B.~Hastings, C.~Wang, N.~Langer, ``Stringent upper limit on Be star fractions produced by binary interaction'', \aap, 633, A165 (2020). 

\item B.~Hastings, N.~Langer, C.~Wang, A.~Schootemeijer, A.P.~Milone), ``The single star path to Be stars'', \aap~\textbf{653}, A144 (2021).

\item U.~Heber, ``Hot Subluminous Stars'', \pasp~\textbf{128}, 082001 (2016).

\item A.~Heger, N.~Langer, S.E.~Woosley, ``Presupernova Evolution of Rotating Massive Stars. I. Numerical Method and Evolution of the Internal Stellar Structure'', \apj~\textbf{528}, 368 (2000).

\item A.~Heger, S.E.~Woosley, H.C.~Spruit, ``Presupernova Evolution of Differentially Rotating Massive Stars Including Magnetic Fields'', \apj~\textbf{626}, 350 (2005).

\item J.~Henneco, F.R.N.~Schneider, E.~Laplace, ``Contact tracing of binary stars: Pathways to stellar mergers'', \aap~\textbf{682}, A169 (2024).

\item B.~Hovis-Afflerbach, Y.~G\"{o}tberg, A.~Schootemeijer, J.~Klencki, A.L.~Strom,
   B.A.~Ludwig, M.R.~Drout, ``The mass distribution of stars stripped in binaries: The effect of metallicity'', Astron. Astrophys. \textbf{697}, A239 (2025).

\item A.~Holas, S.W.~Jones, F.K.~R\"{o}pke,  R.~Pakmor, C.~Fakiola, G.~Leidi, R.~Hirschi, K.J.~Shen, ``Drawing the line between explosion and collapse in electron-capture supernovae: I. Impact of conductive flame speeds and ignition conditions on the explosion mechanism'', Astron. Astrophys. \textbf{707}, A84 (2026).

\item A.M.~Cherepashchuk {\it Close binary stars}, (M.: Physmatlit, 2013), v. 2 (in Russian).

\item L.~Chomiuk, M.S.~Povich, ``Toward a Unification of Star Formation Rate Determinations in the Milky Way and Other Galaxies'', Astron. J. \textbf{142}, 197 (2011).

\item B.~Chaboyer, J.-P.~Zahn, ``Effect of horizontal turbulent diffusion on transport by meridional circulation'', \aap~\textbf{253}, 173 (1992).

\item Y.~Shao and X.-D.~Li, ``Population Synthesis of Galactic Be-star Binaries with a Helium-star Companion'', Astrophys. J. \textbf{908}, 67 (2021).

\item M.M.~Shara, A.F.J.~Moffat, J.~Gerke, D.~Zurek, K.~Stanonik, R.~Doyon, E.~Artigau, L.~Drissen, ``A Near-Infrared Survey of the Inner Galactic Plane for Wolf-Rayet Stars. I. Methods and First Results: 41 New WR Stars'', \aj~\textbf{138}, 402 (2009).

\item V.~Schaffenroth, I.~Pelisoli, B.N.~Barlow, S.~Geier, T.~Kupfer, ``Hot subdwarfs in close binaries observed from space. I. Orbital, atmospheric, and absolute parameters, and the nature of their companions'', \aap~\textbf{666}, A182 (2022).

\item T.~Shenar, J.~Bodensteiner, M.~Abdul-Masih, M.~Fabry, L.~Mahy, P.~Marchant, G.~Banyard, D.M.~Bowman, et al.,  ``The ``hidden'' companion in LB-1 unveiled by spectral disentangling'', Astron. Astrophys. \textbf{639}, L6 (2020a).

\item T.~Shenar, A.~Gilkis, J.S.~Vink, H.~Sana, A.A.C.~Sander, ``The young massive SMC cluster NGC 330 seen by MUSE. I. Observations and stellar content'',
   Astron.~Astrophys. \textbf{634}, A51 (2020b).

\item J.~H. Shiode, E. Quataert, ``Setting the Stage for Circumstellar Interaction in Core-Collapse Supernovae. II. Wave-driven Mass Loss in Supernova Progenitors'', \apj~\textbf{780}, 96 (2014).

\item F.R.N.~Schneider, M.Y.~Lau, F.K.~R\"{o}pke, ``Stellar mergers and common-envelope evolution'', Encyclopedia of Astrophysics \textbf{2}, 349 (2026).

\item С.~Sch\"{u}rmann, N.~Langer, ``Exploring the boundary between stable mass transfer and L$_2$ overflow in close binary evolution'', \aap~\textbf{691}, A174 (2024).

\item C.~Sch\"{u}rmann, N.~Langer, X.~Xu, C.~Wang, ``The spins of stripped B stars support magnetic internal angular momentum transport'', \aap~\textbf{667}, A122 (2022).

\item C.~Sch\"{u}rmann, X.-T.~Xu, N.~Langer, D.~Lennon, M.U.~Kruckow, J.~Antoniadis, F.~Haberl, A.~Herrero, et al., ``Populations of evolved massive binary stars in the Small Magellanic Cloud: II. Predictions from rapid binary evolution'', Astron. Astrophys. \textbf{704}, A219 (2025).

\item K.~El-Badry, C.~Conroy, E.~Quataert, H.-W.~Rix, J.~Labadie-Bartz,
   T.~Jayasinghe, T.~Thompson, P.~Cargile, et al., ``Birth of a Be star: an APOGEE search for Be stars forming through binary mass transfer'', MNRAS \textbf{516}, 3602 (2022). 

\item E.~\"{O}pik, ``Statistical Studies of Double Stars: On the Distribution of Relative Luminosities and Distances of Double Stars in the Harvard Revised Photometry North of Declination $-31^\circ$'', Publ. Tartu Astrofizica Observatory \textbf{25}, 1 (1924).

\item A.~Ercolino, H.~Jin, N.~Langer, L.~Dessart, ``Mass-transferring binary stars as progenitors of interacting hydrogen-free supernovae'', Astron. Astrophys. \textbf{696}, A103 (2025).

\item L.R.~Yungelson, ``Evolution of the secondary component of a close binary system'', Nauchnye Informatsii \textbf{27}, 93 (1973), in Russian.

\item L.~Yungelson, A.~Kuranov, K.~Postnov, M.~Kuranova, L.M.~Oskinova, W.-R.~Hamann, ``Elusive hot stripped helium stars in the Galaxy. I. Evolutionary stellar models in the gap between subdwarfs and Wolf-Rayet stars'', Astron. Astrophys. \textbf{683}, A37 (2024).
\end{enumerate}
\end{spacing}
}

\label{lastpage}
\end{document}